\documentclass{emulateapj}

 \shorttitle{Comparing the SNe host galaxies}
 \shortauthors{Shao et al.}

\begin{document}

\title{Comparing the Host Galaxies of Type Ia, Type II and Type Ibc Supernovae}

\author{
 X. Shao\altaffilmark{1,2,3,5}, Y. C. Liang\altaffilmark{1}, M. Dennefeld\altaffilmark{4},
 X. Y. Chen\altaffilmark{1,3}, G. H. Zhong\altaffilmark{1,3}, F. Hammer\altaffilmark{5},
 L. C. Deng\altaffilmark{1}, H. Flores\altaffilmark{5}, B. Zhang\altaffilmark{1,2}, W. B. Shi\altaffilmark{1,6} and L. Zhou\altaffilmark{1,3,6}
} \email{xshao@bao.ac.cn,ycliang@bao.ac.cn}

\begin{abstract}
We compare the host galaxies of 902 supernovae, including 
SNe Ia, SNe II and SNe Ibc, which are selected by cross-matching 
the Asiago Supernova Catalog with the SDSS Data Release 7. 
We further selected 213 galaxies by requiring the 
light fraction of spectral observations $>$15\%, which could 
represent well the global properties of the galaxies. Among them, 135 galaxies appear on the
Baldwin-Phillips-Terlevich diagram, which allows us to compare 
the hosts in terms of star-forming, AGNs (including composites, LINERs and Seyfert 2s)
and ``Absorp" (their related emission-lines are weak or non-existence) galaxies. 
The diagrams related to parameters D$_n$(4000),
H$\delta_A$, stellar masses, SFRs and
specific SFRs for the SNe hosts show that almost all 
SNe II and most of SNe Ibc occur in SF galaxies, which have a wide 
range of stellar mass and low D$_n$(4000). The SNe Ia
hosts as SF galaxies follow similar trends. A significant fraction of SNe Ia occurs 
in AGNs and Absorp galaxies, which are massive and have high
D$_n$(4000). The stellar population analysis from
spectral synthesis fitting shows that the
hosts of SNe II have a younger stellar population than hosts of SNe Ia. 
These results are compared with those of 
the 689 comparison galaxies where the SDSS fiber 
captures less than 15\% of the total light. These comparison 
galaxies appear biased towards higher 12+log(O/H) ($\sim$0.1dex) at a given
stellar mass. Therefore, we believe the aperture effect should be
kept in mind when the properties of the hosts for different
types of SNe are discussed.
\end{abstract}


 \altaffiltext{1}{ Key Laboratory of Optical Astronomy, National Astronomical
  Observatories, Chinese Academy of Sciences, 20A Datun Road, Chaoyang District, Beijing, 100012,
  China}
\altaffiltext{2}{Department of Physics, Hebei Normal University,
Shijiazhuang 050016, China}
 \altaffiltext{3}{Graduate University
of the Chinese Academy of
  Sciences, 19A Yuquan Road, Shijingshan District, 100049, Beijing,
  China}
 \altaffiltext{4}{Institut d'Astrophysique de Paris, CNRS, and Universite P. et
M. Curie, 98bis Bd Arago, F-75014 Paris, France }
 \altaffiltext{5}{GEPI,Observatoire de Paris-Meudon, 92195 Meudon, France}
\altaffiltext{6}{Shandong Provincial Key Laboratory of Optical Astronomy
     and Solar-Terrestrial Environment, School of Space Science and
     Physics, Shandong University at Weihai, Weihai 264209, China}

\date{Received  / Accepted }

%
\section{Introduction}\label{Sect:intro}

Supernovae (SNe) are classified into various types (II, Ib, Ic and
Ia) according to the presence or absence of various features in
their spectra (\citet{1997ARA&A..35..309F} and reference therein). The presence
or absence of hydrogen distinguishes type II from type I SNe,
respectively. Among the type I's, the presence of Si lines
characterizes types Ia, while the presence of He lines distinguishes
types Ib from types Ic's (\citet{2002AJ....124..417H, Turatto2003}, and references therein).

SNe Ia are observed in all types of galaxies (ellipticals,
irregulars, spirals), suggesting that they are somehow connected to
the evolution of less-massive stars (e.g., \citet{1979AJ.....84..985O, 
1990PASP..102.1318V, 1994ApJ...423L..31D, 1999A&A...351..459C}). 
It is widely accepted that the progenitors of Ia SNe are
carbon-oxygen white dwarfs (CO WD's), which have accreted material up to
the Chandrasekhar limit \citep{1931ApJ....74...81C} from non-degenerate
companion stars) in a single-degenerate (SD) model \citep{1973ApJ...186.1007W, 
1982ApJ...253..798N}, or come from a double-degenerate (DD) model, which involves the merger of two CO WDs 
\citep{1984ApJS...54..335I, 1984ApJ...277..355W}.

However, type II, Ib and Ic SNe are only found in star-forming
galaxies, indicating that they are the product of evolution in
massive stars from the gravitational collapse of their Fe
cores. Thus, type II and  type Ibc (which include types Ib, Ic and
Ib/c) SNe are also called core collapse SNe (CC-SNe). CC-SNe are thought to
arise from stars with initial masses $>8M_{\odot}$: this value results from the
 agreement between direct detections of progenitors \citep{2009ARA&A..47...63S}, and the maximum observed mass for white dwarfs (WD's)
\citep{2009ApJ...693..355W, 2009MNRAS.399..559A}.

Many studies have tried to understand
 how various types of  supernovae behave,
including mass of progenitors, effect of environments and the
relations between the host properties and the SNe themselves etc.,
but these were generally only based on small samples (\citet{1995AJ....109....1H, 
1996AJ....112.2398H, 2000AJ....120.1479H, 
2005ApJ...634..210G, 2008ApJ...685..752G, 2009ApJ...707.1449N} for
SNe Ia; \citet{2009MNRAS.399..559A, 2010ApJ...717..342H} for
CCSN).

 The studied sample has been greatly extended in recent years, especially
benefitting from the successful projects associated with SDSS, both for galaxies
\citep{2002AJ....124.1810S, 2003MNRAS.346.1055K, 2003MNRAS.341...33K, 2004MNRAS.351.1151B, 2004ApJ...613..898T}
 and supernovae (e.g. the SDSS-II
Supernova Survey: \citet{2008AJ....135..338F, 2008AJ....135.1766Z, 2010MNRAS.401.2331L, 
2010ApJ...708..661D, 2009ApJ...704..687C}). The enlarged
sample makes it possible to carefully compare the properties of
 host galaxies of supernova explosions.  Some comparison studies have been made
that examine the differences between galaxies that host SNe Ia and CC-SNe, separately. \citet{2010ApJ...724..502H} 
investigated the color, luminosity and
environments of SNe Ia host galaxies in Stripe 82 of the SDSS-II
Supernova Survey centered on the celestial equator. \citet{2012ApJ...759..107K} 
examined the host galaxies of core-collapse supernovae where
they separately inspected colors at the sites of the explosions,
the chemical abundances and specific star formation rates for hosts of SNe II,
SNe IIn, SNe IIb, SNe Ib and SNe Ic.

However, there are few works that combine SNe Ia and CC-SNe together to compare their hosts. This is interesting
and useful for understanding the environments where SNe explode,
especially because some SNe Ia hosts are also star-forming
galaxies, like those for SNe II and SNe Ibc. Some researches have even tried
to do so, but they have only focused on some limited aspects of the
properties, for example, \citet{2008ApJ...673..999P} mostly discussed the
metallicities of galaxies, \citet{2012A&A...544A..81H} mostly reported the creation of their database from SDSS-DR8
and presented some measurements from images.
Therefore, many more comparisons are needed for studying SNe host
galaxies, such as stellar populations, stellar mass and star formation rates (SFRs) and so
on. These properties are very important for understanding the
characteristics of supernova hosts.

In this work, we take into account SNe Ia, SNe II and SNe Ibc (the later two as CC-SNe) together to
compare the properties of their hosts. In particular,
we will use a stricter selection criterion to select the objects,
for which  the 3 arcsec fiber spectra of SDSS can represent the global
properties of the galaxies better. We believe it is important to show the global properties of supernova host galaxies
since it is often difficult to acquire the local properties
 at the sites of an SN explosion.

Our idea can be summarized as follows: 1) Comparing the host
galaxies of all kinds of supernovae, including both Type Ia and
CC-SNe (SNe II and Ibc). 2) We will compare them in terms of parameters that describe many properties,
including stellar masses, star formation rates, specific
star formation rates, D$_n$(4000), H$\delta_A$ and gas-phase oxygen
abundances.
3) We also run spectral synthesis analysis for the optical spectra
and obtain the light weighted average ages for the host galaxies. 4)
The hosts  can be carefully checked following the classification from their high quality
emission-line ratios, which can diagnose the hosts on the BPT diagram \citep{1981PASP...93....5B}. Then  star forming (SF), AGN and Absorption line
galaxies acting as hosts can be compared in terms of these property relations. 5) Since the sample is large, it is
possible for us to manage a good sub-sample that can represent the global
properties of the host galaxies, for which the 3 arcsec SDSS fiber
observations can cover $>$15\% light. This minimizes
the cases that the fiber only record a small part of the global light
of the hosts. 6) Then we can also carefully discuss
the aperture effect of the SDSS fiber observations by comparing the
two sub-samples (the ones that have light fraction  $>$15\%  and the others that
have light fraction $\leq$15\%); this approach gives clearer results for stellar
mass-metallicity relations.
7) The properties of SNe hosts mentioned above will also be compared with the main sample of galaxies from SDSS-DR7.

This paper is organized as follows. We describe the sample selection
in Sect.~2, which demonstrates how we select the 213 sample galaxies with
light fraction higher than 15\% from the SDSS fiber observations,
and the selection for the comparative sample galaxies are also mentioned.
The parameters describing properties and their relations are shown in Sect.~3 for these
hosts. Results of the stellar population analysis are presented
in Sect.~4. Discussions are presented in Sect.~5, where we show the
results of comparisons with the 689 comparative sample galaxies, the stellar mass-metallicity relation and the
aperture effect/bias.
 Conclusions are given in Sect.~6. Throughout the paper, a
cosmological model with H$_0$ = 70 km s$^{-1}$ Mpc$^{-1}$,
$\Omega_M$= 0.3 and $\Omega_{\Lambda}$= 0.7 is adopted.

\section{Sample selection}\label{Sect:sample}

We select the SNe and their host galaxies by cross-matching the Asiago Supernova Catalog (ASC)
with the SDSS DR7 main galaxy sample (MGS), only retaining spectral observations of the SNe host galaxies with good
quality.

\subsection{The Asiago Supernova Catalog (ASC)}

The ASC was first published in 1984 by Barbon et al., who assembled
all the pertinent information on the 568 Supernovae (SNe) discovered
from 1885 up to 1983, as well as some  parameters associated with their host
galaxies. The catalog was subsequently updated for newly discovered SNe. \citet{1989A&AS...81..421B}, \citet{1999A&AS..139..531B} 
and their group made the following updates. Up to the end of 2013, the Asiago supernova catalog includes 6312
SNe up to SNe 2013hx (http://graspa.oapd.inaf.it).

In this work, we adopted the RA and DEC of the SNe host galaxies in
the ASC table, to enable cross-matching with the SDSS-DR7 MGS
galaxies.

\subsection{The SDSS main galaxy sample catalog (MGS)}

The SDSS is the most ambitious astronomical survey ever undertaken
in imaging and spectroscopy \citep{2000AJ....120.1579Y, 2002AJ....123..485S, 
2003AJ....126.2081A, 2004AJ....128..502A}. The imaging data are done in
drift scan mode and are 95\% complete for the surveyed area for point sources at 22.0,
22.2, 22.2, 21.3 and 20.5 in five bands ($u$, $g$, $r$, $i$ and
$z$) respectively. The spectra are flux- and wavelength-calibrated
from 3800 to 9200 \AA~at $R\approx1800$.

The sample used in this work is selected from the
SDSS-DR7 MGS \citep{2002AJ....124.1810S}, which comprises galaxies
with  $r$-band Petrosian magnitude 14.5$<$$r\leq$17.77 (corrected
for foreground Galactic extinction using the reddening maps of
\citet{1998ApJ...500..525S}) and $r$-band Petrosian half-light surface
brightness $\mu_{50}\leq$24.5 mag arcsec$^{-2}$.

The parameters of the galaxies in SDSS-DR7 have been derived and
published by the MPA/JHU
group\footnote{http://www.mpa-garching.mpg.de/SDSS/}. We adopt their
emission-line measurements and some properties of the galaxies,
such as D$_n$(4000), H$\delta_A$, stellar masses, star formation
rates, metallicities etc. for the present work.

\subsection{Cross-correlations of ASC and SDSS-MGS}

To select the working sample of SNe host galaxies, we
cross-correlated the coordinates of SNe host galaxies from
ASC and the coordinates of spectral observations of the SDSS
galaxies in their MGS with the following criteria.

\begin{figure}
\centering
\includegraphics [width=7.0cm, height=6.0cm] {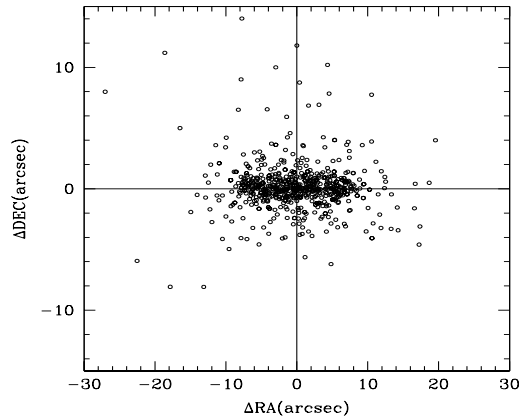}
\caption {The discrepancy between the original coordinates of the 687 SNe host galaxies from the ASC
and those from the ICRS from SIMBAD
which have the ICRS coordinates.   } \label{fig.radec}
\end{figure}

\begin{enumerate}

\item {\sl Selecting those with well-defined SNe types} ---
We firstly select the supernovae and their host galaxies from the updated ASC (up to SN 2013Y, 6105 samples in total).
Then only those having well-defined types of SNe,
e.g., type Ia, II and Ib/c, are further selected. This is 4,934 samples. It is worth noting here that the ASC
contains events since 1885, and that a precise sub-classification,
particularly of type I's, did not exist at the beginning of research about SNe.

\item {\sl First cross-matching with 30$^{{\prime}{\prime}}$ radius} ---
In the ASC, we notice that the coordinates of the SNe host galaxies
are not accurate enough in both RA (h:m:s)
and DEC (d:m:s) since both terms of ``s" only down to integers without decimals. This
could cause obvious discrepancy when matching the two catalogs, which may be up to
15 arcseconds for RA (if the discrepancy up to $\pm$1s) and a bit less for DEC.

To minimize this problem, we adopt two steps for  matching
coordinates to select the final working sample. In the first, we
adopt 30 arcsec as the matching radius to cross-correlate the
coordinates of the 4,934 SNe host galaxies in ASC with the SDSS DR7
MGS (698,260 entries). This radius is large enough for considering the
low precision of both RA and DEC. This value is half the radius for
matching used by \citet{2008ApJ...673..999P}. With this 1,105 host galaxies are
selected, by removing the duplicated and multiple spectral
observations, 1,041 host galaxies are retained.

\item {\sl Coordinate corrections} ---
In order to improve the reliability of cross-matching between ASC and
SDSS, we try to improve the accuracy of the coordinates of the SNe
hosts in ASC. To do so, we retrieve the International Celestial
Reference System (ICRS) coordinates (epoch in J2000) from SIMBAD
(http://simbad.u-strasbg.fr) for the hosts by searching them by
designation. The ICRS coordinates have two or three decimals for ``s"
terms in RA and DEC, which could improve the accuracy of matching between ASC and SDSS catalogs for SNe host galaxies.
687 (among 1,041) host galaxies are obtained their ICRS coordinates from SIMBAD, but the other 354 ones could not
be found there, so we had to keep their original coordinates
from ASC.

To show the necessity and significance of the improvement in such coordinate precision
for the SNe host galaxies, in Fig.~\ref{fig.radec} we
plot the difference between the coordinates of the 687 hosts from the
ASC and those of the ICRS ones, which shows the low precision of the coordinates in ASC.
Now we have a new table for 1,041 SNe hosts, and 687 of them have updated RA and DEC from ICRS.

\item {\sl Second cross-matching with 15$^{{\prime}{\prime}}$ radius} --- We
prefer a smaller radius for matching the two catalogs since it will help to avoid the mis-matching cases.
Therefore, we adopt 15
arcsec as the matching radius to redo the cross-correlation for the 1,041 SNe host galaxies from
ASC and SDSS-DR7 MGS. Among them, 687 objects had revised
RA-DEC as ICRS coordinates from SIMBAD as mentioned above. After
this step, 980 objects were obtained (where 672 are from the 687 sample and 308 are
from the 354 sample).

Hopefully, the 15 arcsec matching radius here can help to select as
many interesting objects as possible, and to avoid the
mistaken cases of cross-matching.

\end{enumerate}

\subsection {Light fraction criterion for global properties from fiber spectral observations}

The aperture of the fiber is 3 arcsec in the SDSS observations.
In this study, we mostly focus on the global properties of the host
galaxies from SDSS spectra, thus we further select the
sample galaxies which have more than 15\% of their light
covered by the fiber observations.
This light fraction criterion ($>$15\%) could help us to select the cases in which 
the 3 arcsec fiber observations of SDSS cover most of the light of the whole galaxy 
and thus retrieve the global properties of the SNe host galaxies.

 To judge this,
  one simple
   and accurate way is to compare the ``fiber" and
   ``petrosian" magnitudes of the SDSS galaxies.
   The fiber mag is a measurement of
   the light going down the fiber and the Petrosian mag is a good
   estimate of the total magnitude. Thus, we adopt the formula below
     to estimate how much light was covered by the fiber observations in r-band \citep{2010MNRAS.409..213L}:
\begin{equation}
   light\_fraction = 10^{(-0.4*(fiber\_{mag} - petro\_{mag})_r)}.
\label{apar}
\end{equation}

\begin{figure}
\centering
\includegraphics [width=7.0cm, height=6.0cm] {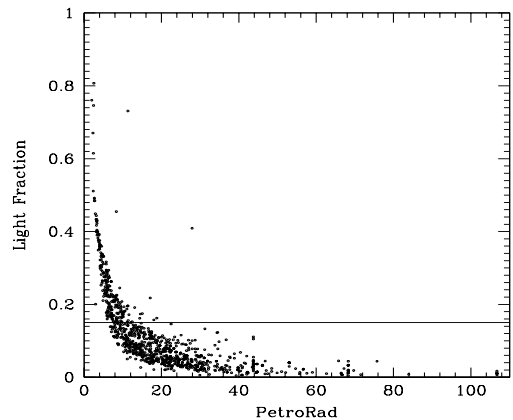}
\caption {Relation between light fraction and Petrosian radius (in
arcsec) in r-band for the 980  SNe host galaxies.  } \label{fig.lf}
\end{figure}

 Fig.~\ref{fig.lf} shows the relations between the calculated light
fractions and the Petrosian radius in r-band for the selected 980 SNe
hosts.

 From Fig.~\ref{fig.lf} we can see that for
a large part of the SNe hosts, the 3 arcsec fiber of SDSS cannot cover more than 10\% of
the light from the whole galaxy. Here a bit stricter criterion is used to select our sample
galaxies, which have light fraction larger than 0.15 in the spectral observations. To choose
0.15 of light fraction here, we expect not only a reasonable sample of
galaxies are selected for studies, of which the SDSS spectra are representative of the whole
galaxy, but also we won't lose too many objects with large size. Although we lose a large
fraction of the initial sample by
performing such a light fraction cut, we believe it is necessary to
guarantee that the spectral observations are able to represent the global
light of the galaxies, rather than a small region inside the galaxy.
We will discuss the effects of lower light fraction  with the comparative sample in Sect.~5, 
where the host galaxies with light fraction lower than 0.15 will be taken for comparison and then the aperture effect will be shown clearly.

The horizonal line on Fig.~\ref{fig.lf} shows this light fraction
cut. Above the 0.15 line, there are 243 objects.
It is clear that this selection biases against close,
large galaxy hosts.  These 243
galaxies are our main working sample in this work. The remaining
737 galaxies (from the total of 980 objects) having lower light fractions will be taken as comparison and are specially discussed
in Sect.~\ref{lowlf}.

\begin{figure}
\centering
\includegraphics [width=7.0cm, height=6.0cm] {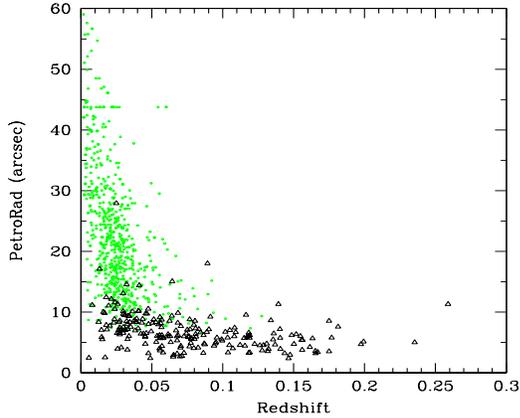}
\caption {Relation between redshift and  Petrosian radius (in
arcsec) in r-band for the 980  SNe host galaxies.  The black
triangles refer to the 213 galaxies having light fraction larger
than 15\%, and the green circles refer to the 689 galaixes having
lower light fraction.} \label{fig.Pz}
\end{figure}

Our sample galaxies are plot in the Petrosian
radius vs. redshift relation in Fig.~\ref{fig.Pz}. We see that our
main working sample galaxies, denoted by the triangles, have small Petrosian radius.

\subsection {Further criteria}
\label{furcri}

We do further careful checks for the selected sample galaxies.

\begin{figure}
\centering
\includegraphics [width=7.0cm, height=6.0cm] {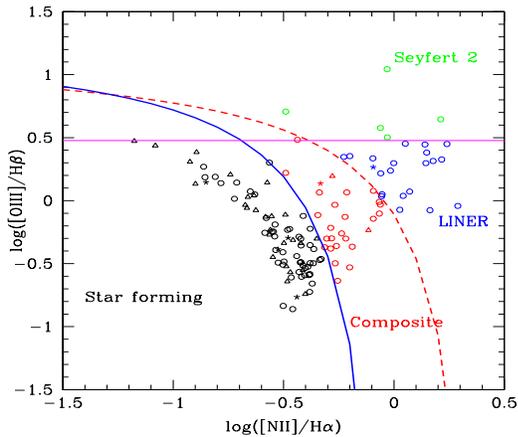}
\caption {The BPT diagram for the 135 SNe hosts (among the 213) with
good quality
 measurements of the four relevant emission lines. In the later sections, we will take SF as the star-forming galaxies, and
 AGNs for the combined sample of Composites, Seyfert 2s and LINERs. The three diagnostic lines are taken from Kauffmann et al. (2003a, the solid curve),
 Kewley et al, (2001, the dashed curve) and Shuder et al. (1981, the narrow solid horizonal
 line). Here the triangles refer to hosts of SNe II, the stars for hosts of SNe Ibc, and the circles
 refer to hosts of SNe Ia.}
\label{fig.BPT}
\end{figure}

\begin{enumerate}

\item
 {\sl Image checking by eye} ---
 In order to guarantee the correction for host-identification between supernovae
 and galaxies with SDSS observation, we check the 243 samples
 case by case and remove 14 cases which have misidentification.
 For example, the SNe explosion locates
 in a faint galaxy, which was not taken spectrum by SDSS observation. Instead the SDSS
 takes spectrum on a bright galaxies quite close to it. After this
 step, 229 hosts are left.

\item
{\sl Spectral observations and quality control} --- We downloaded
the SDSS 1D spectra of these 229 SNe host galaxies. Some of them have
to be removed since interruption appear in their spectral energy
distribution (SED). After this cleaning, we
have 213 SNe host galaxies left, whose S/N (median value per pixel of
the whole spectrum provided in the MPA/JHU catalog) are larger than
5.

In Table~\ref{tab.lis}, we list 6 examples of these 213 objects: the
table includes coordinates (RA and DEC of SNe hosts from ASC, RA and
DEC from SDSS), PID-MJD-FID numbers of the SDSS spectral
observations, Petrion radius in r-band, light fraction,
 redshift, and the types and names of the supernovae. The table with the
entire 213 galaxies will be available in electronic form.

\item
{\sl Three sub-samples associated with emission-line ratios} --- Since we
are working on the properties of host galaxies of supernovae, it
will be interesting to check their
emission or absorption lines.
In the total sample of 213 SNe host galaxies,  135
 have good quality observations in all four
emission lines [NII]6583, H$\alpha$, [OIII]5007 and H$\beta$ with an
S/N better than 3$\sigma$. They can be plotted on the BPT diagram \citep{1981PASP...93....5B} as shown
in Fig.~\ref{fig.BPT}.  The remaining 78 objects have a
lower S/N in  these four emission lines, or only display some of
them or none at all. Then the sample galaxies can be divided into
three sub-samples:

1) 82 star-forming (SF) galaxies identified by their emission-line
ratios in BPT;

2) 53 AGNs (including composites, LINERs and Seyfert 2s);

3) 78 absorption-line and weak emission-line galaxies (named as
``Absorp" simply) which cannot appear on the BPT diagram due to the
absence of some or all of the four emission lines mentioned above.

\end{enumerate}

In Sect.~\ref{secBPT} we will plot the
relations of some parameters associated with properties of the hosts of different
types of supernovae, by marking their hosts as SF, AGN and Absorp.

Table~\ref{tabN} shows the numbers of the hosts of the different types of SNe
according to the galaxies appearing on the BPT diagram or not, as well
as the total numbers of each type of SNe. We can see
that among 169 SNe Ia hosts, 49 are SF
galaxies, 49 are AGNs and 71 are Absorp galaxies.

\subsection {The comparative sample with low light fraction}
\label{com.sam}

For those galaxies whose light fraction is lower than 15\%,
we will take them as a comparative sample and perform a similar analysis.
With the light fraction cut and further selecting criteria as described
in Sect.~\ref{furcri}, this comparative sample includes 689
galaxies. All the details about this part of galaxies
will be given in Sect.~\ref{lowlf}, where we will show the difference
between 213 samples and 689 samples.

\begin{table*}
\begin{center}
\caption{Basic information of six example galaxies, including all
types of SNe. The coordinates of host galaxies in ASC (in hhmmss for
RA and ddmmss for DEC) and SDSS-MGS (in degree) are given.
PID-MJD-FID are the 3 IDs to take the SDSS spectra. }
\label{tab.lis} \tiny
\begin{tabular}{lcccccccccc}
\hline
   No.  &  RA (ASC)           & DEC (ASC)           &   RA (SDSS)         &  DEC (SDSS)         &   PID-MJD-FID       & Petrion  & light  & redshift    &    Type  & name of SN  \\
        &                    &                &                    &           &         & radius & fraction &    &    &   \\  \hline
   111     &      105741         &       573648     &     164.42208862  &     57.61346436    &     0949-52427-109      &      6.375  &   0.193  &   0.080    &     Ia  &   2010 bg   \\
    99     &      230548         &       141956     &     346.45358276  &     14.33142471    &     0742-52263-590      &      7.429  &   0.209  &   0.108    &     Ia  &   2012 ff   \\
   153     &      131523         &       462509     &     198.84954834  &     46.42040253    &     1461-53062-166      &      8.732  &   0.162  &   0.056    &     II  &   2009 ct   \\
   130     &      163214         &       383920     &     248.05726624  &     38.65555954    &     1173-52790-537      &      8.355  &   0.150  &   0.039    &     II  &   2012 ct   \\
   183     &      111229         &       312305     &     168.12567139  &     31.38496208    &     2092-53460-516      &      6.309  &   0.176  &   0.027    &     Ib  &   2011 bp  \\
   159     &      122450         &       082557     &     186.20838928  &      8.43370342    &     1626-53472-419      &      7.143  &   0.174  &   0.090    &     Ic  &   2009 bh   \\    \hline
\end{tabular}
\end{center}
$Note: $ {\footnotesize Here we give only a few lines,  with the
meaning of all the columns. The full table will be available in the
electronic version. The revised RA and DEC for some samples are
given if we have them. The Nos. in first Column follow the full table.
Here we present examples including all SNe Ia, II, Ib and Ic. }
\end{table*}

\begin{table}
\begin{center}
\caption{The numbers of the different types of SNe among the
galaxies appearing or not on the BPT diagram, as well as the total
numbers.  } \label{tabN}
\begin{tabular}{|c|c|c|ccc|}
  \hline
 Samples & Galaxies   & Total & SN Ia & SN II & SN Ibc \\ \hline
 SF & Star-forming &  82   & 49    & 28  & 5 \\  \hline
          &  Composite    &  27   & 24    &  2  & 1 \\
 AGN   & LINER       &  21   & 20    &  0  & 1  \\
          & Seyfert 2   &   5   &  5    &  0  & 0 \\ \hline
 Absorp & Absorp. \& WE  & 78  &  71   &  4  & 3 \\ \hline
   Total  &               & 213   & 169 & 34  & 10 \\ \hline
\end{tabular}
\end{center}
$Note:$ {Here ``Absorp." refers to absorption-line galaxies, and
``WE" refers to weak emission-line galaxies that have not been
detected all the four emission-lines in the BPT diagram.}
\end{table}

\section{Relations of some parameters of properties for the 213 supernova host galaxies}
\label{secBPT}

\begin{figure*}
\centering
\includegraphics [width=4.0cm] {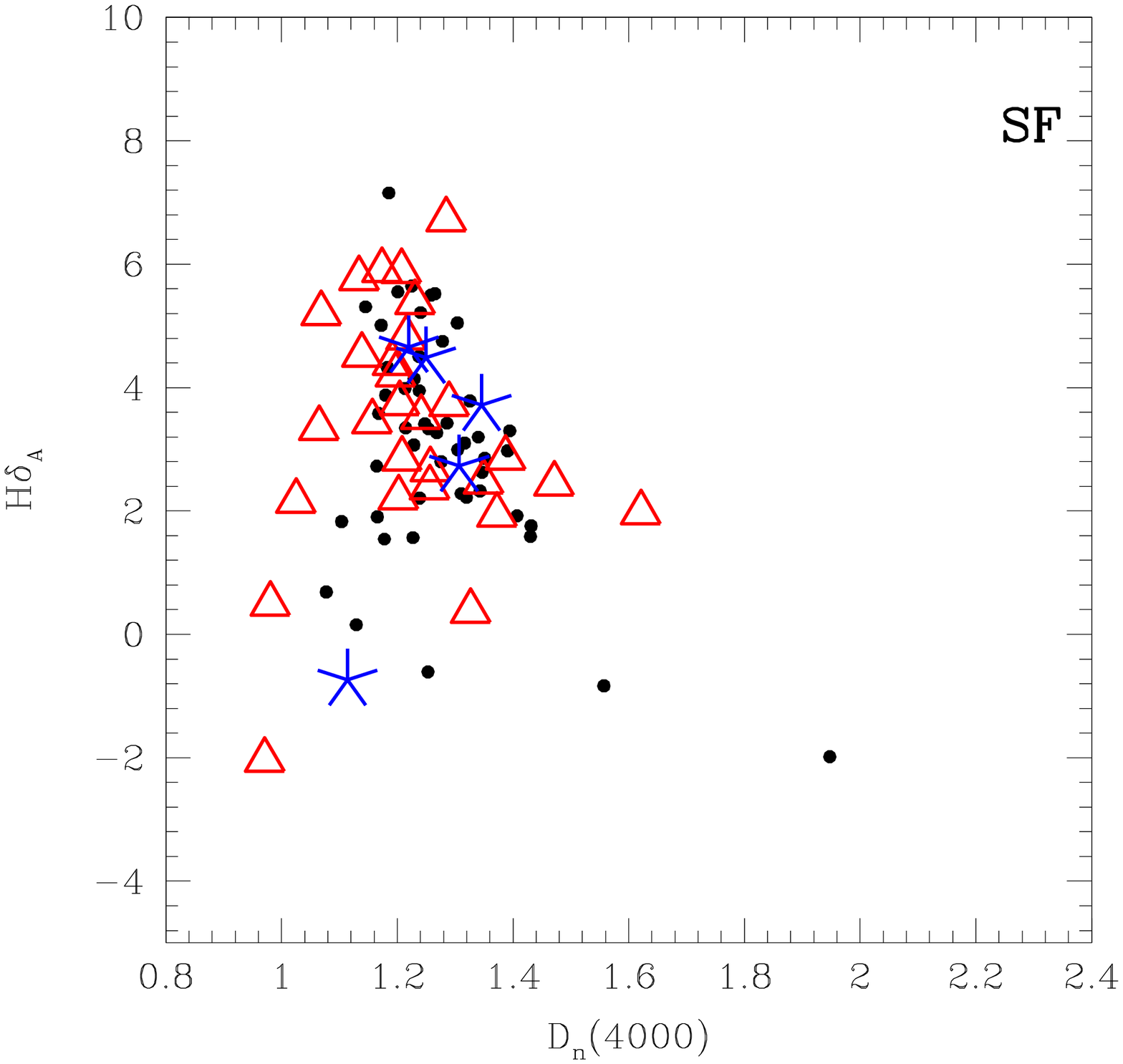}
\includegraphics [width=4.0cm] {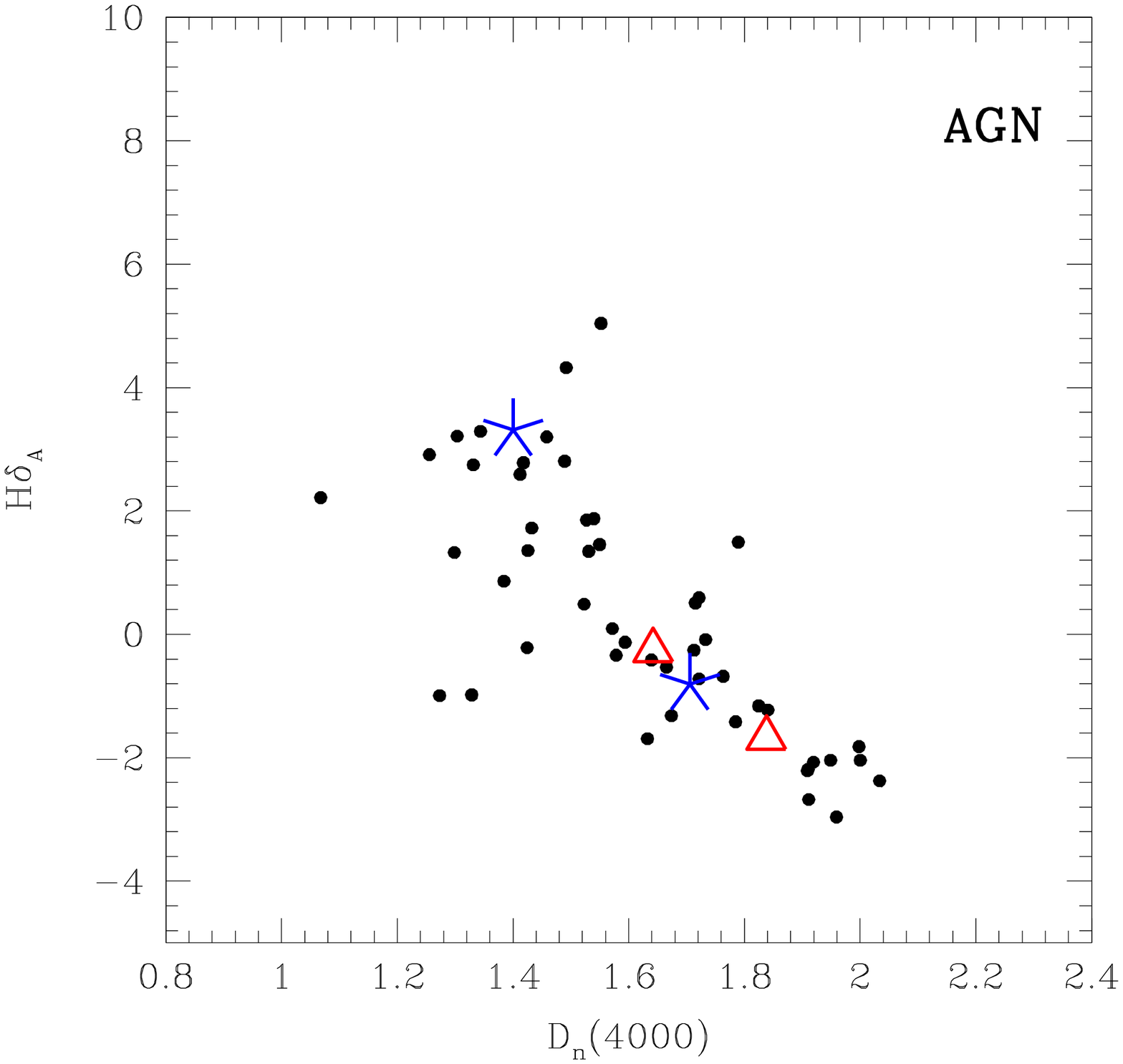}
\includegraphics [width=4.0cm] {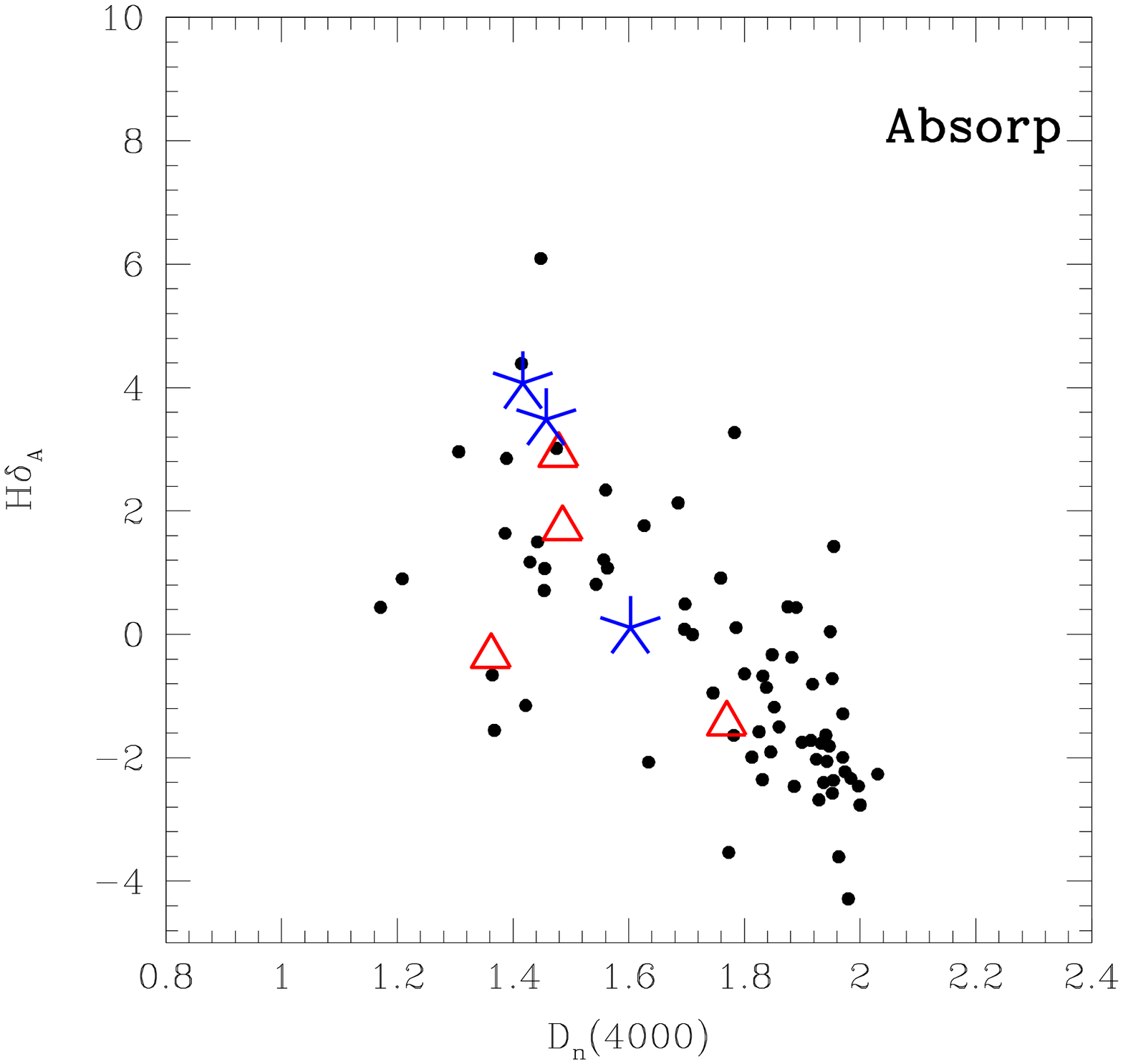}
\includegraphics [width=4.0cm] {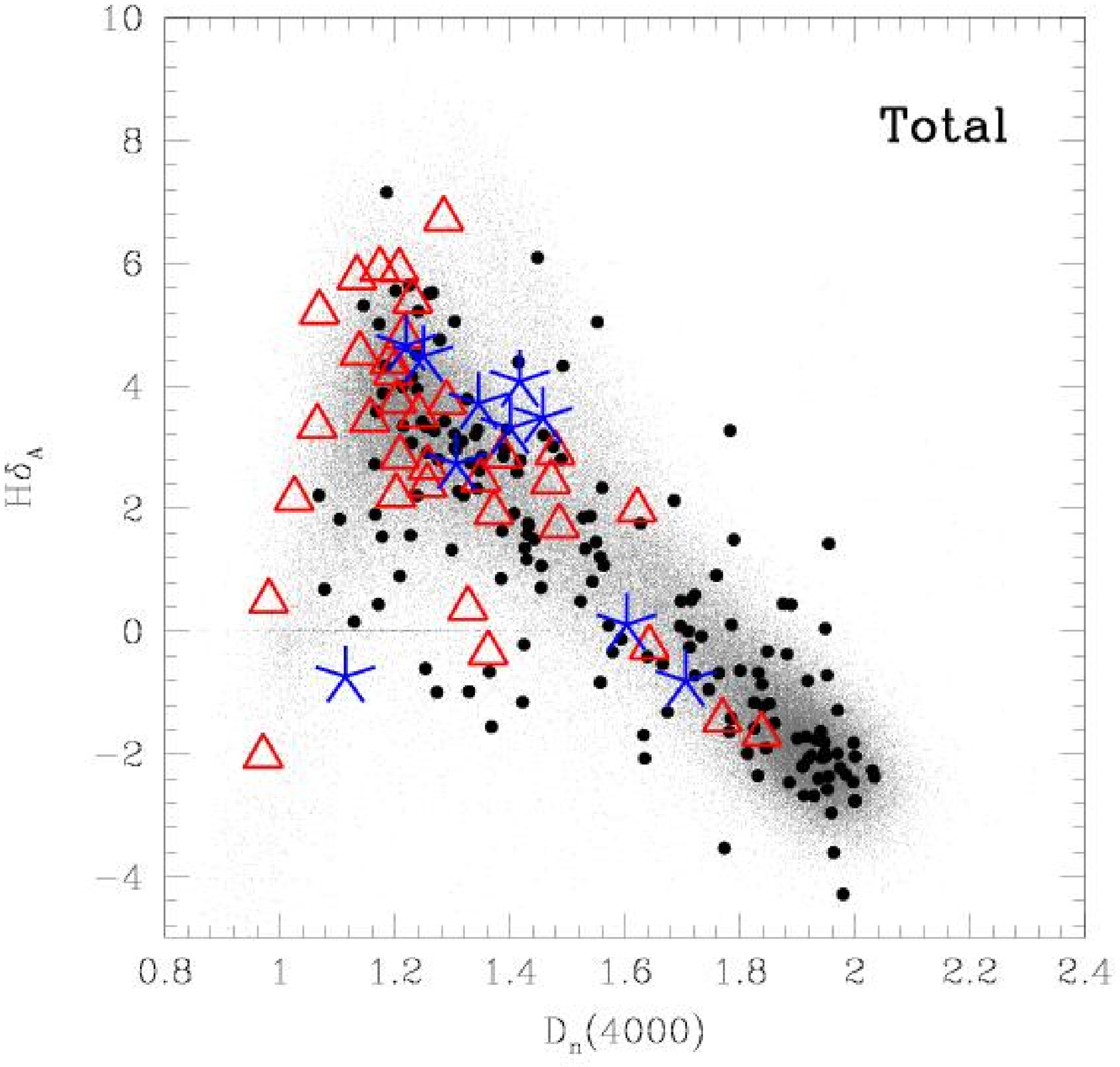}
\includegraphics [width=4.0cm] {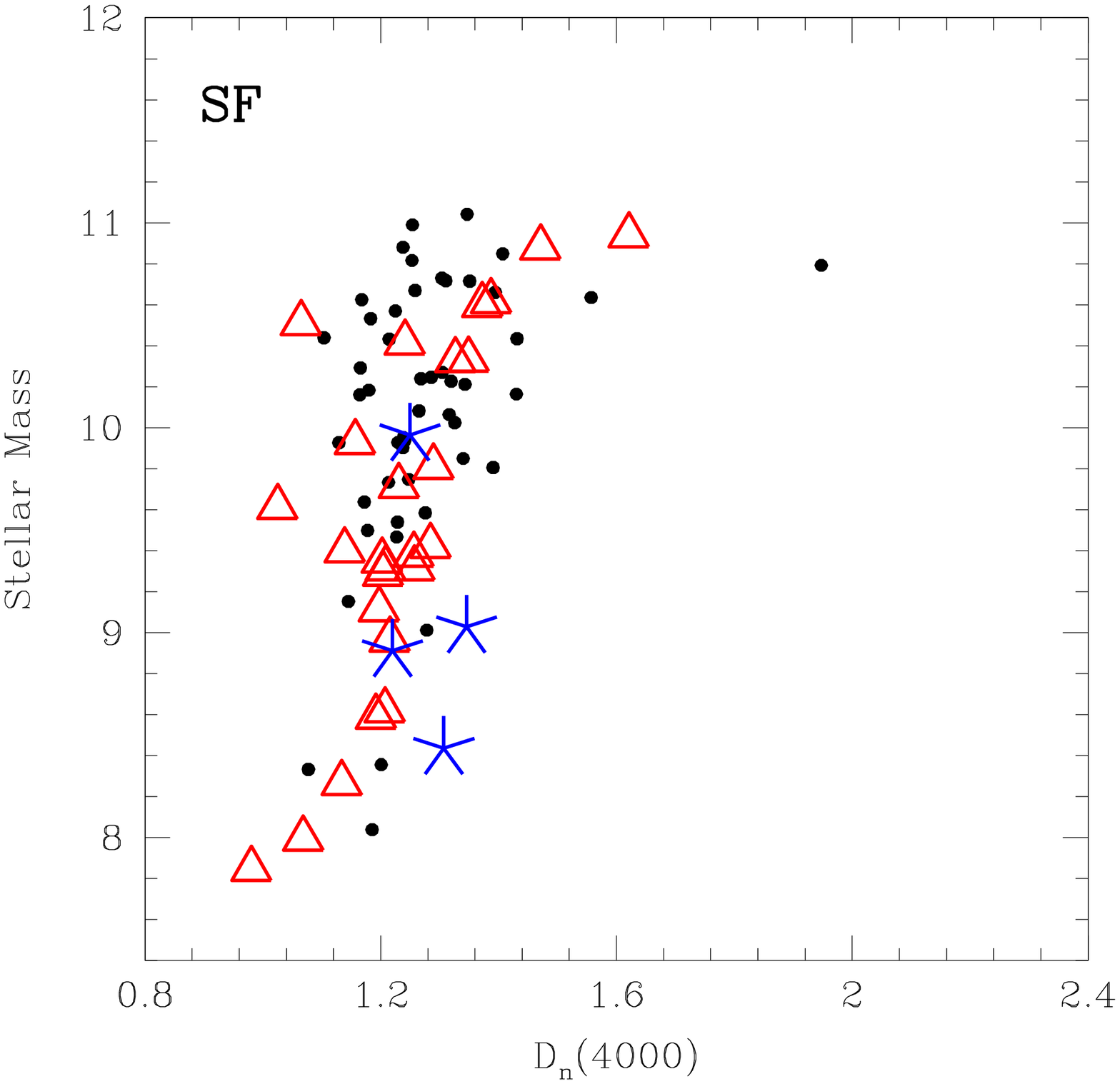}
\includegraphics [width=4.0cm] {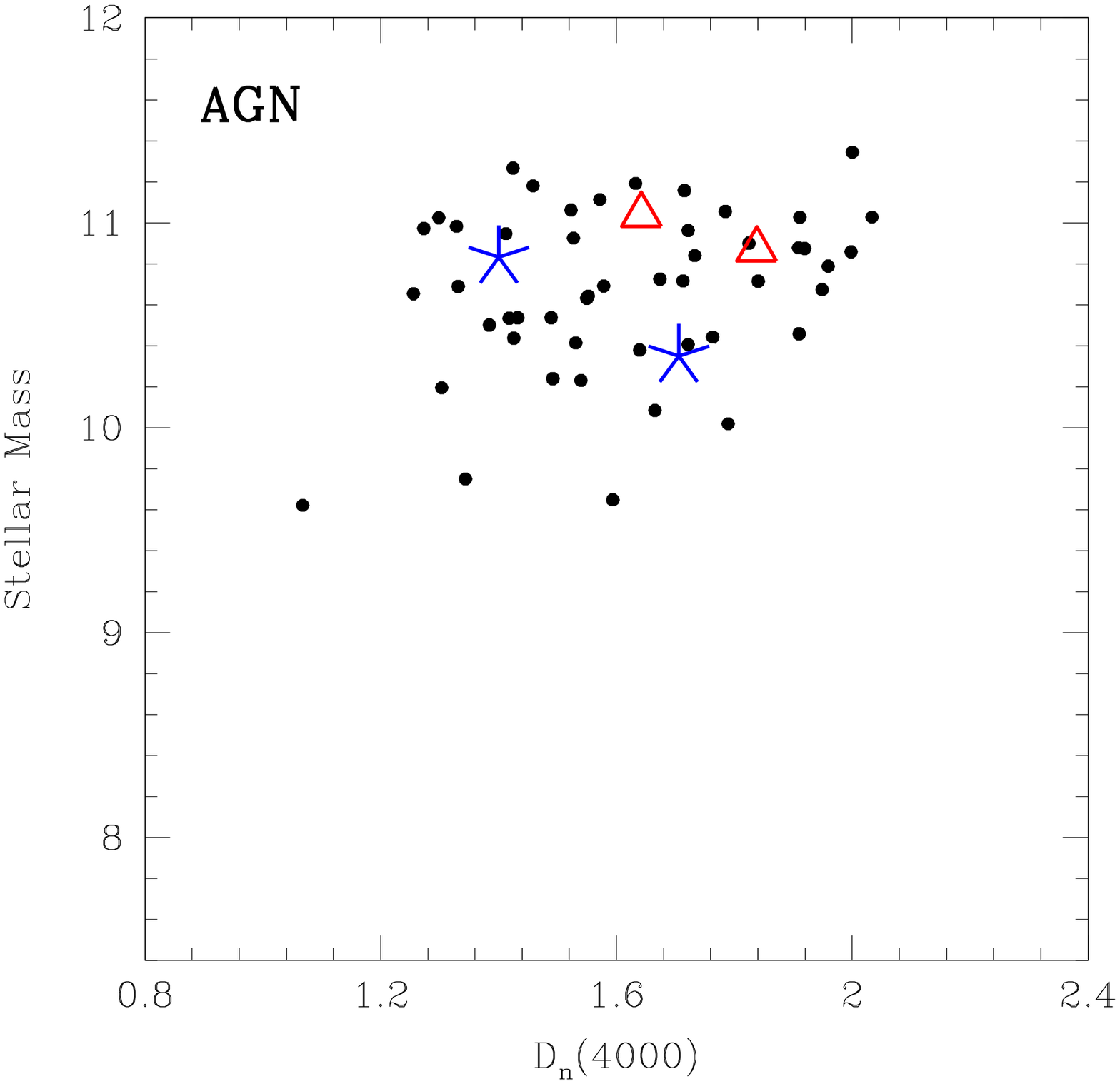}
\includegraphics [width=4.0cm] {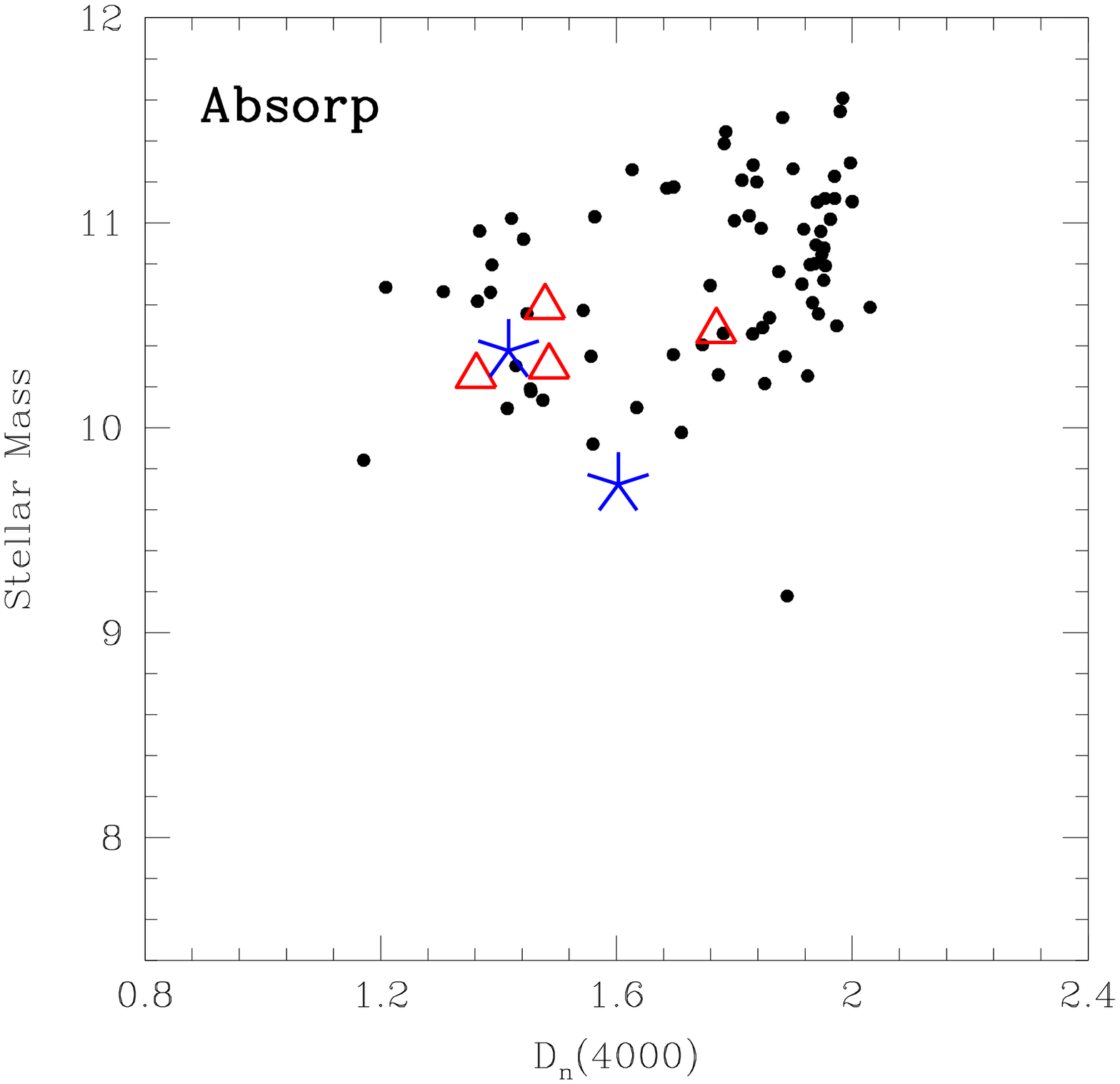}
\includegraphics [width=4.0cm] {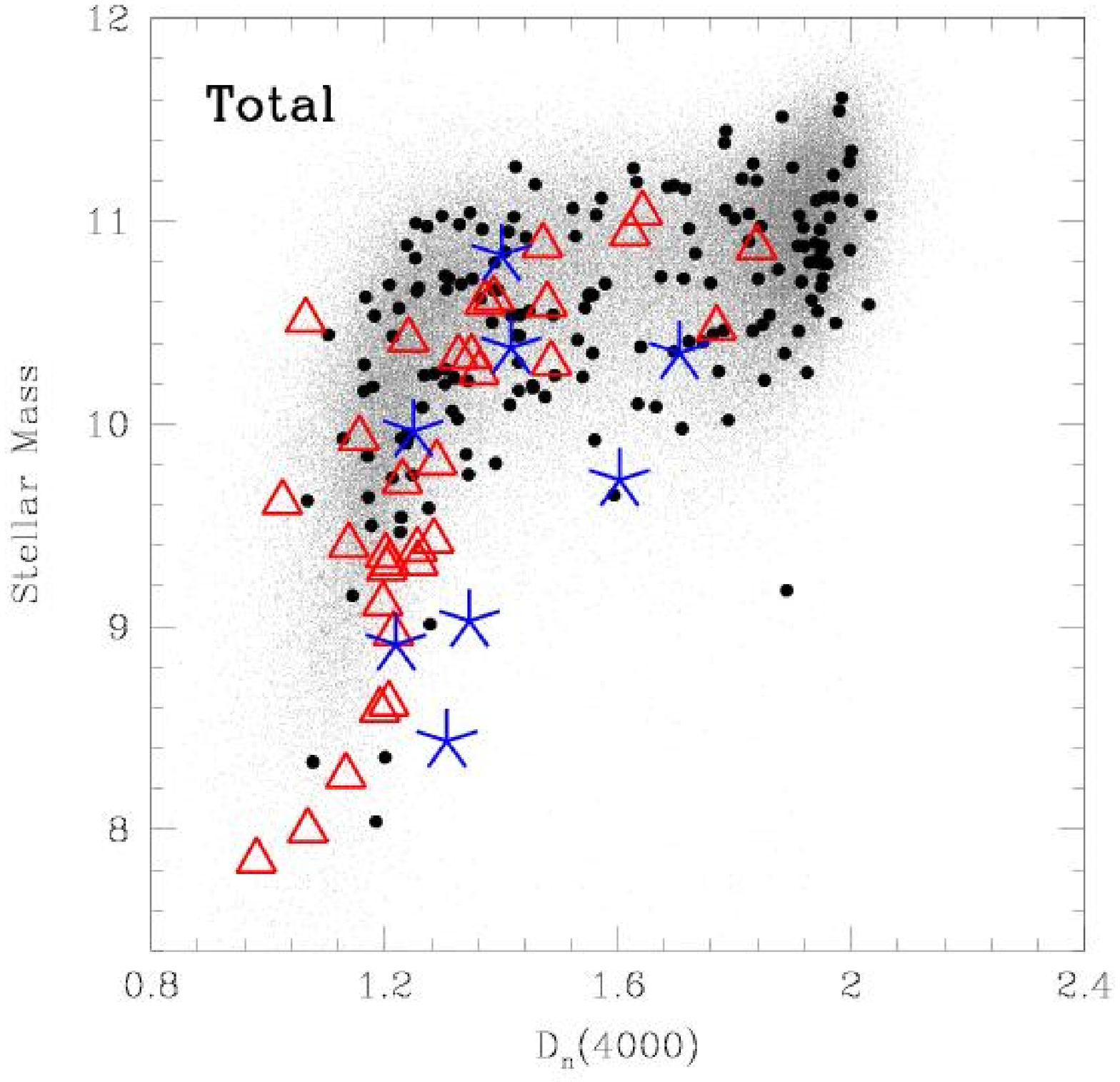}
\includegraphics [width=4.0cm] {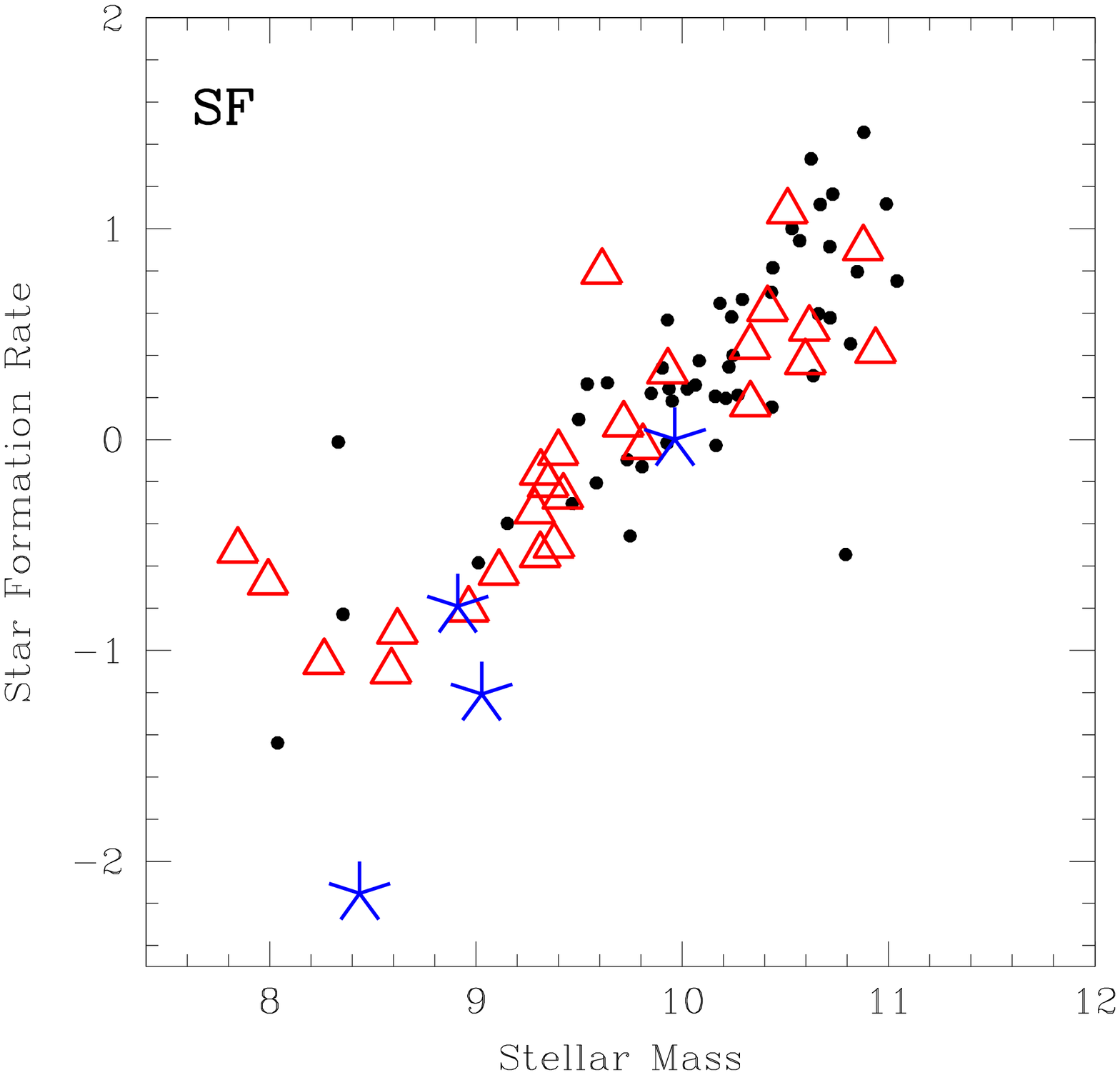}
\includegraphics [width=4.0cm] {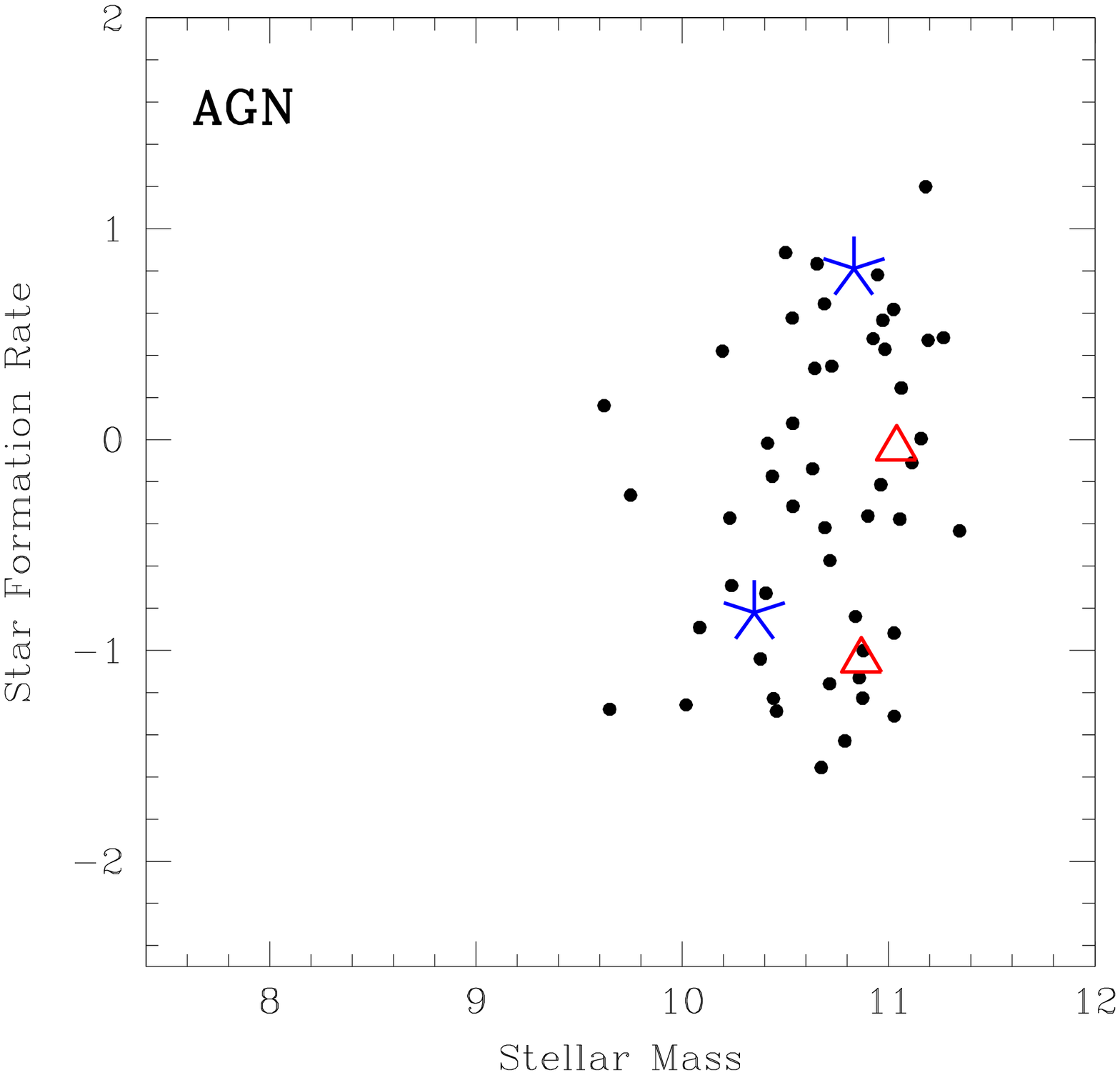}
\includegraphics [width=4.0cm] {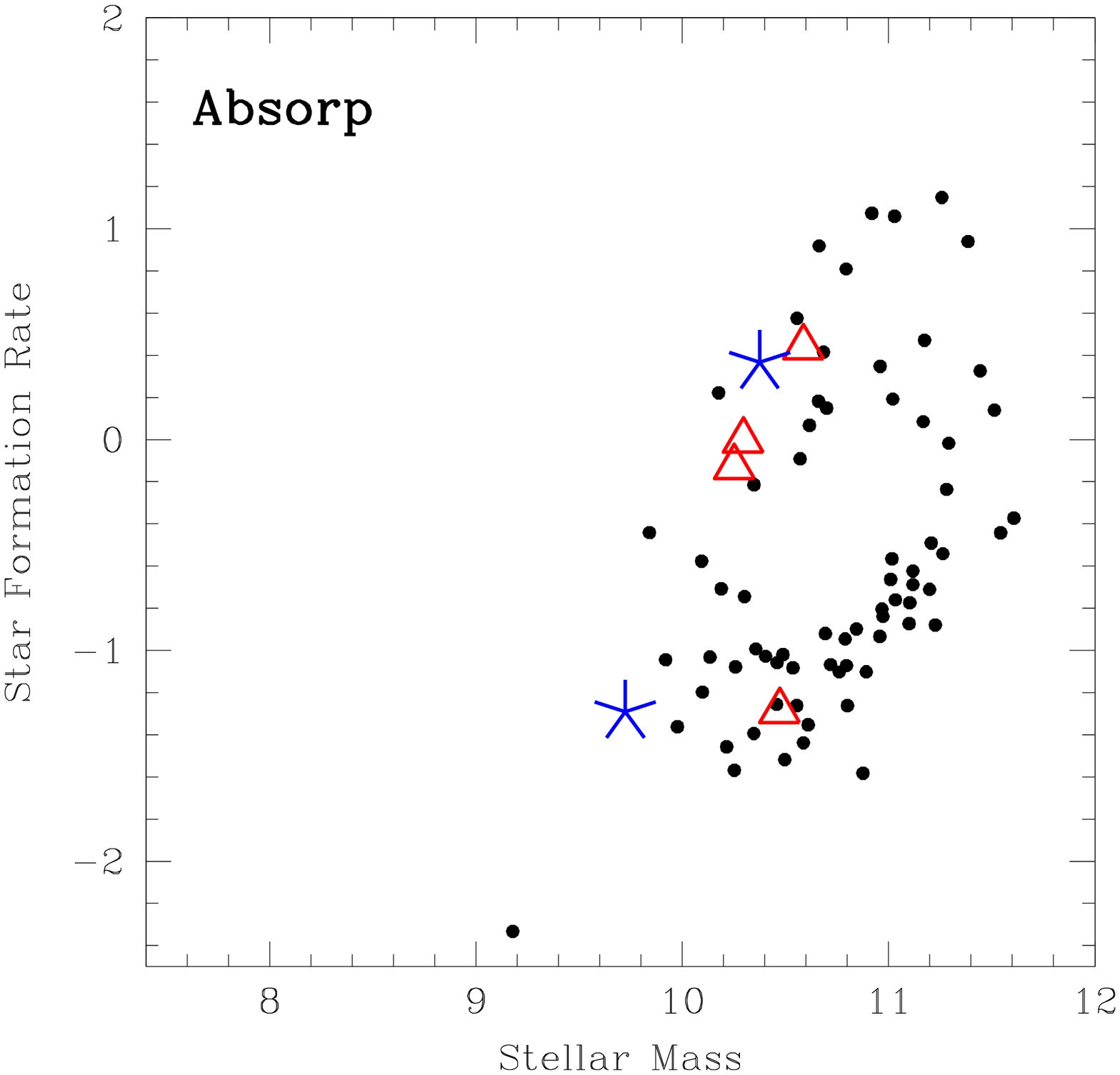}
\includegraphics [width=4.0cm] {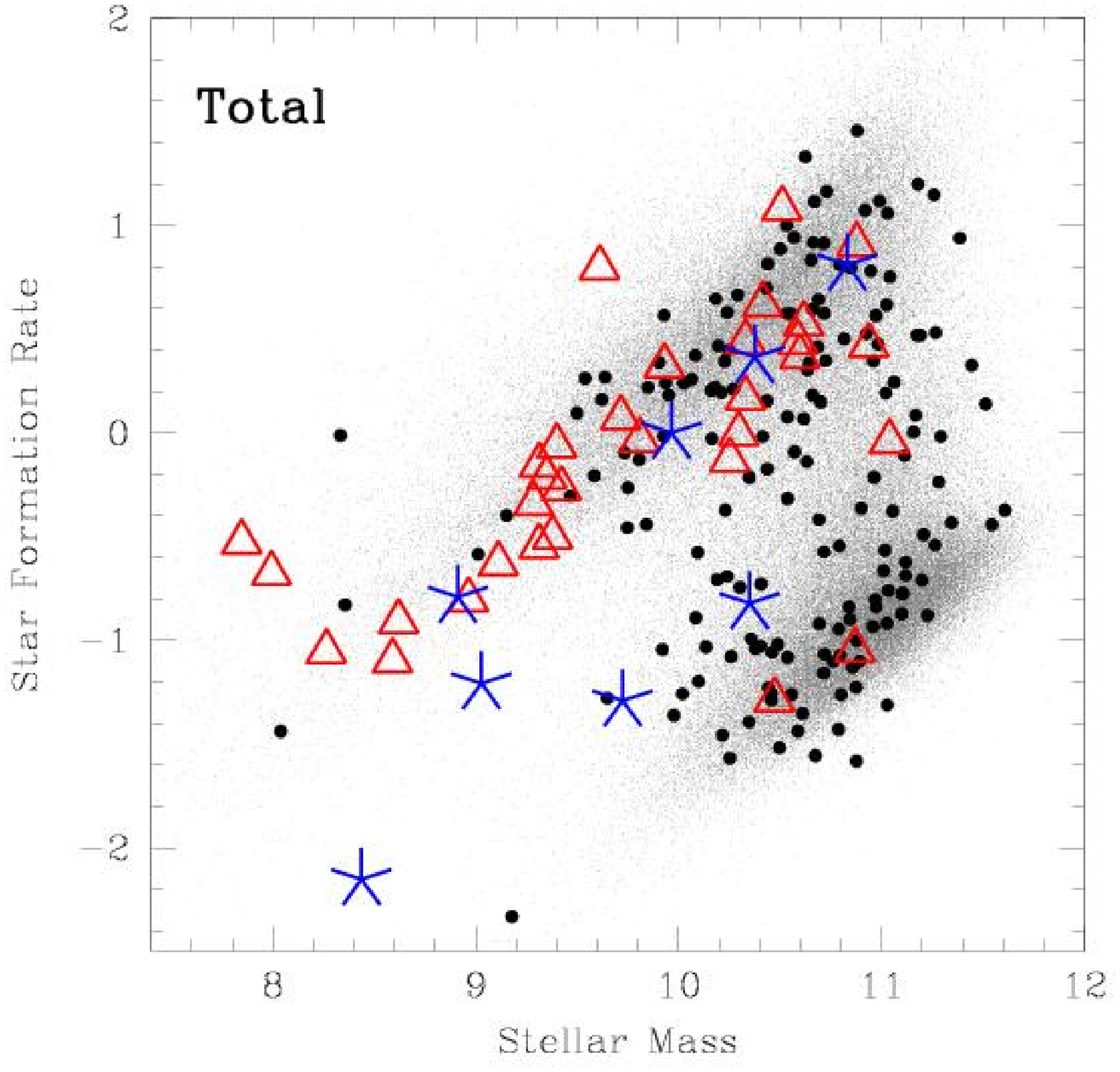}
\includegraphics [width=4.0cm] {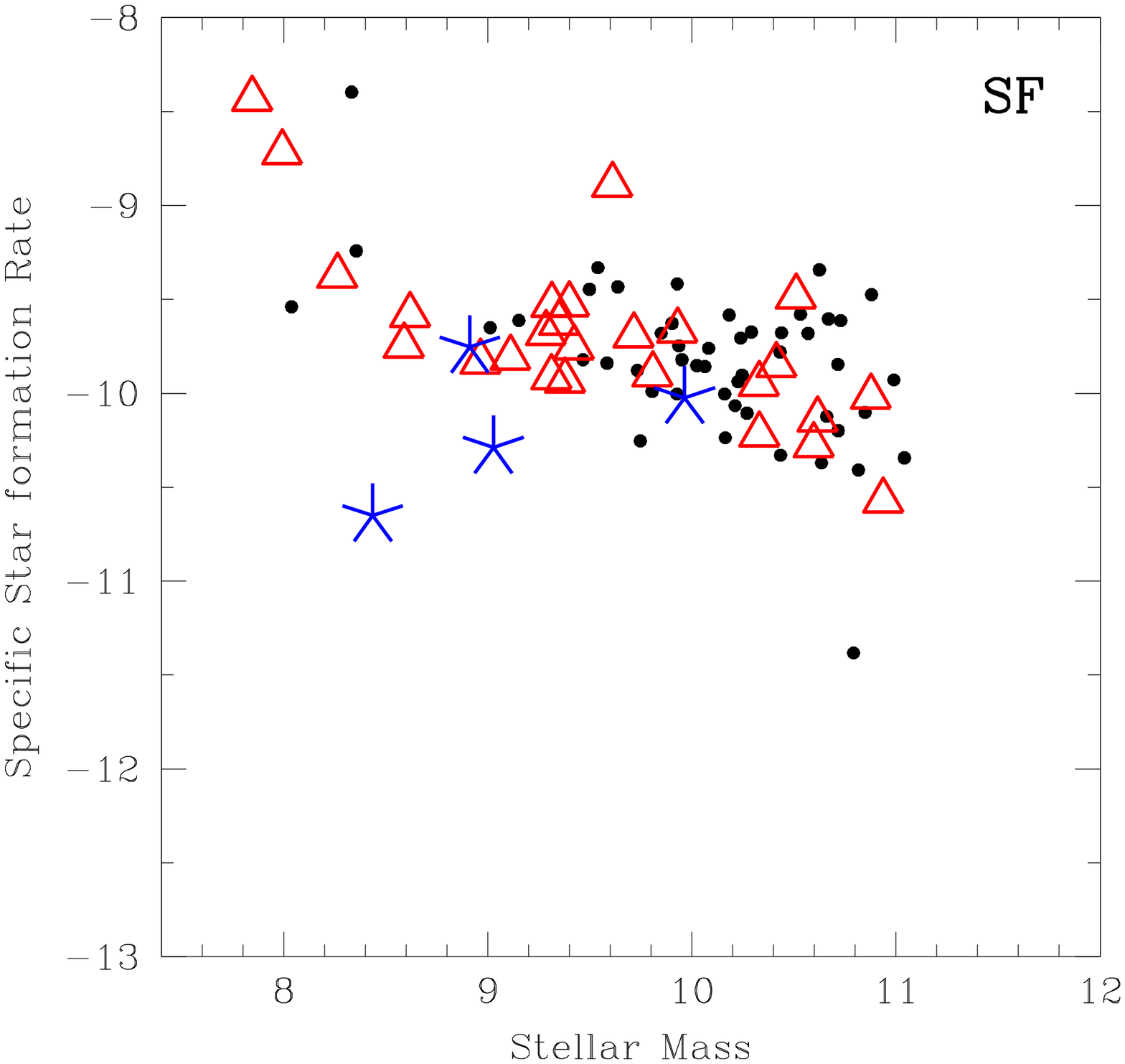}
\includegraphics [width=4.0cm] {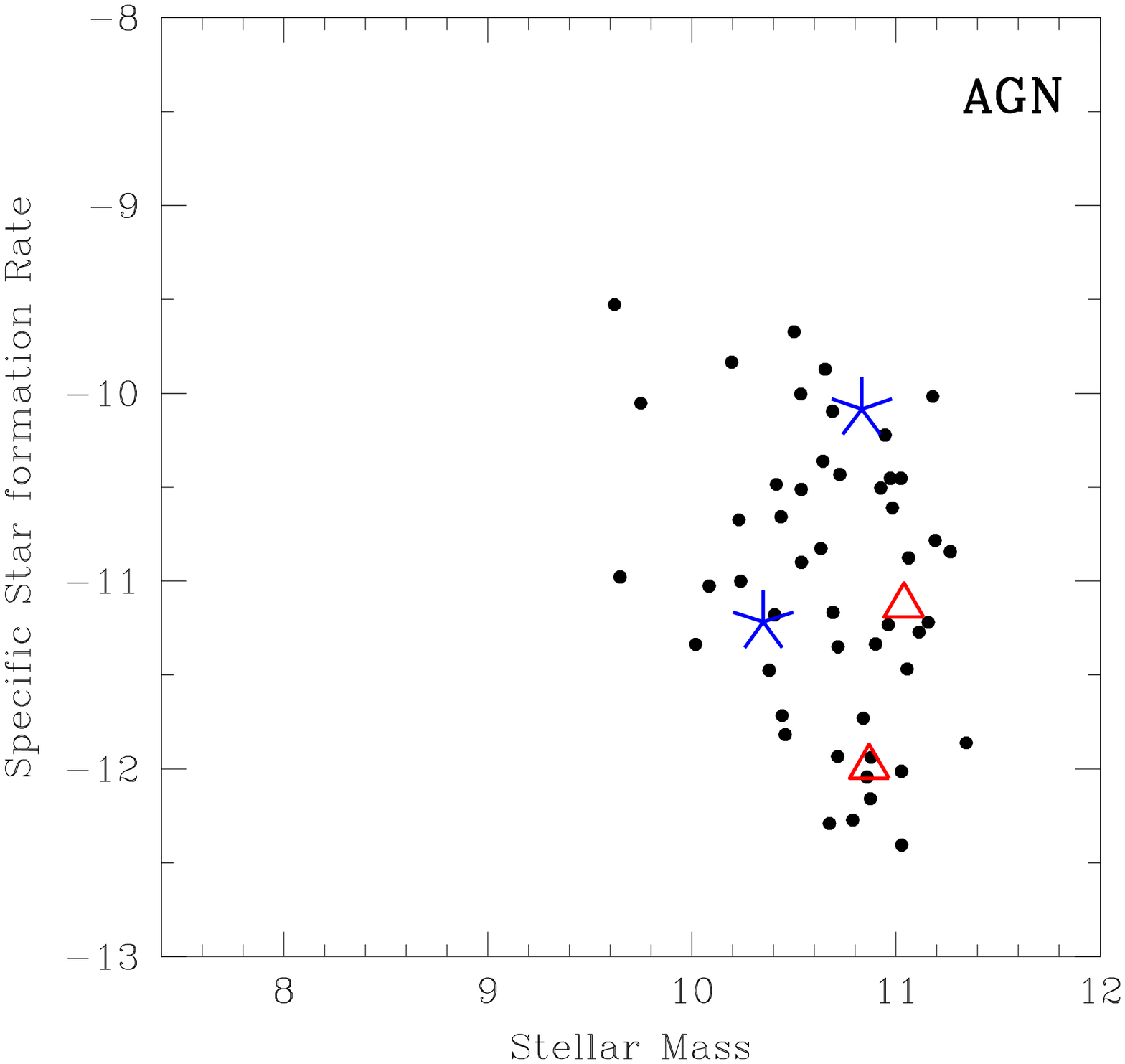}
\includegraphics [width=4.0cm] {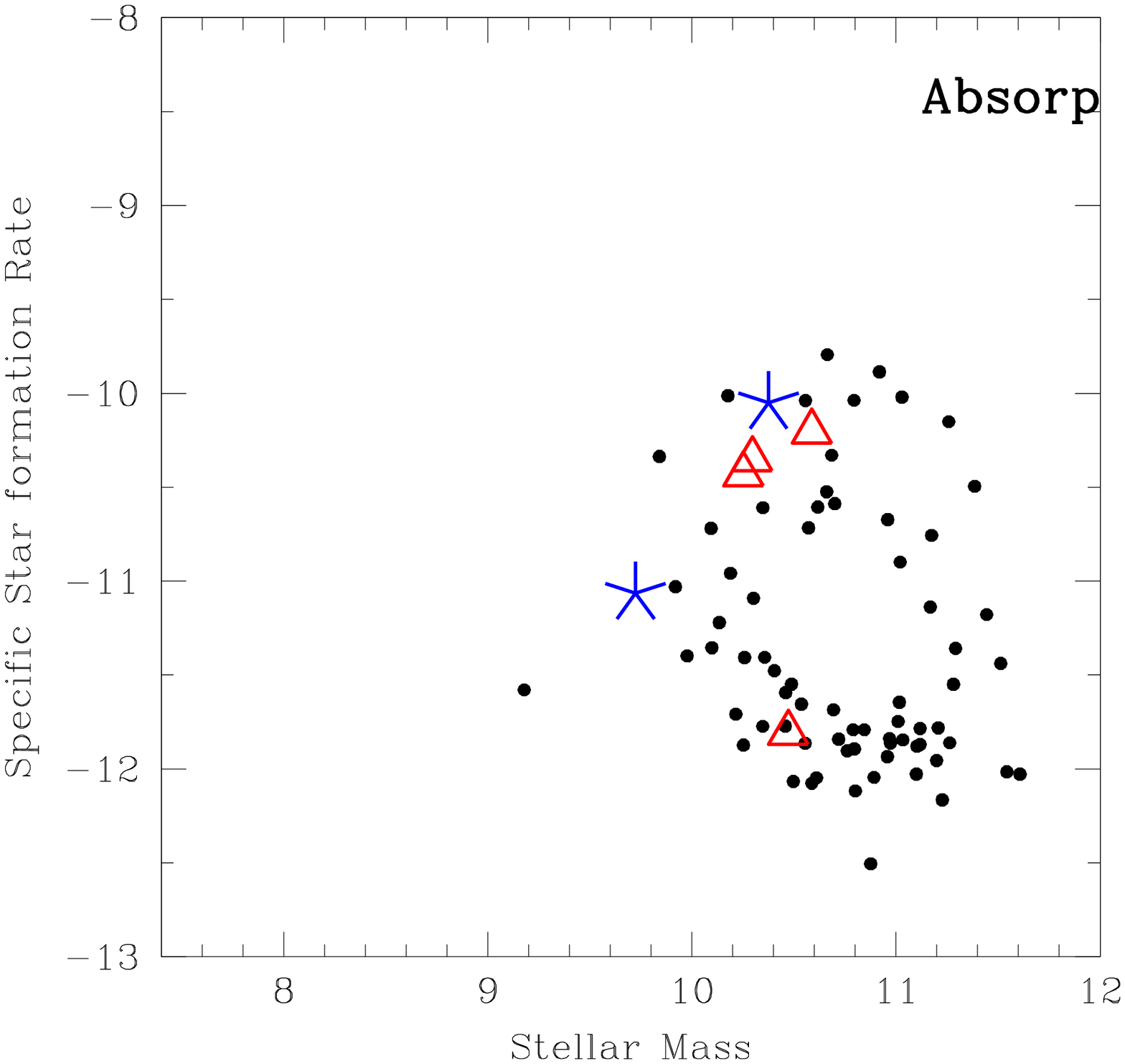}
\includegraphics [width=4.0cm] {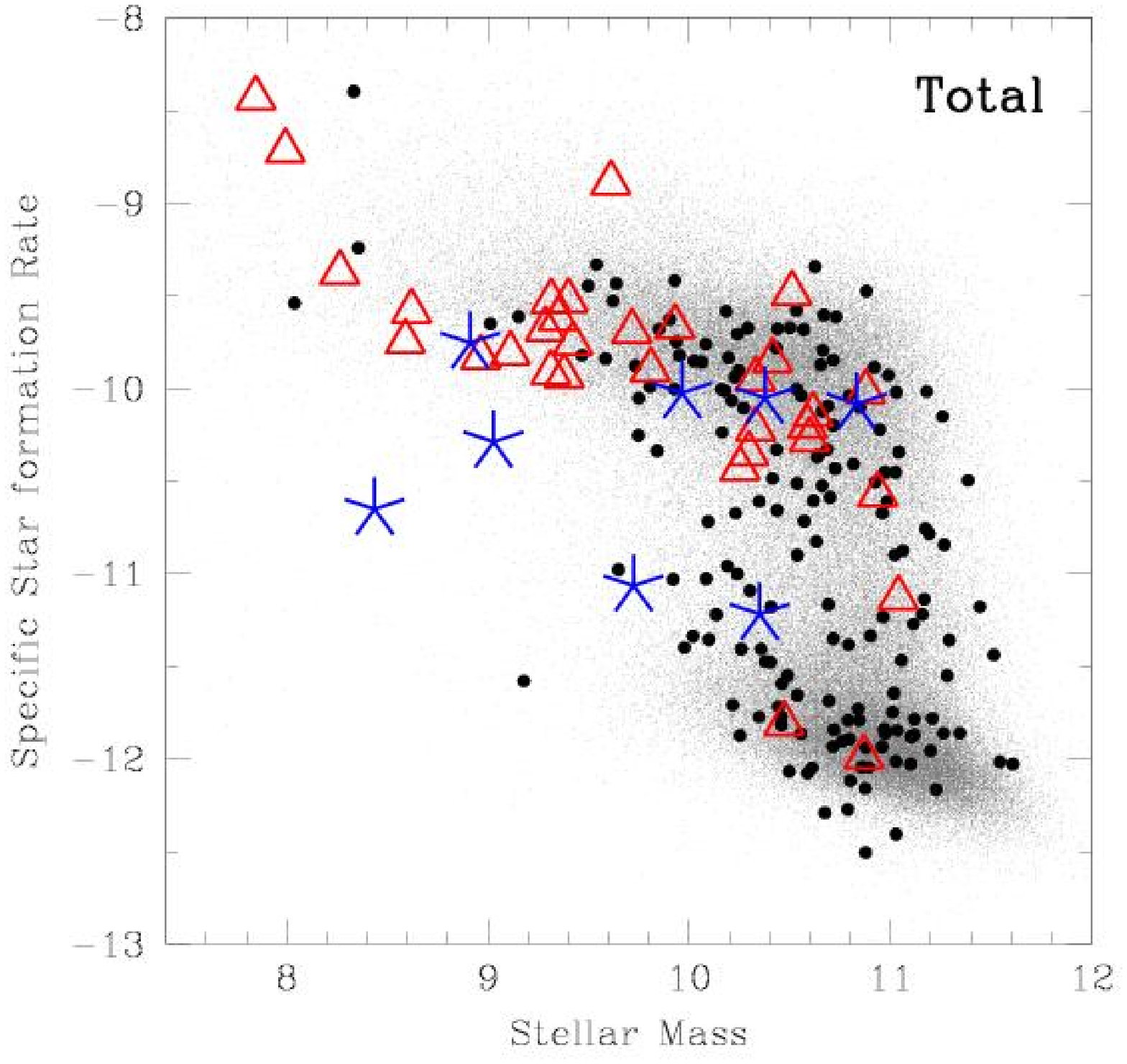}
\caption {The relations among several property parameters for the
213 SNe host galaxies: the D$_n$(4000) vs. H$\delta_A$, and stellar
mass vs. D$_n$(4000), stellar mass vs. SFR and sSFR on each line,
respectively. The star-forming (SF) galaxies, AGNs (composite,
LINER, Seyfert 2), absorption and weak emission line (Absorp(+WE))
galaxies and the complete sample of them (Total) are shown in panels from left
to right. In all these panels, the red triangles refer to the hosts
of SNe II, the blue stars for hosts of SNe Ibc, and the black filled
circles for hosts of SNe Ia. These SNe host galaxies are compared
with the SDSS main galaxy sample of galaxies (the dotted background)
in the last panel of each line.} \label{fig.iamges}
\end{figure*}

Some parameters associated with properties can indicate the evolutionary status and
star-forming history of the galaxies. In this section, we plot some relations between
properties of the 213 hosts for different types of SNe as
shown in Fig.~\ref{fig.iamges}. The parameters describing
sample galaxies are taken from the MPA/JHU database.

D$_n$(4000): The break at 4000\AA~ is a strong interruption in
the optical spectrum. Two narrow
continuum bands (3850-3950 and 4000-4100 \AA, narrower than the
first definition by \citet{1983ApJ...273..105B}), as introduced by \citet{1999ApJ...527...54B}, were used to estimate this parameter. With increasing ages
of the stellar populations of galaxies, the D$_n$(4000) values are
increasing as well, indicating a
 larger fraction of older populations.

H$\delta_A$: A strong H$\delta$ absorption line arises in galaxies
that experienced a burst of star formation that ended  about 0.1-1
Gyr ago. The peak of H$\delta$ absorption occurs once hot O and B stars, which have weak
intrinsic absorption, have terminated their evolution. The optical
light from the galaxies is then dominated by late-B to early-F
stars. \citet{1997ApJS..111..377W} defined an H$\delta_A$ index,
using  a central bandpass (4083-4122A in the MPA/JHU database)
bracketed by two pseudo-continuum band-passes.

  The stellar masses of the SNe host
galaxies are taken from \citet{2003MNRAS.341...33K} and \citet{2005MNRAS.362...41G} 
The SFRs of them are taken from \citet{2004MNRAS.351.1151B} 
(for the AGNs and absorption-line galaxies, they used the
measured D$_n$(4000) value to estimate the SFRs and denote this as
SFR$_d$).

Figure~\ref{fig.iamges} presents some relations between properties of the SNe
hosts. The D$_n$(4000) vs. H$\delta_A$, stellar mass vs.
D$_n$(4000),  stellar mass vs. SFRs and stellar mass vs. sSFRs are given from
the first to fourth lines, respectively. On each line, the
four panels show the relations when their hosts are SF, AGNs
(composites, LINERs, Seyfert 2s), Absorp (absorption and weak
emission line ones), and then all of 213 samples together with the SDSS MGS galaxies for background, respectively.

The interests of these results are as follows. First, since sample of our galaxies is large and
they have good quality of spectral observations, the host galaxies of SNe of different types could be discussed
following their BPT diagram, then the hosts as SFs, AGNs and Absorps are checked. Some of the previous studies often focus on
one kinds of hosts, for example, \citet{2008ApJ...673..999P} only study the SF ones which have metellicities 12+log(O/H) estimated from emission lines;
Second, since we are considering hosts of SNe Ia, SNe II and SNe Ibc all together,
we could compare the properties of hosts with different types of SNe, not only SNe Ia hosts or CC-SNe hosts.

From both Table~\ref{tabN} and Figure~\ref{fig.iamges}, we notice that almost all the hosts of
SNe II (the red triangles) are SF galaxies, and most of the SNe Ibc
also occur in SF galaxies. A significant part of the SNe Ia (the black filled
circles) occur in AGNs and the absorption (and weak emission) line
galaxies. The rest part of SNe Ia occur in SF
galaxies, which means their hosts have young stellar populations.
These are consistent with the observation that SNe Ia can occur in
all kinds of galaxies, from star-forming to passive cases.

Since our sample of host galaxies with different types of SNe can be classified as SF, AGN and Absorp galaxies, we could exhibit
the distributions of some parameters of galaxies in these sub-groups.
The top panels of Fig.~\ref{fig.iamges}
show that the star-forming (SF) host galaxies lie in the top-left
region of the sample, showing they have experienced very recent
star-forming activities and their young populations (with D$_n$(4000)$<$1.4) are dominate.
The AGNs (compsite, LINER and Seyfert 2) and absorption-line
galaxies occupy the region with lower H$\delta_A$ and higher
D$_n$(4000), indicating they are dominated by old stellar populations.
They are much more
massive than the SF ones, log$M/M_{\odot}$ $>$10 generally.
While many of the hosts as SF have both low
(some down to log$M/M_{\odot}$ $\sim$8) and high (10$<$ log$M/M_{\odot}$ $<$11)
stellar masses.

For the all kinds of SNe(SNe Ia,
SNe II and SNe Ibc) hosts as SF, they all span a wide range of mass(8$<$ log$M/M_{\odot}$ $<$11)
and there is no obvious differences among them.
The relations of stellar mass and SFRs
show the increasing SFRs following increasing
stellar masses in a wide range of mass of log$M/M_{\odot}$
$\sim$8-11, for hosts of all types of SNe.
In the bottom panels, the effect of mass was extracted from the
SFRs, where we put the sample galaxies in the relations of
sSFRs versus stellar mass. The
discrepancy between the different SNe hosts is clear; most
SNe II hosts are star-forming galaxies, very few are AGNs or weak
emission line galaxies. The SNe Ia can explode in all kinds of galaxies. The
hosts as AGNs and Absorption galaxies show an obvious discrepancy from
the star-forming hosts. They are more massive and have
low sSFRs.

It is interesting to compare the SNe host galaxies with the
global SDSS MGS galaxies. The last column on each line in
Fig.~\ref{fig.iamges} show that the SNe host galaxies fall well
within the regions of SDSS main galaxies.
From this column, we see the discrepancy between the SNe
II hosts and the SNe Ia hosts. The SNe Ia can explode from star-forming to passive galaxies 
since their hosts fall within all the regions.
Most of SNe II hosts locate in the regions of
young spiral galaxies, suggesting they are dominated by young stellar populations.
In a word, the host galaxies of SNe preferentially occupy some sub-regions of
the diagrams depending on their host properties and SNe types.

Here we also notice that there are four SNe II hosts appearing on the figures of ``Absorp"
galaxies. We further check the images and spectra of
these hosts. We find that three of them (ID: 1461-53062-166, 0377-52145-289, 0387-51791-587) 
are the so-called weak emission-line galaxies. Their
H$\beta$ and/or [OIII]5007 fluxes are under 3$\sigma$, while H$\alpha$ and [NII]6583
can be measured from their spectra. By using the upper limits of the line fluxes,
we roughly estimate  their positions on the BPT diagram. The results show that they 
should belong to star-forming galaxies, but close to the lower end of [OIII]5007/H$\beta$, 
suggesting they may be metal-rich galaxies. Their [OIII]5007 lines become too weak 
since there are many metal ions as coolant in metal-rich environment. 
There is one SNe II host (ID: 2586-54169-158) whose spectrum shows that it is a typical passive galaxy. But when we check its
image, we find that there is a spiral galaxy which is very close to this SN II host. 
We think there may be some uncertainties when determining which 
galaxy is the host of this type II SNe, and this is beyond the research of this paper.
There are still three SNe Ibc hosts appearing on ``Absorp" galaxies and their situations
are very similar with those four SNe II hosts. Two of them (ID: 0391-51782-442, 1626-53472-419) 
are weak emission-line galaxies. But for the other SN Ibc host (ID: 1337-52767-086), it should be an
passive galaxies, from both image and spectrum. This is interesting and should
be further studied in future. 

\section{Stellar populations of the 213 host galaxies from spectral synthesis analysis}
\label{secType}

To retrieve the stellar populations of SNe hosts,
here we consider all the 213 sample galaxies as a representative
sample to study and compare the properties of the hosts, i.e., the Type Ia, Type II and Type Ibc.
Table~\ref{tabN} shows the corresponding numbers of host galaxies as
169, 34 and 10, respectively. Here we will obtain information about
their detailed stellar population by fitting the full
optical spectra using the spectral synthesis method, both on the
continuum and absorption lines.

\subsection{Spectral synthesis method}
\label{Sect:SP}

Spectral synthesis provides an efficient way to retrieve information on
stellar populations of galaxies from observed spectra, which is a
crucial step for a deeper understanding of galaxy formation and
evolution. This is because galaxy spectra contain information on
both the age and the metallicity distributions of their stars, which
in turn reflects the star formation and chemical histories of the
gallaxies.

\label{sec.method} We fit the spectral absorptions lines and
continua of the sample galaxies to study their stellar populations
by using the software
Starlight\footnote{http://www.starlight.ufsc.br} \citep{2005MNRAS.356..270C, 
2007MNRAS.375L..16C, 2006MNRAS.370..721M, 2007MNRAS.381..263A, 2009A&A...495..457C}. 
This software fits an observed spectrum $O_{\lambda}$ with a
model $M_{\lambda}$ that adds up $N_{\ast}$ Simple Stellar
Populations (SSPs) with different ages and metallicities from
different stellar population synthesis models. A Gaussian
distribution centered at velocity $v_{\ast}$ and broadened by
$\sigma_{\ast}$ models the line-of-sight stellar motions. The fit is
carried out with the Metropolis scheme \citep{2001MNRAS.325...60C},
which searches for the minimum
$\chi^{2}=\Sigma_{\lambda}[(O_{\lambda}-M_{\lambda})\omega_{\lambda}]^{2}$,
where $\omega_{\lambda}^{-1}$ is the error in $O_{\lambda}$ except
for masked regions. Pixels that are more than $3\sigma$ away from
the rms $O_{\lambda}-M_{\lambda}$ are given zero weight by the
parameter $'clip'$.

In the outputs of STARLIGHT, one of the most important parameters that
traces the stellar population is the population vector $\vec{x}$. The
component $x_{j} (j=1,...,N_{\ast})$ represents the fractional
contribution of the SSP with age $t_{j}$ and metallicity $Z_{j}$ to
the model flux at the normalization wavelength
$\lambda_{0}=4020\AA$. Another important parameter,
the mass fraction $\mu_{j}$, has a similar meaning. For the
uncertainties in the fitting results,
Starlight group has carefully checked the reliability of
 this software by analyzing the ``stellar populations"
 of fake galaxies made with known SSPs (see Fig.4 in \citet{2005MNRAS.356..270C},
 and Fig.1 in \citet{2004ApJ...605..105C}).
\citet{2005MNRAS.356..270C} presented their errorbars centered on
the mean values obtained  by fitting 20 realizations of each of 65
test galaxies. Their three condensed populations are recovered well
by Starlight, with uncertainties smaller than 0.05 (young: $t <
10^{8}$), 0.1 (intermediate: 10$^8 < t < 10^9$ ) and 0.1 (old: $t >
10^9$ ) for S/N $>$ 10.

In this work, the optical spectra of the SNe host galaxies were fit
by using the Starlight code. We use 45 SSPs from \citet{2003MNRAS.344.1000B} 
(BC03), including 15 different ages from 1 Myr to 13 Gyr (i.e.
1, 3, 5, 10, 25, 40, 100, 280, 640, 900 Myr and 1.4, 2.5, 5, 11, 13
Gyr and 3 metallicities (i.e. 0.2, 1, and 2.5 $Z_\odot$); the
stellar evolutionary tracks of Padova 1994 \citep{1993A&AS...97..851A, 
1996A&AS..117..113G}; the Initial Mass Function (IMF) of \citet{2003PASP..115..763C}; 
and the extinction law of \citet{1989ApJ...345..245C} with $R_V =
$ 3.1. The Galactic extinctions are corrected with the reddening map
of \citet{1998ApJ...500..525S}, then shifted to the rest frame. The range
of the spectra is from 3700 to 7800 \AA~ with a step of 1 \AA~ and
normalized to the median flux in the 4010 to 4060 \AA~ region.
During spectral synthesis fitting, we exclude the emission lines,
the sky lines and four windows (5870-5905 \AA, 6845-6945
\AA,7550-7725 \AA,7165-7210 \AA),  as done in \citet{2009A&A...495..457C, 2010A&A...515A.101C}. 
For LINERs and Seyfert 2s,  a power law contribution has
been added following \citet{2010A&A...515A.101C}.

 The Starlight code will result in a contribution to the percentage of
 each SSP at a given age and metallicity to the whole SED of the
 galaxy. To see the general trend, the SSPs are put in 3 bins: young populations
with age $<$0.2 Gyr, intermediate populations with age 0.2-2 Gyr and
old populations with age $>$2 Gyr (following \citet{2010A&A...515A.101C}). The
stellar populations at three metallicities (i.e. 0.2, 1 and 2.5
$Z_\odot$) are also obtained.

\subsection{The results from Starlight}
\label{sec4.2}

We have done the spectral synthesis analysis for all the 213
individual galaxies. In Fig.~\ref{fig.fitting}, we show an example of the fitting and results.

This figure consists of four parts of plots: the top left one displays
the synthesis spectrum (red line), the observed spectrum (black
line), and the error spectrum (blue line); the bottom left one shows
the residual spectrum, where the green lines represent masked
regions as given by the SDSS flag; the right panel shows the
fractional contribution in light (top) and mass (bottom) from the 45
SSPs with different ages. We list the resulting six parameters in
the top right corners, namely ${\chi_\lambda}^2$, i.e. the reduced
${\chi}^2$; the mean relative difference between synthesis and
observed spectra $\Delta_\lambda$; the S/N in the region of
4730-4780 \AA; the $V$-band extinction; the velocity $v_\star$ and
the velocity dispersion $\sigma_\star$.

In Fig.~\ref{fig.KSsp}, we compare the stellar populations of hosts of different types of
supernovae, where we separate the 213 sample galaxies into three groups:
hosts of SNe Ia, SNe II and SNe Ibc, respectively. The top panel of
Fig.~\ref{fig.KSsp} presents the KS test
for the young populations ( $<$0.2 Gyr) of the hosts of three types of
supernovae. The middle panel of Fig.~\ref{fig.KSsp} presents the
KS test for the metal-rich populations ($Z\sim$ 2.5$Z_\odot$) of the hosts of three types of supernovae, and the
bottom panel for the
light-weighted mean ages $<logt_*>_L$ of the hosts of three types of
supernovae.

The top panel of Fig.~\ref{fig.KSsp} shows that the hosts of SNe II have more young stellar
populations than hosts of SNe Ia. It is difficult for us to distinguish the hosts of SNe Ibc from
the hosts of both SNe II and SNe Ia, partly owing to the small number of SNe Ibc hosts.
The possibilities of two hosts being drawn from the same distribution are $3.8 \times 10^{-5}$
(Ia-II), 0.17 (Ia-Ibc) and 0.42 (II-Ibc).
We should also notice that there is still a fraction of SNe Ia
hosts which have large young stellar populations, suggesting that
SNe Ia can also explode in star forming galaxies.
The middle panel shows
that the hosts of SNe Ia have more metal-rich stellar population than hosts
of SNe II. The differences between hosts of SNe Ibc and hosts of SNe Ia, as well as SNe II, 
are still not obvious. The possibilities of two hosts being drawn from the same distribution are
$1.2 \times 10^{-4}$ (Ia-II), 0.56 (Ia-Ibc) and 0.15 (II-Ibc).
The bottom panel shows that the hosts of SNe Ia
have older ages than hosts of SNe II. Hosts of SNe Ibc are not distinguished from them obviously.
The possibilities of two hosts being drawn from the same distribution are
$3.8 \times 10^{-5}$ (Ia-II), 0.051 (Ia-Ibc) and 0.50 (II-Ibc).

{From all 3 figures, we notice that the differences between SNe Ia hosts
and SNe II hosts are significant, both for age and metallicities. The differences 
between the hosts of SNe Ibc and hosts of the other two types of SNe are not 
significant enough for us to get a strong conclusion.

\begin{figure}
\centering
\includegraphics [width=8.5cm] {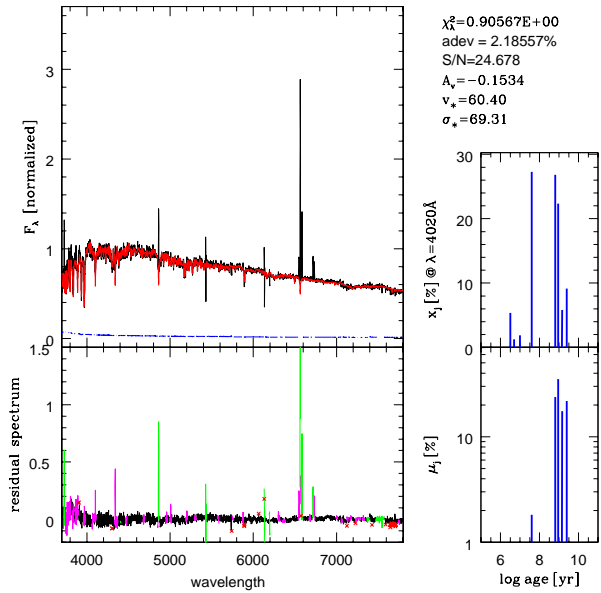}
\caption {The spectral synthesis results for one example of SNe Ia
host galaxies (PID, MJD, FID as 366, 52017, 485 respectively) by
using Starlight with 45 SSPs from \citet{2003MNRAS.344.1000B}.  We
have at top left: the synthesis spectrum (red line), the observed
spectrum (black line) and the error spectrum (blue line); at bottom
left: the residual spectrum, the green lines representing masked
regions as given by the SDSS flag; on the right: the contributed
fractions of light (top) and mass (bottom) as a function of the 15
ages of SSPs. The resulting six parameters are listed in the top
right corners. (Please see the online color version for more details.)
} \label{fig.fitting}
\end{figure}

\begin{figure}
\input epsf
\includegraphics [width=7.50cm]{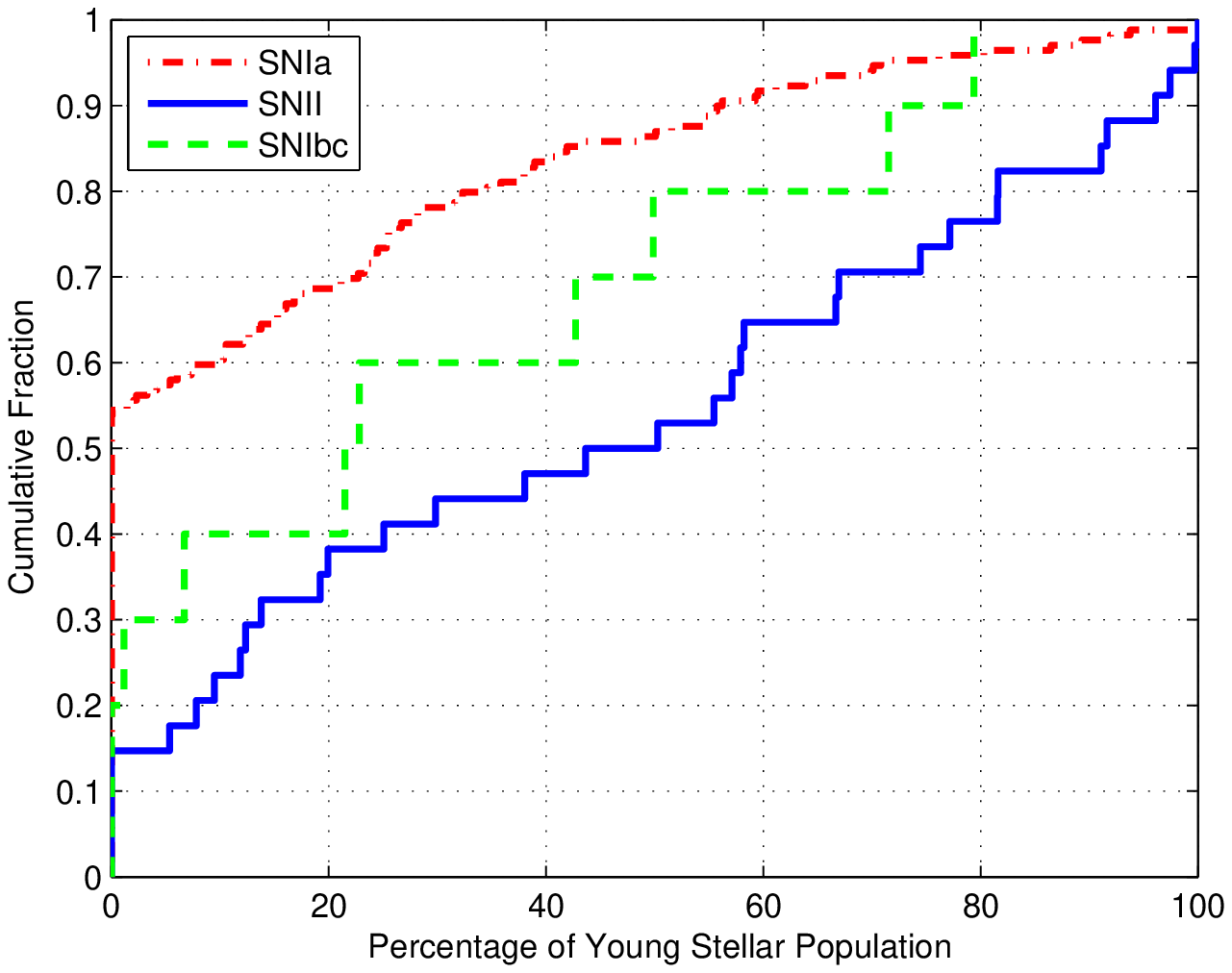}
\includegraphics [width=7.50cm]{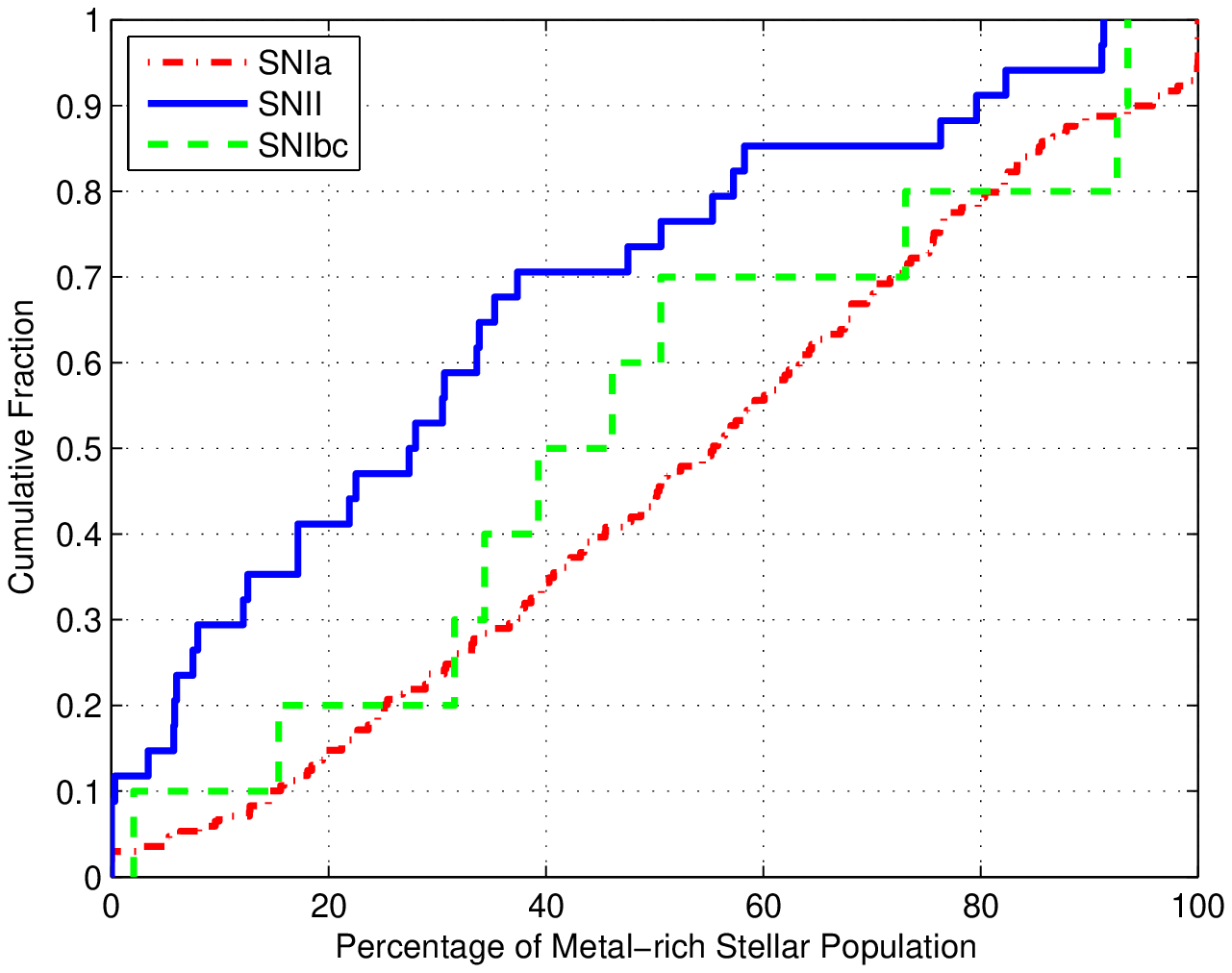}
\includegraphics [width=7.50cm]{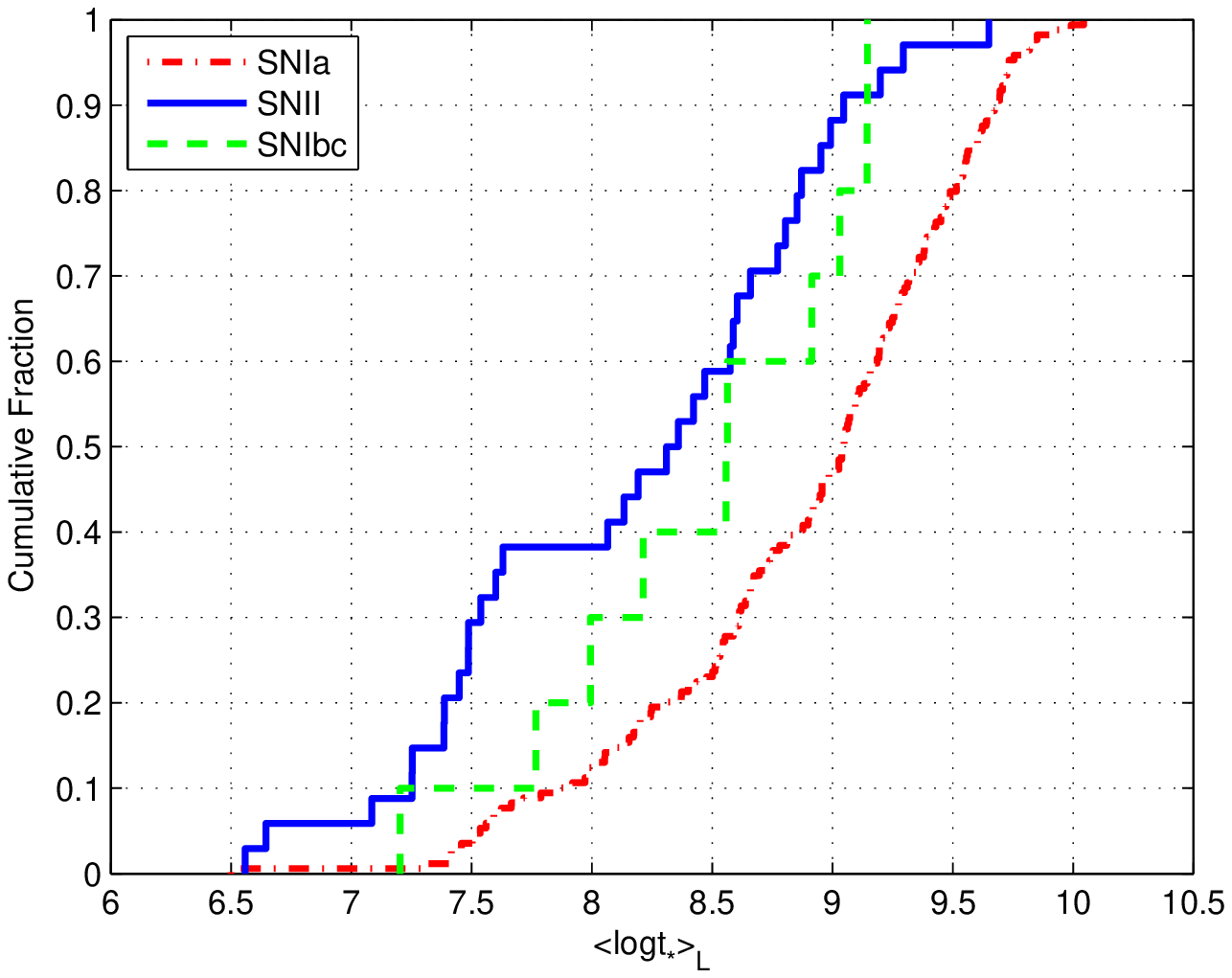}
\caption{The cumulative fraction from the KS test for the stellar
populations of the three types of supernovae in the 213 sample
galaxies: (a) for the young population ($<$0.2 Gyr), (b) for the
metal rich population ($Z\sim$ 2.5$Z_\odot$), (c) for the
light-weighted mean ages $<logt_*>_L$ of the three types of
supernovae. The dot-dashed line is for hosts of SNe Ia, the solid
line is for hosts of SNe II, and the dashed line is for hosts of SNe
Ibc.} \label{fig.KSsp}
\end{figure}

\section{Discussions}\label{Sect:Discussion}

\subsection{The comparative sample with low light fraction}
\label{lowlf}

In the main part of this work, we present the properties of SNe host
galaxies which have SDSS fiber observations covering at least 15\% of the
light from the galaxy as our main working sample. For the remaining part that has a lower light fraction,
we take them as comparison sample as mentioned in
Sect.~\ref{com.sam}. Their characteristic relations are given in
Fig.~\ref{fig.com689} following Fig.~\ref{fig.iamges}. The sample
galaxies are given in Table~\ref{tabN689} following the same items
in Table~\ref{tabN}.

Table~\ref{tabN689} shows that, in the 689 galaxies, the SFs are 50\%,
AGNs are 41\% and Absorption galaxies are 9\%. In the 213 galaxies
shown in Table~\ref{tabN}, the SFs are 38.5\%, AGNs are 24.9\% and
Absorption cases are 36.6\%. This means the comparative sample
includes less Absorption cases as the hosts of SNe Ia.
The increase of the AGN fraction can be understood since the nuclei region
are covered in the fiber when targeting the large galaxies.
This also shows the aperture effect of the fiber observations.
Therefore, we believe the criterion of light fraction  $>$ 0.15 is necessary to present the global properties of host galaxies of SNe.

For these 689 hosts from the comparative sample, we only show two sets
of plots in Fig.~\ref{fig.com689}, following Fig.~\ref{fig.iamges} for the main working sample.
As a result of the aperture effect, when
compare Fig.~\ref{fig.com689} with Fig.~\ref{fig.iamges}, we can see
some low SFRs case at a given stellar mass are
added in the 689 galaxies, which makes the data more scattered.
Owing to adding some large spiral galaxies,
whose central regions are just covered by SDSS, the SFRs of the 689 sample are lower than those of 213
samples. This phenomenon can be obviously seen when we compare the 689 sample with
the SDSS main galaxies in the fourth column of Fig.~\ref{fig.com689}, and compare with the 213 galaxies in Fig.~\ref{fig.iamges}.
The aperture effects are shown then.

Therefore, we believe the criterion of light fraction $>$ 0.15 is necessary to better present the global properties of host galaxies of SNe.

\begin{table}
{\centering \caption{The numbers of
different types of SNe among the galaxies appearing or not on the
BPT diagram, as well as the total numbers, for the 689 galaxies.
The meanings are the same as in Fig.\ref{fig.iamges}.}
\label{tabN689}
\begin{tabular}{|c|c|c|ccc|}
  \hline
 Samples & Galaxies   & Total & SN Ia & SN II & SN Ibc \\ \hline
 SF & Star-forming &  345   & 114    & 177  & 54 \\  \hline
          &  Composite    &  135   & 66    &  54  & 15 \\
 AGN   & LINER       &  121   & 68    &  46  & 7  \\
          & Seyfert 2   &   25   &  11    &  12  & 2  \\ \hline
 Absorp & Absorp. \& WE  & 63  &  40   &  19  & 4 \\ \hline
   Total  &               & 689   & 299 & 308  & 82 \\ \hline
\end{tabular}}
\end{table}

\begin{figure*}
\centering
\includegraphics [width=4.0cm] {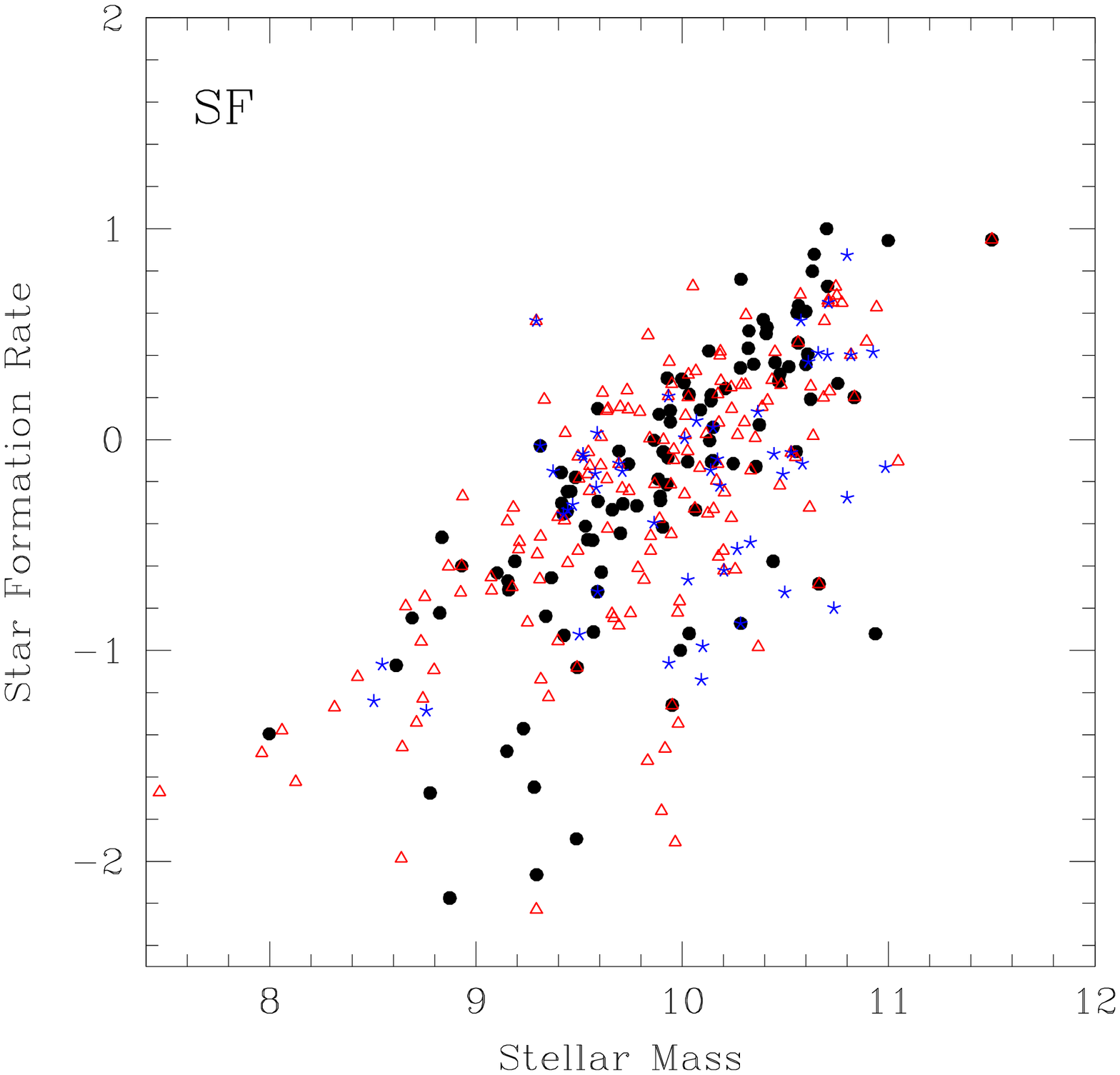}
\includegraphics [width=4.0cm] {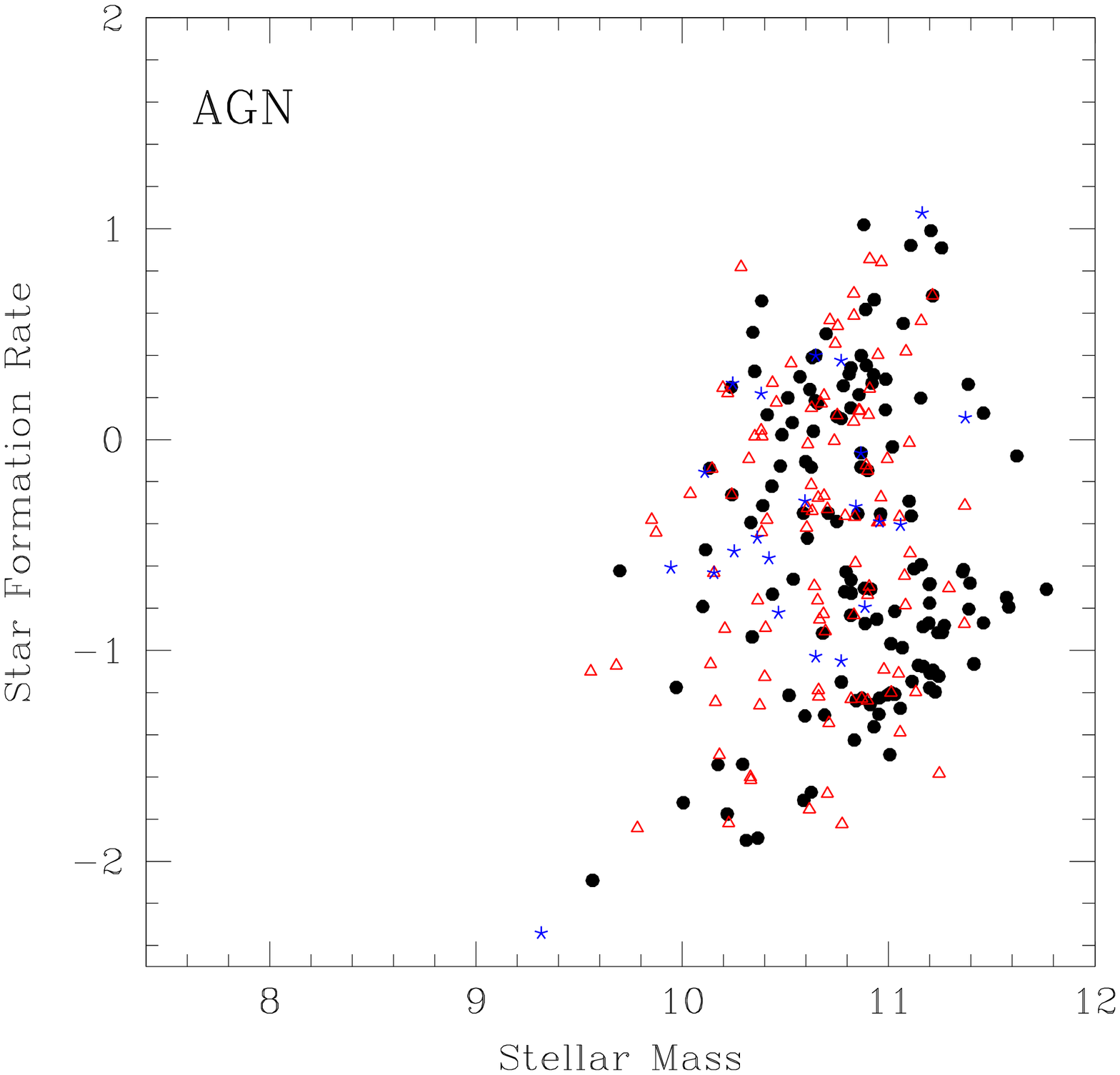}
\includegraphics [width=4.0cm] {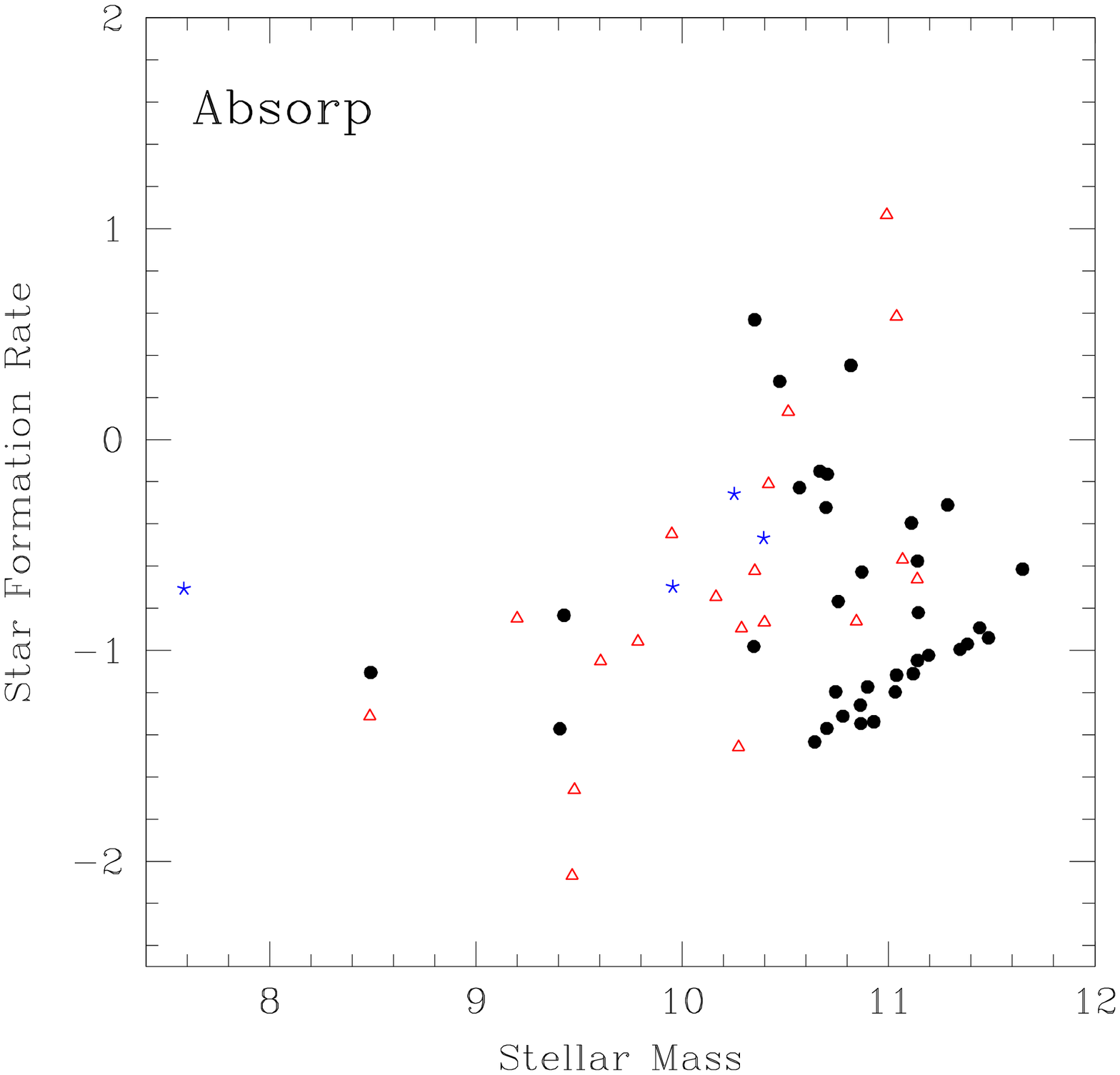}
\includegraphics [width=4.0cm] {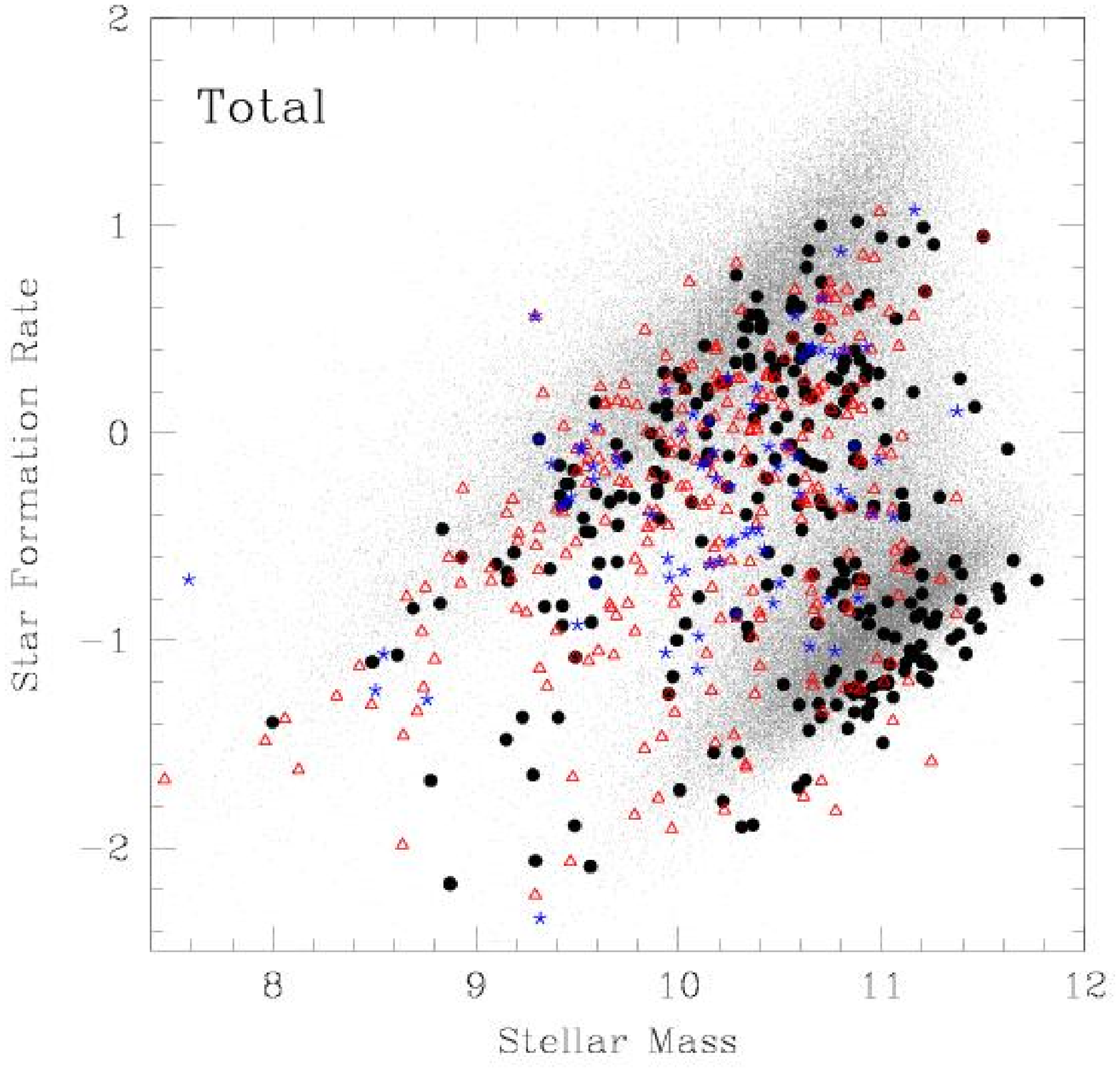} \\
\includegraphics [width=4.0cm] {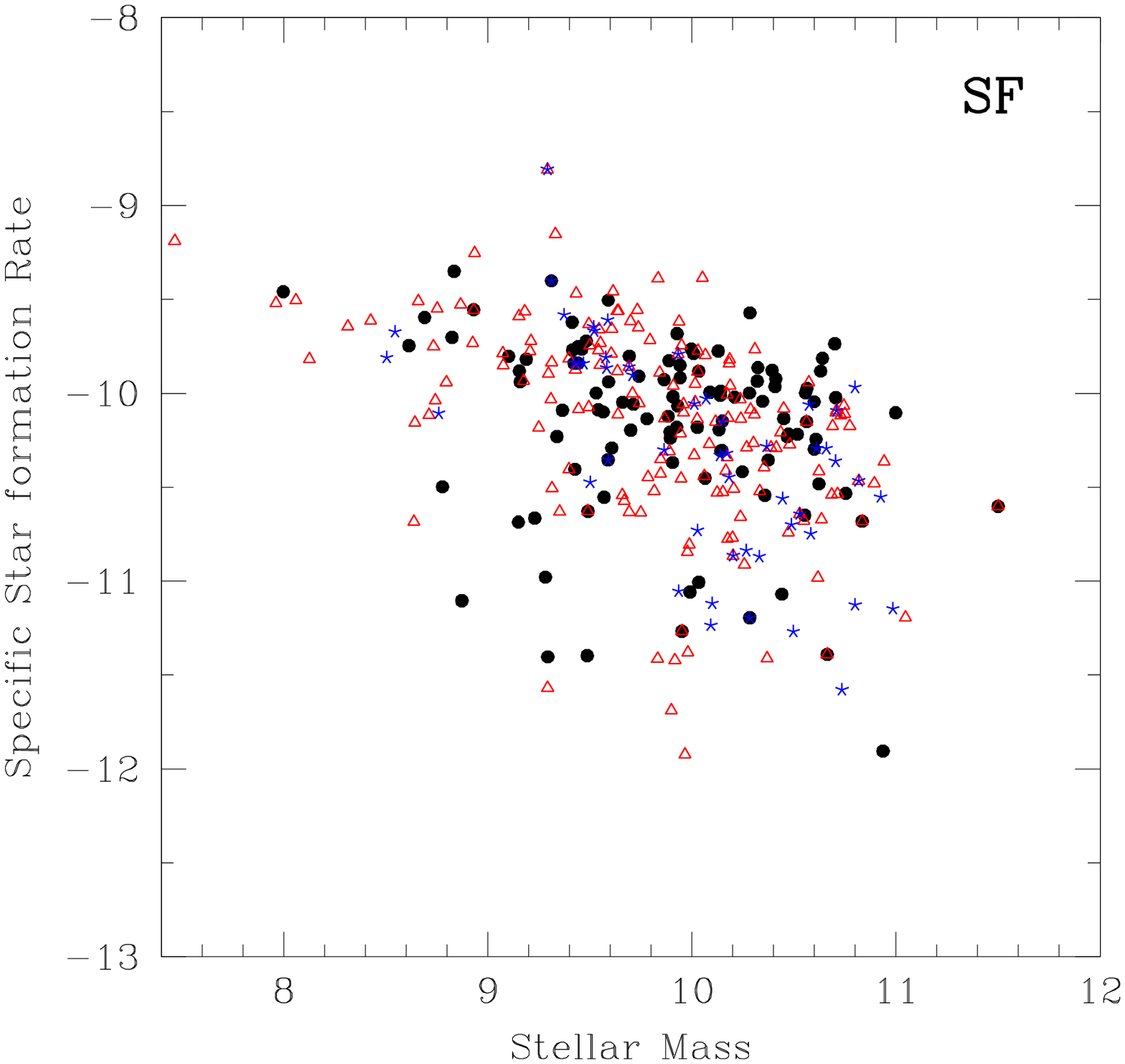}
\includegraphics [width=4.0cm] {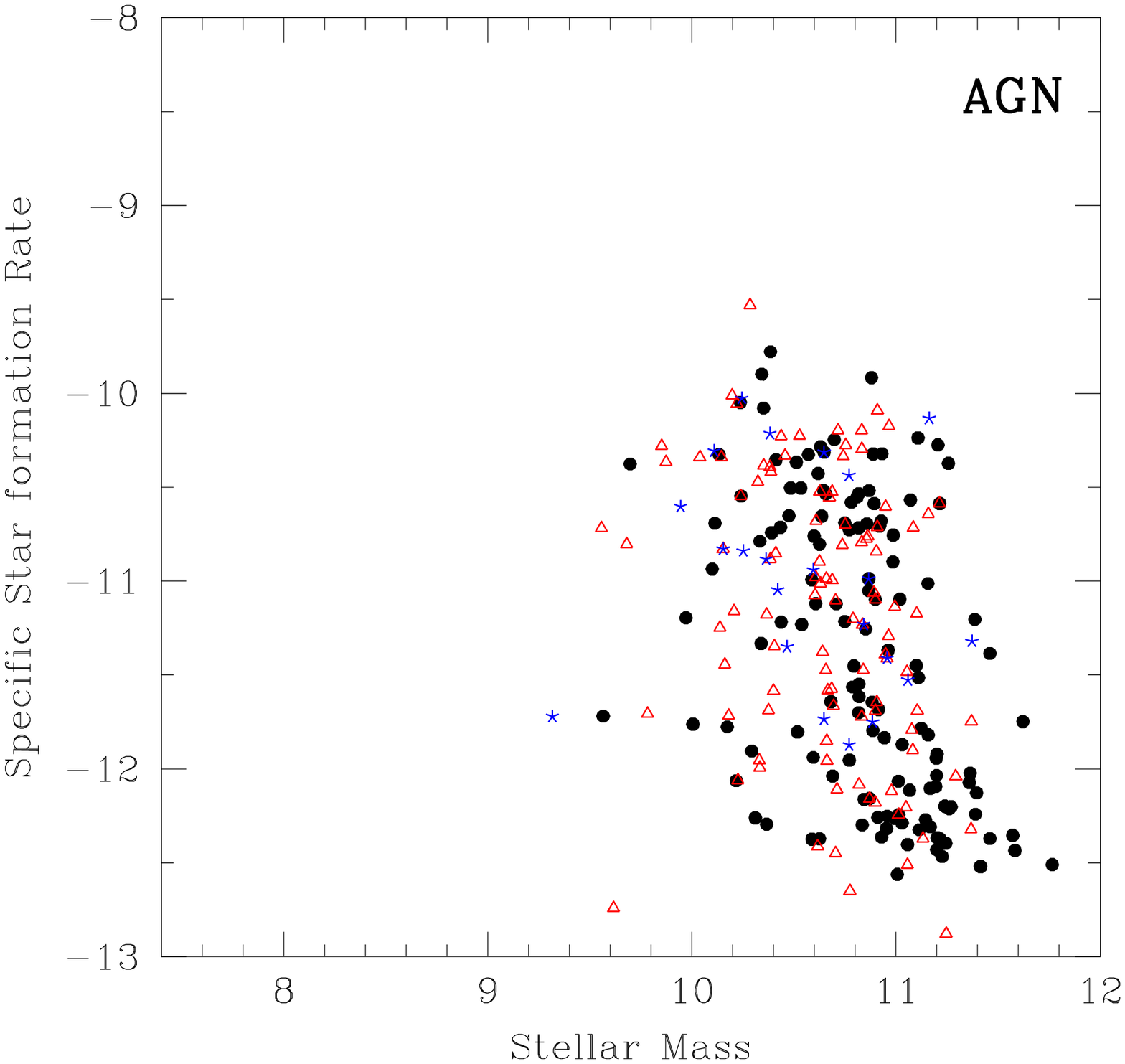}
\includegraphics [width=4.0cm] {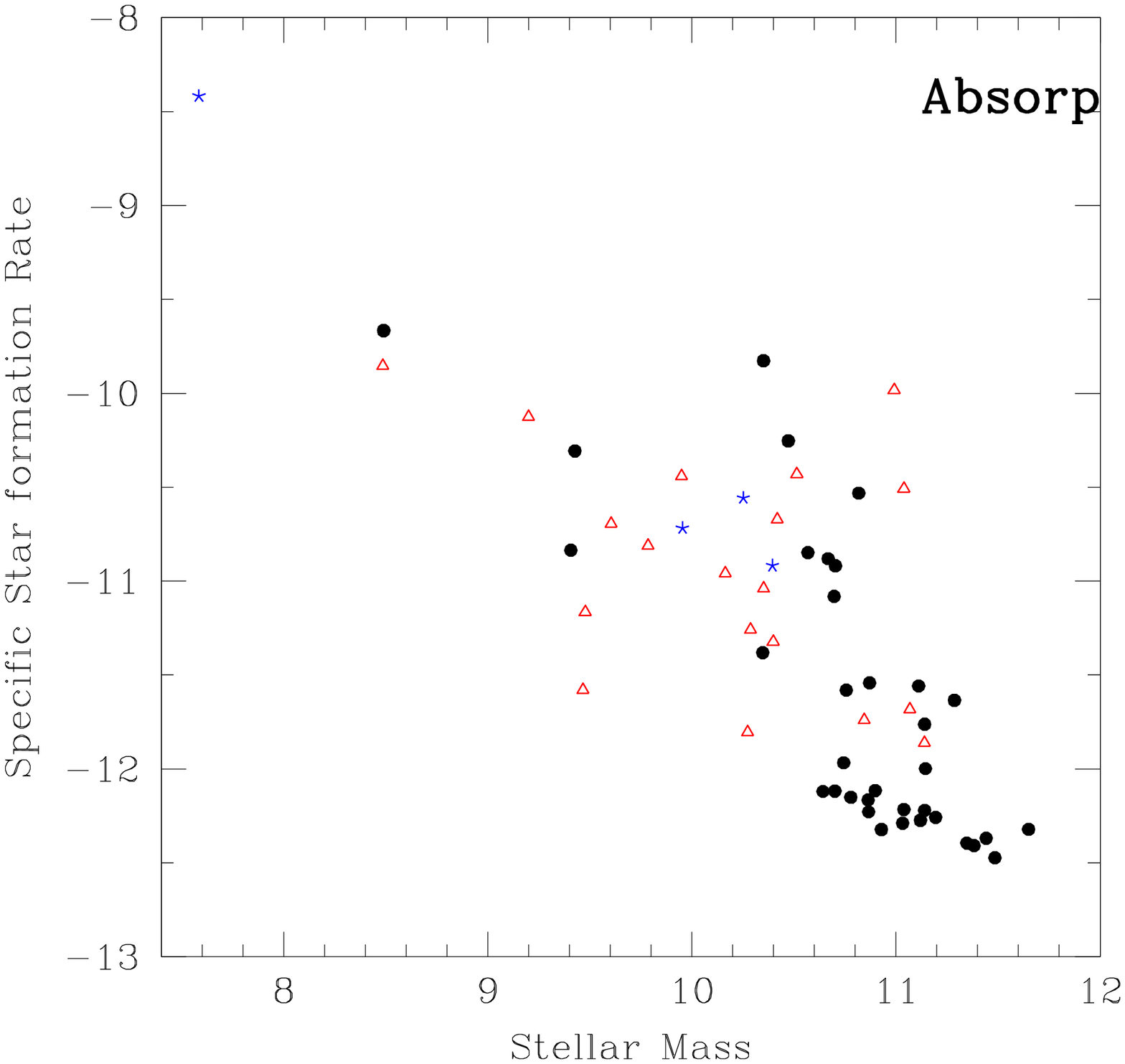}
\includegraphics [width=4.0cm] {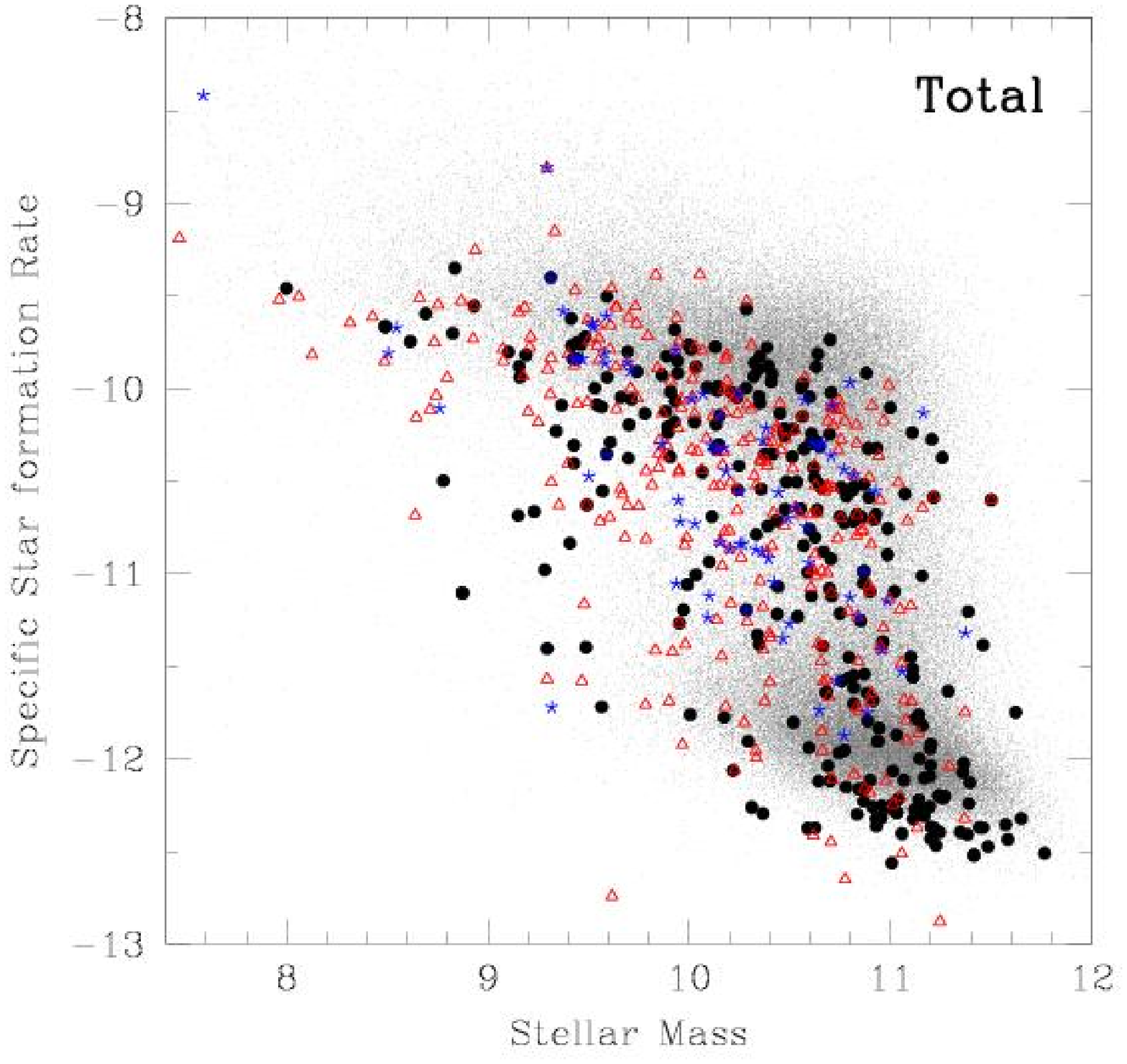}
\caption {The relations of stellar masses vs. SFRs and sSFRs for the
689 comparison sample. The symbols are the same as in
Fig.~\ref{fig.iamges}. } \label{fig.com689}
\end{figure*}

\subsection{The gas-phase metallicities and the aperture effect/bias}

The metallicities and stellar masses are fundamental parameters to
indicate the evolution status and history of galaxies.
Therefore, it is worth obtaining the relation of stellar mass vs. metallicity (MZR)
for the SNe host galaxies. This analysis is performed for both the 213 galaxies and the 689 comparison galaxies.
Here, gas-phase oxygen abundances are taken into account, thus only star forming galaxies are considered.
There are 74 from the 213 and 314 from the 689 samples.
Fig.~\ref{fig.MZR} shows that the 74 galaxies locate closer to the
MGS galaxies, but the 314 objects show about 0.1 dex higher
12+log(O/H) at a given stellar mass. The details are given as follows.

The top panels of Fig.~\ref{fig.MZR}
show the MZR of 74 SF galaxies (among 213 samples) and the KS test of the cumulative
fraction of 12+log(O/H) for hosts of SNe Ia, II and Ibc. The hosts of
SNe Ia are more metal-rich than the hosts of SNe II (Figs.~\ref{fig.MZR}a,b,c) generally.
The lower panels of Fig.~\ref{fig.MZR} are for the 314 SF galaxies (among 689 samples).
Comparing the solid lines in Fig.~\ref{fig.MZR}c and
Fig.~\ref{fig.MZR}f, it shows that these 314 galaxies are more
metal-rich than the 74 galaxies in our main working sample,
which shows the aperture effect clearly, namely, the fiber observations
cover more in central parts for these 314 host galaxies, and thus they are more metal-rich
than the 74 galaxies mentioned above at a given stellar mass.

When comparing Fig.~\ref{fig.MZR}a and Fig.~\ref{fig.MZR}d, we can see that the
points and dashed line (the 2 order polynomial fit to the MZR
for the 314 host galaxies) are generally 0.1\,dex higher than the
solid line (the 2 order polynomial fitting to the MZR of the 74
galaxies, which are shown in Fig.~\ref{fig.MZR}a), at a given
stellar mass. The main galaxy sample from SDSS of
\citet{2004ApJ...613..898T} are also added here as the dashed-line. We can see that
the dashed-line is close to the solid line, which suggests the distribution of 74 galaxies
is consistent with that of main galaxies.
From  Fig.~\ref{fig.MZR}b and Fig.~\ref{fig.MZR}e, for the
comparison with the SDSS main galaxy sample as the background, it
shows that the 74 galaxies are distributed nicely among the SDSS
galaxies, but the 314 galaxies bias toward the more metal-rich region at a
given stellar mass. These results are also the evidences that reducing
the aperture bias is very necessary to present the global properties of SNe hosts.

In Fig.~\ref{fig.MZR}f, it is the KS test
of the oxygen abundances of the 314 SNe host galaxies with a lower
light fraction. The possibilities of SNe Ia hosts and
SNe II hosts being drawn from the same distribution are 0.25.
This difference is less obvious than in Fig.~\ref{fig.MZR}c, in which
the possibilities of SNe Ia hosts and
SNe II hosts being drawn from the same distribution are 0.005.
We hold that the difference of significance is from the aperture bias, which we have discussed
above.

The difference between SNe Ia hosts and SNe Ibc hosts are not obvious in
Fig.~\ref{fig.MZR}c (significance as 0.9) perhaps part of the reason is the small number of the sample.
In Fig.~\ref{fig.MZR}f, it shows that there are no significant differences between different SN types, 
and the KS possibilities of SNe Ibc hosts and
SNe Ia hosts being drawn from the same distribution are 0.29 here.
This result is
similar to Fig.3 in \citet{2008ApJ...673..999P}, \citet{2003A&A...406..259P} and \citet{2009A&A...503..137B}. This is understandable
since both their samples and ours are focusing on star forming galaxies. This is a bit different from the results in our Sect.4.2, where not only star forming galaxies are considered, but also some AGNs and Absorption galaxies are included.

\begin{figure*}
\input epsf
\centering
\includegraphics [width=5.0cm]{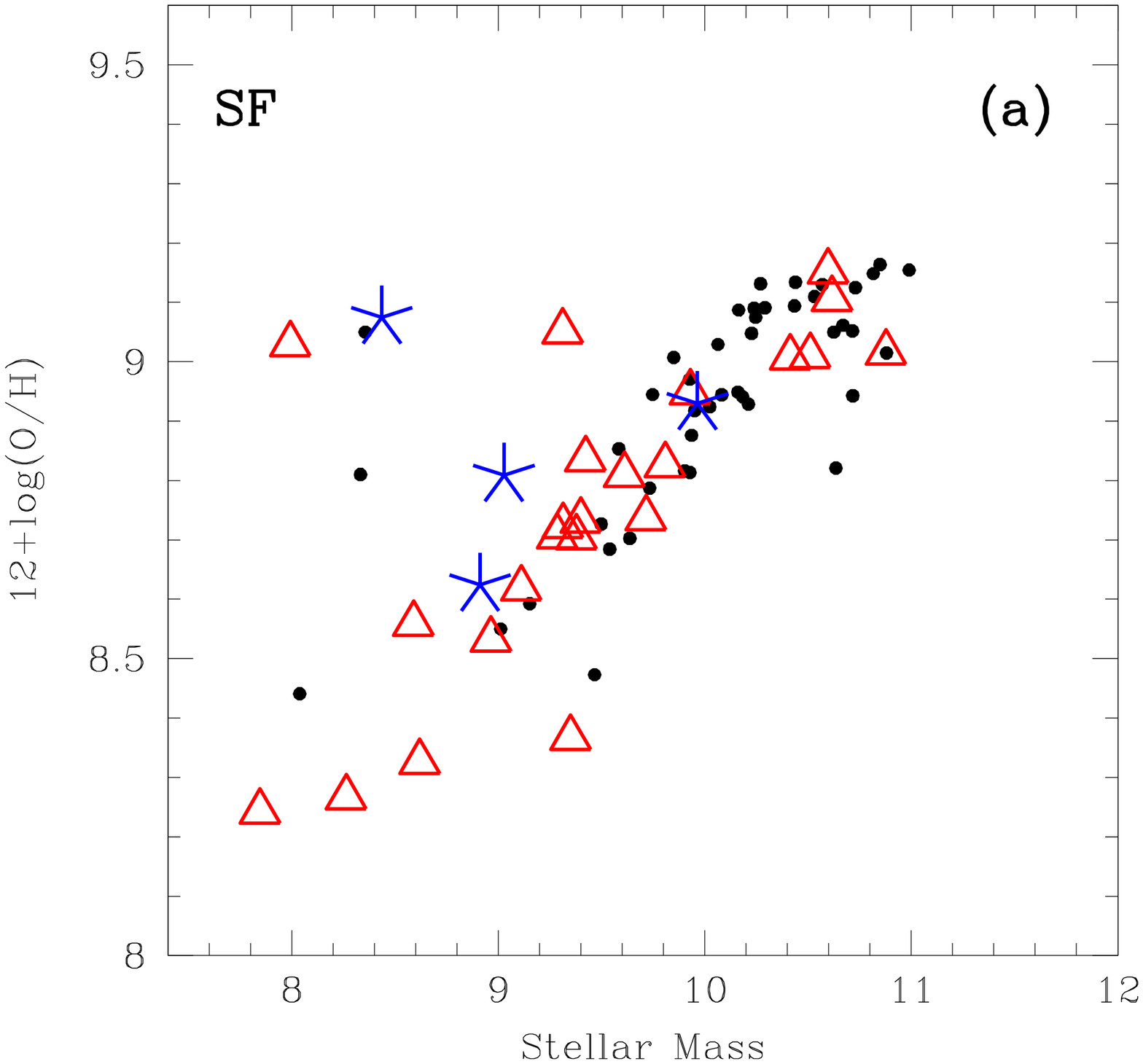}
\includegraphics [width=5.0cm]{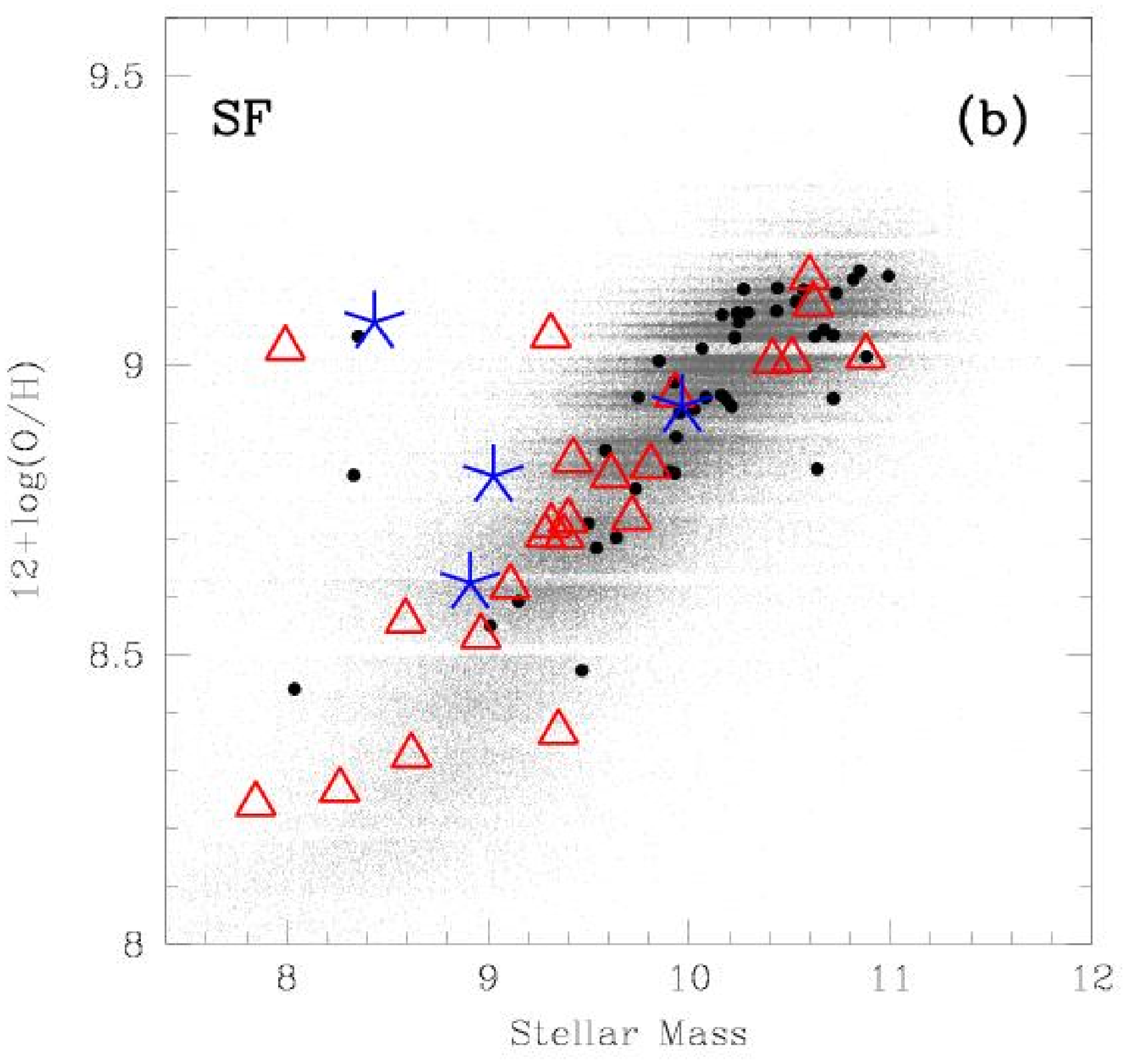}
\includegraphics [width=6.0cm]{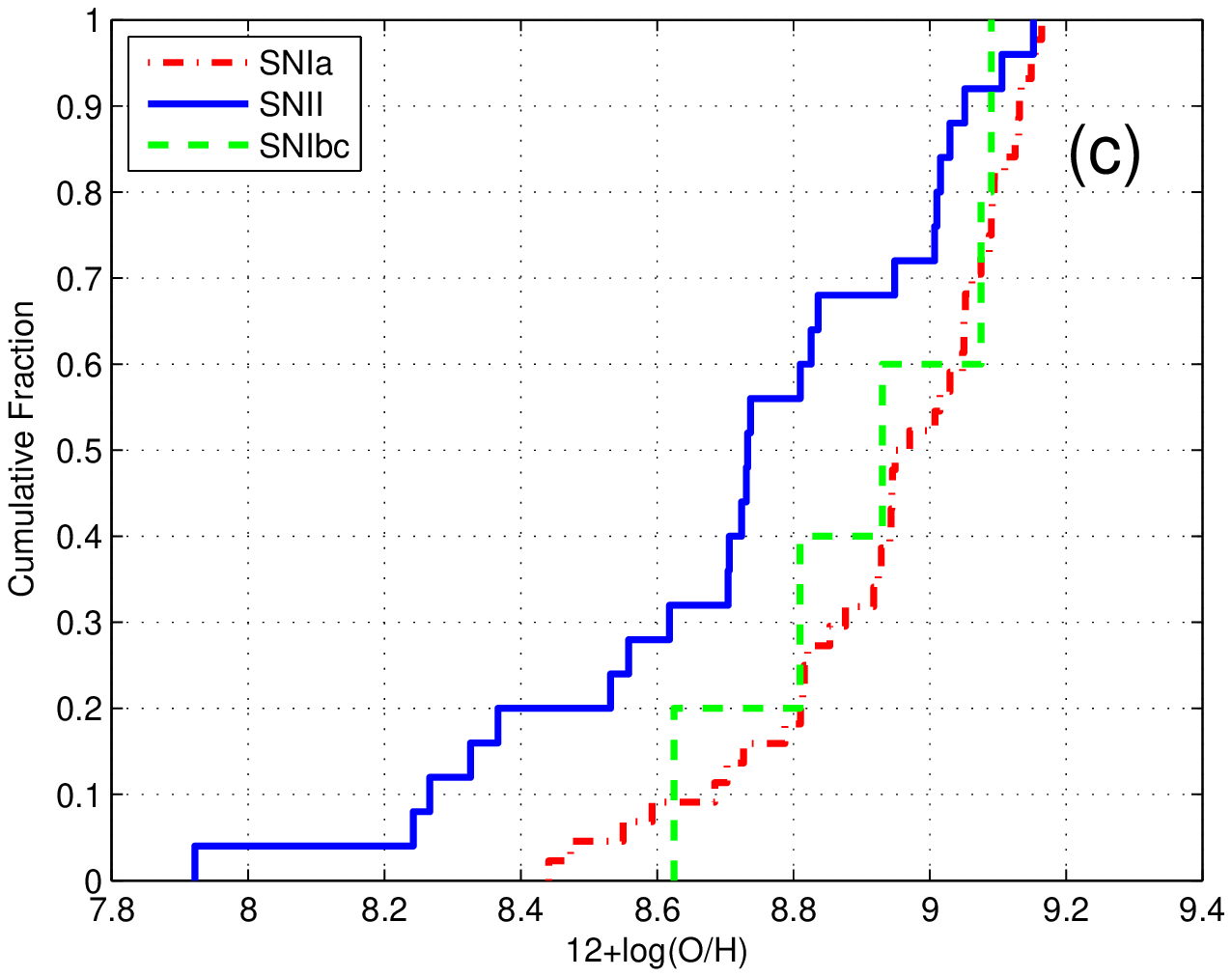} \\
\includegraphics [width=5.0cm]{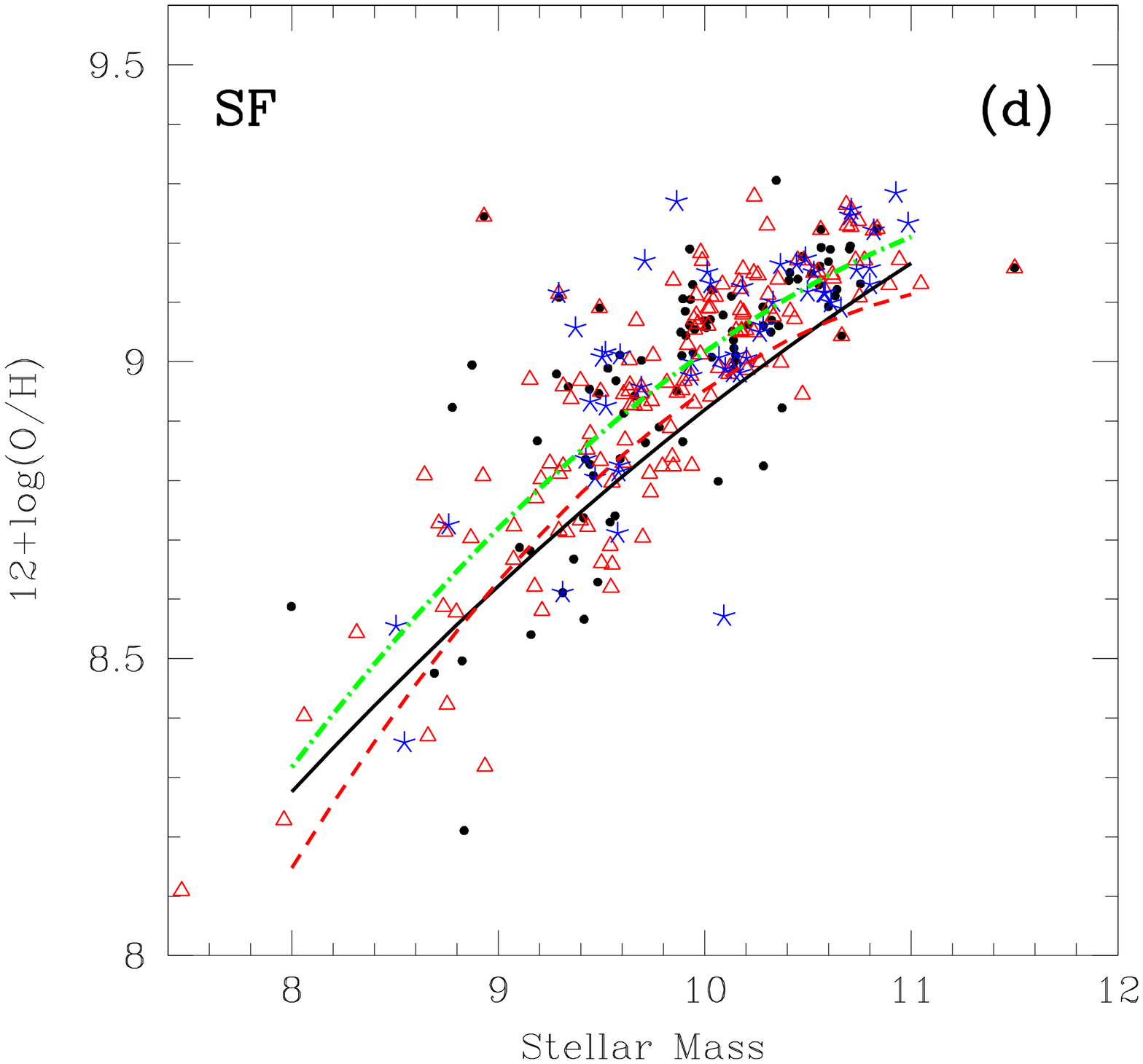}
\includegraphics [width=5.0cm]{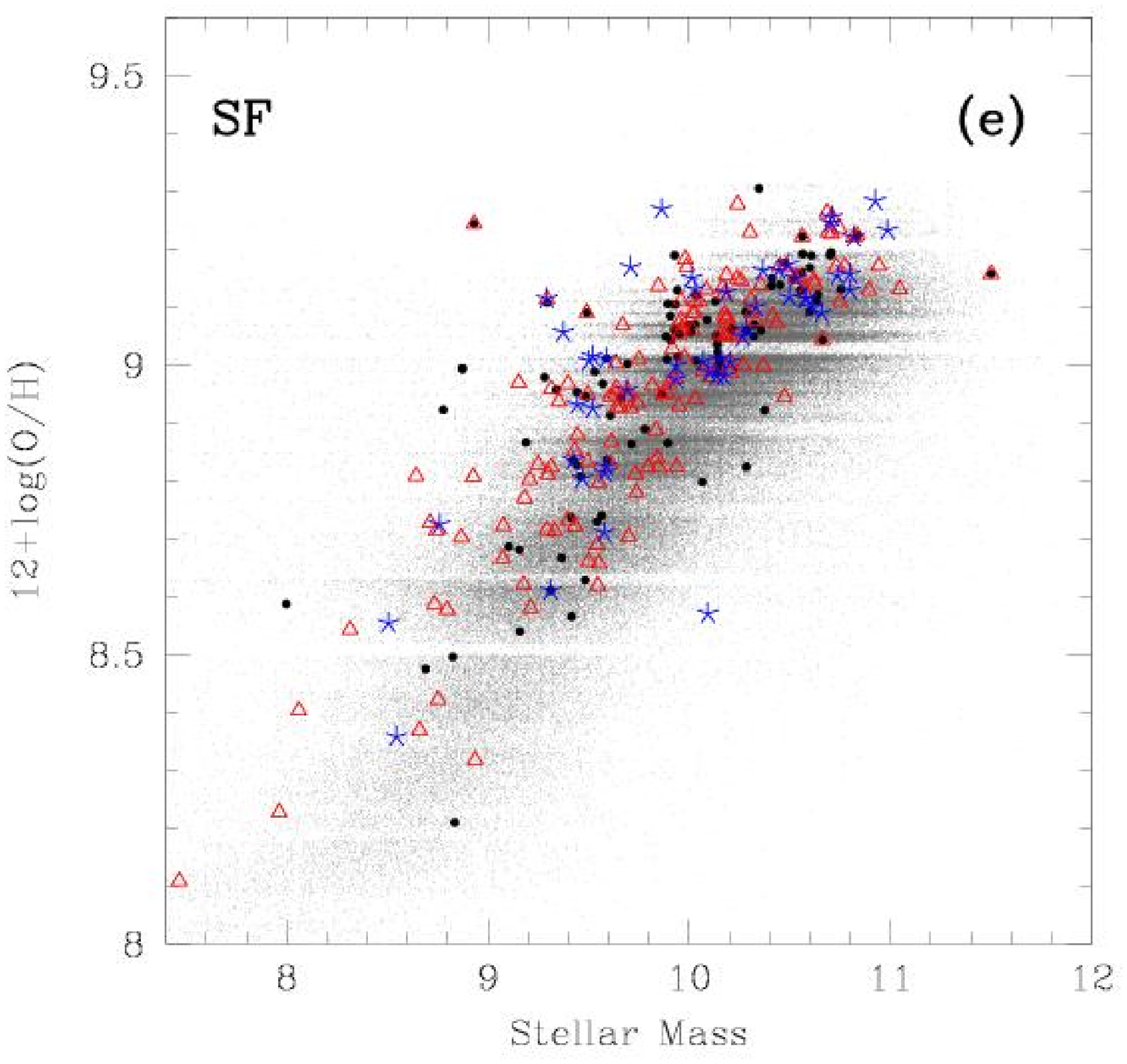}
\includegraphics [width=6.0cm]{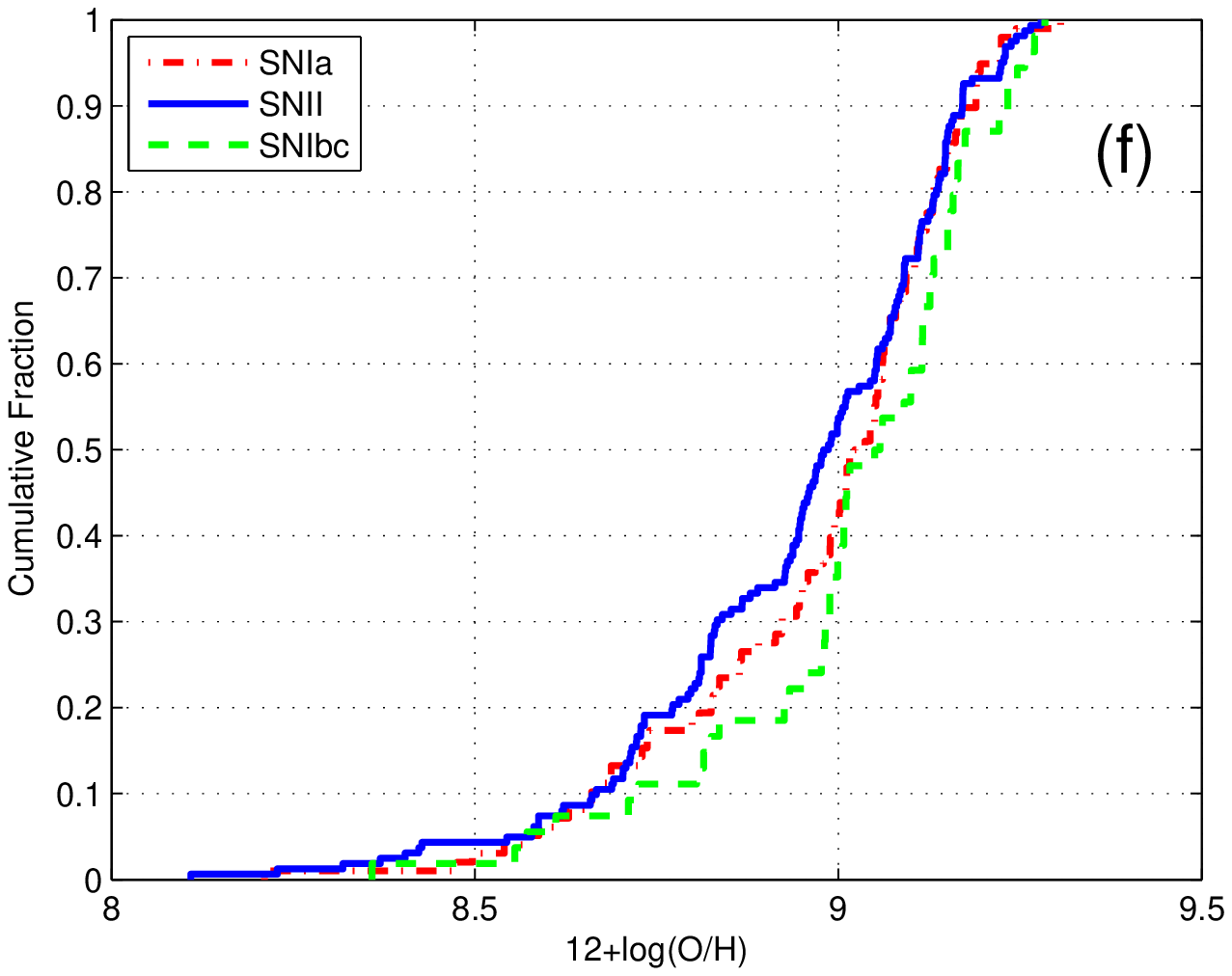}
\caption{The mass-metallicity relations of the SNe host galaxies: the
top panels for the 74 SF galaxies (among the 213 ones with higher
light fraction than 15\% in the SDSS fiber observations),  and the
bottom panels are for the 314 SF
 galaxies (among 689 ones that have a low light fraction). The triangles are hosts of SNe II,
 the stars for SNe Ibc and the filled circles for SNe Ia. In the KS test of the cumulative fraction,
 the solid line is for hosts of SNe II, the dashed line is for SNe Ibc and the dot-dashed line is for SNe Ia.
 In Fig.d, the thick dot-dashed line is for the polynomial fit for the 314 host galaxies,
 the solid line is for the 74 hosts, the objects in Fig.a, and the dashed line is for MGS of SDSS. The symbols are same as in Fig.~\ref{fig.iamges}. (Please see the online color version for details)} \label{fig.MZR}
\end{figure*}

\section{Conclusions}

In this work, we selected 902 (213+689) SNe of different types to study and
compare the properties of their host galaxies. It is an
improvement for such comparison studies since here both SNe Ia and CC-SNe
(SNe II and SNe Ibc) are considered together, and we will consider the hosts as SFs, AGNs and Absorption galaxies rather than only star forming galaxies. The sample was obtained by
cross-correlating the Asiago Supernova Catalog with the SDSS DR7
Main Galaxy Sample. In particular, we use a stricter criterion to select the sub-sample
of 213 galaxies for detailed studies by requesting the 3 arcsec SDSS
fiber observations covering at least 15\% of the light of the whole
galaxies, so that the spectra can represent the global properties of
the whole galaxies. The remaining 689 galaxies with a lower light fraction
of spectral observations are taken as a comparison sub-sample. Then
the aperture effect/bias are shown clearly by comparing these two
sub-samples.
The sample includes Type Ia, Type II and Type Ibc SNe
hosts, so we
 could compare the environments and properties of these
different types of hosts at the same time.

Thanks to the SDSS for observing the high quality optical spectra and the MPA/JHU group for publishing the property parameters of the
 galaxies, we can then compare and study the sample galaxies in detailed
 stellar population analysis and in some
 interesting relations.

We summarize our results as follows.

\begin{enumerate}

\item We further classify the sample galaxies
by their emission-line ratios on BPT diagram. Among the 213 sample galaxies, 135 of
them can be plotted on the BPT diagram, including 82 star-forming
(SF) galaxies and 53 AGNs (including composites, LINERs and Seyfert
2s). The other 78 cannot appear on BPT diagram as absorption (and
weak emission) line galaxies (named as Absorp) (See Table~\ref{tabN}).

\item As shown in Fig.~\ref{fig.iamges}, almost all the type II SNe
occur in star-forming (SF) galaxies, and only very few are in the AGNs
and weak emission-line galaxies. Most of the SNe Ibc are also in SF
galaxies. The major part of the Type Ia SNe occur in AGNs and
Absorp galaxies, and about one third in SF galaxies. The
host galaxies as SF have a wide range of stellar mass, from
log(M/M$_{\odot}$)$\sim$8 to 11. However, the host galaxies as AGNs and Absorp
galaxies are massive, mostly with log(M/M$_{\odot}$)$>$10, and
the Absorp host galaxies are even more massive, up to 11.6.

\item When we put all
these SNe host galaxies together in the relations of D$_n$(4000) vs.
H$\delta_A$, stellar mass vs. D$_n$(4000), stellar mass vs. SFRs and
sSFRs, two groups for SF and a significant part of Absorp \& AGNs can
been shown for the hosts, but the remaining part of the AGN \& Absorp is in
the middle and even mixed with the corner of the SF galaxies. This
is especially clear in the relation of stellar masses vs. SFRs and
sSFRs as shown in Fig.~\ref{fig.iamges}.
Thus, the SNe host galaxies fall well within the
global SDSS sample, but preferentially occupy some sub-regions of
the diagrams depending on the properties of their hosts.

\item The KS test of the cumulative fraction
for stellar population analysis from spectral
synthesis fitting shows that the hosts of type
II SNe have a younger stellar population and
are younger than hosts of Type Ia SNe. The hosts
of SNe Ia have more metal-rich stellar pop-
ulations and are more metal-rich than hosts
of SNe II. The differences between the hosts
of SNe Ibc and the hosts of other two types
of SNe are not obvious.

\item The same relations of parameters describing properties have been made for the comparison sample with 689 galaxies.
The AGN fraction of the sample is higher
than that of the 213 objects since only nuclei region
are covered in the fiber targeting the large galaxies.

\item The stellar mass-metallicity relations of the star forming galaxies in the two
sub-samples are also presented.
In the MZR, our main working sample of the galaxies with a higher light fraction
of spectral observations are closer to the SDSS MGS galaxies, but the comparison
sample is about 0.1 dex higher in 12+log(O/H) at a given stellar mass. This confirms
that the aperture effect of spectral observations should be taken into account when
we try to understand the properties of SN host galaxies.
The KS test for the comparative sample (Fig.~\ref{fig.MZR}f) shows that 
there are no significant differences between hosts of different SN types.
We should keep in mind that only star forming galaxies are considered here.

\end{enumerate}

 In this
work we have concentrated on the properties of the SNe hosts.  However, we could not give more restrictions on the progenitors of supernovae, and the
relations between the host properties and properties of the SNe
themselves, such as the decline time, the stretch, the SNe peak
luminosity or the SNe rates. These should be discussed in further
studies.

\begin{acknowledgements}
We appreciate our referee who provided very admirable, constructive
and helpful comments and suggestions, which helped to
improve well our work. We also thank James Wicker for improving our English description in the text from the native language.
 We thank Zhanwen Han, Xiangdong Li, Tianmeng Zhang, Bo Wang, Xiaofeng Wang, Yan Gao for helpful discussions. This study is supported by the
Natural Science Foundation of China under grants Nos. 10933001,
11273026, 11273011, 11178013 and the National Basic Research Program of China
(973 program) No. 2007CB815104/06.

\end{acknowledgements}


\begin{table*}
\begin{center}
\caption{On-line data: Basic information of the sample galaxies.
 }
\label{tab-all.lis} \tiny
\begin{tabular}{lllcccccccc}
\hline
   No.  &  RA (ASC)           & DEC (ASC)           &   RA (SDSS)         &  DEC (SDSS)         &   PID-MJD-FID       & Petrion  & light  & redshift    &    Type  & name of SN  \\
     &             &            &            &           &         & radius & fraction &    &    &   \\  \hline
     1     &      101800         &      -000158     &     154.50199890  &     -0.03280395    &     0271-51883-171      &      2.658  &   0.490  &   0.065    &     Ia  &   2000 fx   \\
     2     &      111610         &      -001139     &     169.04255676  &     -0.19220079    &     0279-51984-180      &      3.778  &   0.389  &   0.156    &     Ia  &   1996 R    \\
     3     &      112225         &       011121     &     170.60607910  &      1.18944538    &     0280-51612-409      &      6.564  &   0.182  &   0.029    &     Ia  &   1997 bz   \\
     4     &      124733         &       000557     &     191.88919067  &      0.09919899    &     0291-51928-076      &      3.277  &   0.428  &   0.086    &     Ia  &   2001 kk   \\
     5     &      133955         &       005215     &     204.98309326  &      0.87136489    &     0299-51671-380      &      2.672  &   0.491  &   0.069    &     Ia  &   2005 ac   \\
     6     &      160713         &      -000449     &     241.80650330  &     -0.07878780    &     0344-51693-039      &      9.616  &   0.179  &   0.031    &     II  &   2001 ax   \\
     7     &      173228         &       560425     &     263.11889648  &     56.07372665    &     0358-51818-181      &      4.720  &   0.290  &   0.122    &     Ia  &   2000 gb   \\
     8     &      172822.703     &       573239.00  &     262.09924316  &     57.54539871    &     0358-51818-364      &      7.019  &   0.184  &   0.028    &     II  &   2004 eb   \\
     9     &      172611         &       591831     &     261.54681396  &     59.30865479    &     0366-52017-485      &      7.421  &   0.225  &   0.027    &     Ia  &   2009 ia   \\
    10     &      223507         &      -010637     &     338.78216553  &     -1.10929883    &     0377-52145-289      &      5.047  &   0.222  &   0.095    &     II  &   2007 qv   \\
    11     &      224142         &      -000812     &     340.42520142  &     -0.13693197    &     0377-52145-594      &      3.322  &   0.430  &   0.057    &     Ia  &   2006 py   \\
    12     &      224459         &      -010023     &     341.24508667  &     -1.00638306    &     0378-52146-135      &      4.531  &   0.302  &   0.128    &     Ia  &   2006 nd   \\
    13     &      225942         &      -000049     &     344.92785645  &     -0.01355672    &     0380-51792-192      &      5.915  &   0.233  &   0.045    &     Ia  &   2005 ku   \\
    14     &      231154         &      -003441     &     347.97653198  &     -0.57908654    &     0381-51811-106      &      9.227  &   0.245  &   0.091    &     Ia  &   2007 pd   \\
    15     &      231729         &       002545     &     349.37374878  &      0.42967373    &     0382-51816-486      &      3.337  &   0.403  &   0.119    &     Ia  &   2005 fh   \\
    16     &      232640         &      -005024     &     351.66729736  &     -0.84061658    &     0383-51818-057      &      4.148  &   0.300  &   0.082    &     Ia  &   2006 fy   \\
    17     &      232439         &      -004306     &     351.16223145  &     -0.71791536    &     0383-51818-134      &      6.191  &   0.246  &   0.148    &     Ia  &   2006 ju   \\
    18     &      232447         &       005640     &     351.19976807  &      0.94429779    &     0383-51818-445      &      6.072  &   0.289  &   0.118    &     Ia  &   2007 lv   \\
    19     &      232749         &       002726     &     351.95654297  &      0.45790046    &     0383-51818-581      &      5.753  &   0.177  &   0.118    &     II  &   2007 pg   \\
    20     &      232807         &       005122     &     352.03250122  &      0.85817730    &     0383-51818-618      &      6.188  &   0.272  &   0.119    &     Ia  &   2006 ol   \\
    21     &      233947         &       001218     &     354.94631958  &      0.20583645    &     0385-51877-427      &      6.176  &   0.186  &   0.068    &     Ia  &   2006 hq   \\
    22     &      235420         &      -005503     &     358.58627319  &     -0.91723460    &     0386-51788-002      &      7.028  &   0.222  &   0.105    &     Ia  &   2007 om   \\
    23     &      000348         &       002131     &       0.95135558  &      0.35961452    &     0387-51791-587      &      5.300  &   0.229  &   0.099    &     II  &   2007 nx   \\
    24     &      001124         &       004209     &       2.85239959  &      0.70202297    &     0388-51793-584      &      5.126  &   0.276  &   0.199    &     Ia  &   2006 jz   \\
    25     &      001641         &      -002528     &       4.17437315  &     -0.42516112    &     0389-51795-151      &      4.547  &   0.335  &   0.104    &     Ia  &   2006 fz   \\
    26     &      002635         &      -001807     &       6.64762497  &     -0.30328667    &     0390-51900-033      &      5.012  &   0.310  &   0.235    &     Ia  &   2006 ff   \\
    27     &      002137         &      -010035     &       5.40788269  &     -1.01061416    &     0390-51900-133      &      6.640  &   0.250  &   0.084    &     Ia  &   2006 er   \\
    28     &      002000         &      -003729     &       5.00321960  &     -0.62535346    &     0390-51900-237      &      5.332  &   0.240  &   0.067    &     II  &   2006 fq   \\
    29     &      002244         &      -002844     &       5.68314219  &     -0.47940356    &     0391-51782-316      &      5.992  &   0.180  &   0.108    &     Ia  &   2007 px   \\
    30     &      002741         &       011359     &       6.92455101  &      1.23239827    &     0391-51782-442      &      8.554  &   0.153  &   0.080    &     Ib  &   2007 qx   \\
    31     &      004900         &      -001926     &      12.25123310  &     -0.32327026    &     0393-51794-034      &      5.384  &   0.229  &   0.156    &     Ia  &   2007 lr   \\
    32     &      004624         &       000012     &      11.60083675  &      0.00237388    &     0393-51794-106      &      9.534  &   0.186  &   0.116    &     Ia  &   2006 gf   \\
    33     &      005252         &      -000431     &      13.21838474  &     -0.07436865    &     0394-51913-177      &      4.455  &   0.307  &   0.115    &     II  &   1996 bg   \\
    34     &      004909         &       003549     &      12.28801918  &      0.59661305    &     0394-51913-389      &      5.722  &   0.263  &   0.115    &     Ia  &   2007 lc   \\
    35     &      005924         &       000009     &      14.85048771  &      0.00266656    &     0395-51783-511      &      6.059  &   0.226  &   0.062    &     Ia  &   2005 ho   \\
    36     &      011026         &      -010404     &      17.60987854  &     -1.06876671    &     0397-51794-206      &      6.037  &   0.244  &   0.154    &     II  &   2006 gd   \\
    37     &      011058         &       001634     &      17.74193001  &      0.27615604    &     0397-51794-477      &      2.925  &   0.449  &   0.065    &     Ia  &   2005 kt   \\
    38     &      011357         &       002218     &      18.48825455  &      0.37141645    &     0397-51794-550      &      6.034  &   0.224  &   0.045    &     Ia  &   2006 hx   \\
    39     &      011337         &       002525     &      18.40657425  &      0.42374301    &     0397-51794-551      &      9.715  &   0.163  &   0.046    &     Ia  &   2006 ne   \\
    40     &      011612         &       004731     &      19.05294991  &      0.79055309    &     0398-51789-362      &      6.201  &   0.204  &   0.086    &     Ia  &   2005 gb   \\
    41     &      011643         &       004740     &      19.18280983  &      0.79359138    &     0398-51789-378      &      5.974  &   0.159  &   0.076    &     Ia  &   2005 ir   \\
    42     &      011502         &       001542     &      18.75957489  &      0.26294789    &     0398-51789-440      &      8.026  &   0.192  &   0.047    &     Ia  &   2004 hu   \\
    43     &      012314         &      -001946     &      20.81234741  &     -0.33023268    &     0399-51817-191      &      8.780  &   0.174  &   0.076    &     Ib  &   2006 jo   \\
    44     &      012137         &       002452     &      20.40770721  &      0.41437322    &     0399-51817-349      &      3.564  &   0.394  &   0.131    &     Ia  &   2006 fv   \\
    45     &      013723         &      -001843     &      24.34909821  &     -0.31172231    &     0400-51820-023      &      5.707  &   0.310  &   0.056    &     Ia  &   2007 ol   \\
    46     &      013441         &      -003619     &      23.67431831  &     -0.60421115    &     0400-51820-164      &      4.035  &   0.363  &   0.079    &     Ia  &   2005 js   \\
    47     &      013936         &      -004531     &      24.90002060  &     -0.75794107    &     0401-51788-284      &      4.233  &   0.363  &   0.161    &     Ia  &   2005 fa   \\
    48     &      015236         &      -001331     &      28.15243721  &     -0.22807825    &     0402-51793-071      &      5.786  &   0.289  &   0.176    &     Ia  &   2002 gp   \\
    49     &      015020         &      -002413     &      27.58425140  &     -0.40386590    &     0402-51793-179      &      3.675  &   0.348  &   0.127    &     Ia  &   2007 rc   \\
    50     &      015840         &      -001456     &      29.66862106  &     -0.24855912    &     0403-51871-110      &      7.308  &   0.160  &   0.080    &     II  &   2007 ll   \\
    51     &      015654         &      -010649     &      29.22780800  &     -1.11471581    &     0403-51871-256      &      8.867  &   0.205  &   0.043    &     Ia  &   2002 gn   \\
    52     &      015358         &      -000533     &      28.49486351  &     -0.09352394    &     0403-51871-307      &      6.715  &   0.170  &   0.088    &     Ia  &   2007 rj   \\
    53     &      020232         &      -010521     &      30.63646698  &     -1.08993196    &     0404-51812-285      &      4.636  &   0.258  &   0.136    &     Ia  &   2007 jw   \\
    54     &      020416         &       003911     &      31.07067490  &      0.65270579    &     0404-51812-372      &      7.091  &   0.170  &   0.075    &     II  &   2001 im   \\
    55     &      020335         &       004710     &      30.89647484  &      0.78468317    &     0404-51812-373      &      6.329  &   0.191  &   0.061    &     II  &   2007 hw   \\
    56     &      020503         &       001028     &      31.26476669  &      0.17514242    &     0404-51812-493      &      7.137  &   0.186  &   0.077    &     Ia  &   2007 mn   \\
    57     &      020719         &       011507     &      31.82991409  &      1.25202501    &     0404-51812-565      &      8.535  &   0.197  &   0.174    &     Ia  &   2006 ia   \\
    58     &      021802         &      -003332     &      34.50863647  &     -0.55963063    &     0405-51816-066      &      3.197  &   0.400  &   0.143    &     Ia  &   2004 ia   \\
    59     &      022824         &       001104     &      37.09963226  &      0.18599379    &     0406-51869-633      &      3.573  &   0.382  &   0.165    &     Ia  &   2007 ok   \\
    60     &      023759         &      -010139     &      39.49538803  &     -1.02751327    &     0408-51821-201      &      5.465  &   0.248  &   0.135    &     Ia  &   2001 eu   \\
    61     &      023526         &       010429     &      38.86076736  &      1.07453907    &     0408-51821-337      &      4.446  &   0.321  &   0.093    &     Ia  &   2005 je   \\
    62     &      025229         &      -010822     &      43.12130737  &     -1.13941860    &     0410-51816-247      &      6.460  &   0.203  &   0.136    &     Ia  &   2007 qr   \\
    63     &      030702         &      -000040     &      46.76201248  &     -0.01128918    &     0411-51817-072      &      4.023  &   0.358  &   0.107    &     Ia  &   2004 il   \\
    64     &      025953         &       010938     &      44.97357178  &      1.16005981    &     0411-51817-322      &      6.006  &   0.178  &   0.072    &     Ia  &   2007 jd   \\
    65     &      030522         &       005130     &      46.34433365  &      0.85972154    &     0411-51817-571      &      6.605  &   0.220  &   0.118    &     Ia  &   2005 fv   \\
    66     &      030851         &      -011024     &      47.21431732  &     -1.17334652    &     0412-52258-210      &      4.517  &   0.359  &   0.126    &     Ia  &   2001 kl   \\
    67     &      032331         &       003960     &      50.88061905  &      0.66727877    &     0413-51929-621      &      3.924  &   0.392  &   0.132    &     Ia  &   2007 mi   \\
    68     &      033012         &      -005828     &      52.55371475  &     -0.97447813    &     0415-51810-281      &      6.976  &   0.164  &   0.067    &     Ia  &   2005 if   \\
    69     &      033602         &       010617     &      54.00661087  &      1.10475719    &     0415-51810-571      &     10.190  &   0.178  &   0.040    &     Ia  &   2007 jh   \\
    70     &      033602         &       010617     &      54.00661087  &      1.10475719    &     0415-51810-571      &     10.190  &   0.178  &   0.040    &     Ia  &   2010 kf   \\
    71     &      033942         &       010532     &      54.92742920  &      1.09268486    &     0416-51811-411      &      7.585  &   0.206  &   0.181    &     Ia  &   2007 hy   \\
    72     &      034044.445     &       010323.84  &      55.18504715  &      1.05656338    &     0416-51811-416      &     10.503  &   0.181  &   0.023    &     Ia  &   2010 jf   \\
    73     &      034310         &       000608     &      55.79183960  &      0.10396237    &     0416-51811-514      &      4.798  &   0.263  &   0.130    &     Ia  &   2007 ia   \\ \hline
\end{tabular}
\end{center}
\end{table*}

\addtocounter{table}{-1}
\begin{table*}
\begin{center}
\caption{-continued} \label{tab-all.lis} \tiny
\begin{tabular}{lllcccccccc}
\hline
   No.  &  RA (ASC)           & DEC (ASC)           &   RA (SDSS)         &  DEC (SDSS)         &   PID-MJD-FID       & Petrion  & light  & redshift    &    Type  & name of SN  \\
     &             &            &            &           &         & radius & fraction &    &    &   \\  \hline
    74     &      010905         &       144516     &      17.27483749  &     14.75579071    &     0422-51811-499      &      8.377  &   0.197  &   0.038    &     Ib  &   2008 fs   \\
    75     &      080313         &       473650     &     120.80256653  &     47.61380005    &     0438-51884-462      &      5.947  &   0.284  &   0.117    &     Ia  &   2000 fy   \\
    76     &      092229         &       575429     &     140.62144470  &     57.90814209    &     0452-51911-319      &      8.549  &   0.153  &   0.062    &     Ia  &   2001 kj   \\
    77     &      022814         &      -083632     &      37.06108475  &     -8.60764790    &     0454-51908-520      &      7.191  &   0.170  &   0.140    &     II  &   2007 tn   \\
    78     &      022904         &      -082414     &      37.26552200  &     -8.40379238    &     0454-51908-559      &      3.933  &   0.369  &   0.140    &     Ia  &   2000 ga   \\
    79     &      031304.193     &      -082354.24  &      48.26750946  &     -8.39855385    &     0459-51924-130      &     10.139  &   0.173  &   0.029    &     Ia  &   2012 eu   \\
    80     &      091138         &      -004254     &     137.90991211  &     -0.71499681    &     0472-51955-247      &      4.138  &   0.305  &   0.070    &     Ia  &   2001 km   \\
    81     &      095153         &       010606     &     147.97111511  &      1.10159636    &     0480-51989-024      &      5.044  &   0.284  &   0.063    &     Ia  &   2001 kp   \\
    82     &      141058         &       645051     &     212.74302673  &     64.84748840    &     0498-51984-102      &      5.747  &   0.294  &   0.140    &     Ia  &   2001 ko   \\
    83     &      132059         &       033556     &     200.24960327  &      3.59741163    &     0526-52312-445      &      5.918  &   0.254  &   0.075    &     II  &   1995 I    \\
    84     &      075646         &       365917     &     119.19379425  &     36.98800659    &     0543-52017-002      &      5.567  &   0.264  &   0.077    &     Ia  &   2001 ks   \\
    85     &      083210         &       471728     &     128.04103088  &     47.29100800    &     0549-51981-379      &      8.624  &   0.213  &   0.133    &     Ia  &   2001 kn   \\
    86     &      103926         &       050524     &     159.86250305  &      5.09011221    &     0577-52367-591      &      3.264  &   0.399  &   0.072    &     Ia  &   2008 iq   \\
    87     &      103929         &       051101     &     159.86888123  &      5.18365812    &     0578-52339-326      &      3.154  &   0.433  &   0.067    &     Ia  &   2006 al   \\
    88     &      110823.953     &       032953.72  &     167.09875488  &      3.50014853    &     0581-52356-130      &     27.924  &   0.409  &   0.024    &     Ia  &   2003 eh   \\
    89     &      161713         &       482828     &     244.30583191  &     48.47438812    &     0622-52054-011      &      3.325  &   0.398  &   0.103    &     Ia  &   2001 kt   \\
    90     &      204757         &      -052425     &     311.98709106  &     -5.40590429    &     0635-52145-489      &      6.541  &   0.167  &   0.046    &     Ia  &   2007 hk   \\
    91     &      001333         &      -101309     &       3.38948441  &    -10.21913052    &     0651-52141-034      &      6.692  &   0.193  &   0.109    &     Ia  &   2002 iu   \\
    92     &      003900.200     &      -090053.20  &       9.75100803  &     -9.01458073    &     0655-52162-368      &      7.745  &   0.210  &   0.020    &     Ia  &   2006 ej   \\
    93     &      004113         &      -090900     &      10.30378246  &     -9.15011024    &     0655-52162-532      &      5.059  &   0.331  &   0.053    &     Ia  &   2009 ly   \\
    94     &      004459         &      -085311     &      11.24682903  &     -8.88971710    &     0656-52148-404      &     11.059  &   0.187  &   0.019    &     Ia  &   2003 im   \\
    95     &      020419.002     &      -084407.24  &      31.07903671  &     -8.73529243    &     0666-52149-638      &     12.432  &   0.152  &   0.017    &     Ia  &   2006 ef   \\
    96     &      215721         &      -075124     &     329.34158325  &     -7.85690928    &     0716-52203-494      &      5.819  &   0.282  &   0.056    &     Ia  &   2002 ep   \\
    97     &      222712         &      -092944     &     336.80273438  &     -9.49496269    &     0720-52206-003      &      8.155  &   0.151  &   0.056    &     Ia  &   1995 T    \\
    98     &      225400         &       143924     &     343.50265503  &     14.65674591    &     0741-52261-425      &      6.949  &   0.194  &   0.038    &     Ia  &   2002 fs   \\
    99     &      230548         &       141956     &     346.45358276  &     14.33142471    &     0742-52263-590      &      7.429  &   0.209  &   0.108    &     Ia  &   2012 ff   \\
   100     &      231225         &       135449     &     348.10827637  &     13.91394138    &     0744-52251-228      &      6.047  &   0.261  &   0.033    &     Ia  &   2007 ob   \\
   101     &      231742         &       135724     &     349.42849731  &     13.95838356    &     0745-52258-297      &      9.153  &   0.164  &   0.055    &     Ia  &   2009 jp   \\
   102     &      232748         &       142827     &     351.95312500  &     14.47530746    &     0746-52238-193      &     14.404  &   0.162  &   0.041    &     Ia  &   2006 da   \\
   103     &      233056.797     &       152925.95  &     352.74032593  &     15.48952103    &     0746-52238-564      &     17.105  &   0.217  &   0.013    &     II  &   2011 ef   \\
   104     &      001911         &       150623     &       4.79587030  &     15.10631943    &     0752-52251-634      &      5.880  &   0.205  &   0.014    &     II  &   1995 ah   \\
   105     &      090932.922     &       501654.84  &     137.38726807  &     50.28185272    &     0766-52247-316      &      9.831  &   0.151  &   0.016    &     Ia  &   2001 G    \\
   106     &      093836.273     &       521742.06  &     144.65115356  &     52.29494476    &     0768-52281-131      &      8.215  &   0.221  &   0.050    &     Ia  &   2003 kz   \\
   107     &      144443         &       585542     &     221.18432617  &     58.92900467    &     0790-52441-620      &      6.428  &   0.199  &   0.044    &     II  &   1999 ay   \\
   108     &      113240.195     &       525701.06  &     173.16769409  &     52.95038605    &     0879-52365-580      &      7.024  &   0.262  &   0.026    &     Ia  &   2002 bn   \\
   109     &      111609         &       552930     &     169.03520203  &     55.49029541    &     0909-52379-214      &      5.677  &   0.258  &   0.057    &     Ia  &   1992 B    \\
   110     &      101857.961     &       462714.72  &     154.74162292  &     46.45409393    &     0944-52614-505      &      7.526  &   0.168  &   0.029    &     Ic  &   2003 ds   \\
   111     &      105741         &       573648     &     164.42208862  &     57.61346436    &     0949-52427-109      &      6.375  &   0.193  &   0.080    &     Ia  &   2010 bg   \\
   112     &      104438         &       574840     &     161.15913391  &     57.81105804    &     0949-52427-310      &      3.274  &   0.428  &   0.118    &     Ia  &   2010 bb   \\
   113     &      203843         &      -002828     &     309.68264771  &     -0.47638249    &     0981-52435-205      &      2.970  &   0.200  &   0.147    &     Ia  &   2006 ex   \\
   114     &      203648         &       000554     &     309.20251465  &      0.09849742    &     0981-52435-334      &      2.383  &   0.511  &   0.146    &     Ia  &   2007 js   \\
   115     &      204711         &      -011526     &     311.79736328  &     -1.25806630    &     0982-52466-198      &      6.175  &   0.221  &   0.057    &     II  &   2007 nw   \\
   116     &      205452         &      -001141     &     313.71878052  &     -0.19580916    &     0983-52443-114      &      3.525  &   0.399  &   0.174    &     Ia  &   2006 ni   \\
   117     &      205209         &      -003040     &     313.03787231  &     -0.51093042    &     0983-52443-183      &      4.430  &   0.281  &   0.070    &     Ia  &   2006 fe   \\
   118     &      205036         &      -002114     &     312.65274048  &     -0.35423487    &     0983-52443-271      &     11.305  &   0.195  &   0.258    &     Ia  &   2005 gh   \\
   119     &      210308         &      -010145     &     315.78494263  &     -1.03117168    &     0984-52442-011      &     11.304  &   0.195  &   0.139    &     Ia  &   2007 hz   \\
   120     &      210959         &       002431     &     317.49566650  &      0.40859288    &     0985-52431-587      &      5.510  &   0.266  &   0.099    &     Ia  &   2006 fs   \\
   121     &      211532         &      -002119     &     318.88247681  &     -0.35459545    &     0986-52443-086      &      4.710  &   0.332  &   0.197    &     Ia  &   2005 kn   \\
   122     &      212837         &       011341     &     322.15667725  &      1.23017097    &     0988-52520-511      &      6.801  &   0.189  &   0.049    &     Ia  &   2006 eq   \\
   123     &      213530         &      -005849     &     323.87850952  &     -0.97963822    &     0989-52468-122      &      3.379  &   0.390  &   0.167    &     Ia  &   2006 fa   \\
   124     &      114811.320     &       545930.19  &     177.04718018  &     54.99180603    &     1016-52759-228      &     11.156  &   0.151  &   0.008    &     II  &   2006 iv   \\
   125     &      131446.547     &       540514.69  &     198.69383240  &     54.08740234    &     1040-52722-488      &      9.380  &   0.204  &   0.033    &     Ia  &   2012 ge   \\
   126     &      141346         &       521317     &     213.44218445  &     52.22148895    &     1045-52725-378      &      5.702  &   0.223  &   0.077    &     Ia  &   2012 dm   \\
   127     &      161743         &       345755     &     244.43147278  &     34.96490860    &     1057-52522-315      &      6.634  &   0.239  &   0.027    &     Ia  &   2006 dw   \\
   128     &      074726         &       265532     &     116.86002350  &     26.92566872    &     1059-52618-003      &      5.498  &   0.232  &   0.015    &     Ic  &   2005 kf   \\
   129     &      161921         &       410523     &     244.84024048  &     41.08987808    &     1171-52753-185      &      8.646  &   0.192  &   0.037    &     Ia  &   2003 lx   \\
   130     &      163214         &       383920     &     248.05726624  &     38.65555954    &     1173-52790-537      &      8.355  &   0.150  &   0.039    &     II  &   2012 ct   \\
   131     &      094628.555     &       454509.12  &     146.61911011  &     45.75260162    &     1202-52672-604      &     10.014  &   0.167  &   0.015    &     Ia  &   2003 jz   \\
   132     &      114154         &       102546     &     175.47506714  &     10.43022156    &     1225-52760-463      &      6.341  &   0.279  &   0.151    &     Ia  &   2009 be   \\
   133     &      123443.547     &       090017.02  &     188.68148804  &      9.00471687    &     1233-52734-250      &      8.235  &   0.224  &   0.043    &     Ia  &   1993 I    \\
   134     &      094530         &       063225     &     146.37681580  &      6.53983545    &     1234-52724-114      &      4.816  &   0.315  &   0.087    &     Ia  &   2009 cj   \\
   135     &      035157         &      -002347     &      57.98822021  &     -0.39662281    &     1242-52901-110      &      3.221  &   0.416  &   0.165    &     II  &   2006 qf   \\
   136     &      091236.164     &       345118.84  &     138.15080261  &     34.85499954    &     1273-52993-477      &      5.963  &   0.171  &   0.061    &     Ia  &   2010 au   \\
   137     &      142556         &       462656     &     216.48419189  &     46.44973755    &     1287-52728-614      &      7.189  &   0.184  &   0.033    &     II  &   2007 fe   \\
   138     &      100544.539     &       101636.23  &     151.43548584  &     10.27677631    &     1308-53053-176      &     11.614  &   0.171  &   0.024    &     Ia  &   2004 ap   \\
   139     &      112851         &       570804     &     172.21507263  &     57.13423920    &     1310-53033-432      &      4.910  &   0.210  &   0.077    &     Ia  &   1999 ce   \\
   140     &      114748.141     &       560106.25  &     176.95053101  &     56.01832962    &     1311-52765-055      &      7.947  &   0.209  &   0.054    &     Ia  &   2000 K    \\
   141     &      134139.609     &       554014.69  &     205.41357422  &     55.66810226    &     1322-52791-198      &     11.339  &   0.730  &   0.025    &     Ic  &   2001 ai   \\
   142     &      162148         &       370341     &     245.45220947  &     37.06080627    &     1337-52767-086      &      6.969  &   0.193  &   0.029    &     Ib  &   2008 fn   \\
   143     &      163320         &       344820     &     248.33615112  &     34.80719757    &     1339-52767-430      &      7.549  &   0.154  &   0.035    &     II  &   1988 Q    \\
   144     &      165222.656     &       304240.00  &     253.09448242  &     30.71103859    &     1343-52790-535      &     10.091  &   0.165  &   0.036    &     Ia  &   2002 di   \\
   145     &      115538         &       442301     &     178.90960693  &     44.38381958    &     1368-53084-097      &      8.224  &   0.154  &   0.023    &     II  &   2006 db   \\
   146     &      115017         &       435745     &     177.57095337  &     43.96262360    &     1368-53084-237      &      2.745  &   0.484  &   0.071    &     Ia  &   2010 kn   \\  \hline
\end{tabular}
\end{center}
\end{table*}

\addtocounter{table}{-1}
\begin{table*}
\begin{center}
\caption{-continued} \label{tab-all.lis} \tiny
\begin{tabular}{lllcccccccc}
\hline
   No.  &  RA (ASC)           & DEC (ASC)           &   RA (SDSS)         &  DEC (SDSS)         &   PID-MJD-FID       & Petrion  & light  & redshift    &    Type  & name of SN  \\
     &             &            &            &           &         & radius & fraction &    &    &   \\  \hline
   147     &      121937         &       460157     &     184.90582275  &     46.03244019    &     1371-52821-325      &      4.400  &   0.252  &   0.057    &     II  &   2010 bd   \\
   148     &      152045         &       364842     &     230.18786621  &     36.81179428    &     1400-53470-234      &      5.800  &   0.293  &   0.103    &     Ia  &   2005 bm   \\
   149     &      165311         &       235754     &     253.29701233  &     23.96512794    &     1424-52912-509      &      3.208  &   0.433  &   0.047    &     Ia  &   2009 fx   \\
   150     &      115345         &       482521     &     178.43844604  &     48.42248535    &     1446-53080-454      &      8.335  &   0.159  &   0.052    &     Ia  &   2008 ac   \\
   151     &      120957         &       470543     &     182.48628235  &     47.09600067    &     1449-53116-070      &      7.033  &   0.228  &   0.031    &     Ia  &   2006 ct   \\
   152     &      122435         &       471416     &     186.14967346  &     47.23744965    &     1451-53117-073      &      5.654  &   0.265  &   0.163    &     Ia  &   2009 co   \\
   153     &      131523         &       462509     &     198.84954834  &     46.42040253    &     1461-53062-166      &      8.732  &   0.162  &   0.056    &     II  &   2009 ct   \\
   154     &      160205         &       294334     &     240.52136230  &     29.72726059    &     1578-53496-421      &      7.051  &   0.175  &   0.014    &     II  &   2007 fz   \\
   155     &      154024         &       325157     &     235.10314941  &     32.86589813    &     1581-53149-470      &      6.074  &   0.267  &   0.053    &     Ia  &   2004 cp   \\
   156     &      102250         &       114211     &     155.70881653  &     11.70301819    &     1598-53033-380      &      6.743  &   0.224  &   0.101    &     Ia  &   2004 cj   \\
   157     &      121621         &       123138     &     184.08850098  &     12.52778912    &     1612-53149-015      &     15.051  &   0.174  &   0.064    &     Ia  &   1990 J    \\
   158     &      121952         &       074349     &     184.96885681  &      7.73122072    &     1625-53140-499      &      8.354  &   0.164  &   0.012    &     II  &   1997 bo   \\
   159     &      122450         &       082557     &     186.20838928  &      8.43370342    &     1626-53472-419      &      7.143  &   0.174  &   0.090    &     Ic  &   2009 bh   \\
   160     &      142355         &       351105     &     215.98291016  &     35.18551636    &     1644-53144-167      &      6.184  &   0.179  &   0.055    &     Ia  &   2009 av   \\
   161     &      145000         &       445505     &     222.49903870  &     44.91716766    &     1675-53466-081      &      5.545  &   0.235  &   0.080    &     Ia  &   2009 fb   \\
   162     &      152748         &       413534     &     231.95225525  &     41.59374619    &     1679-53149-427      &      5.938  &   0.168  &   0.081    &     Ia  &   2009 fc   \\
   163     &      160209         &       364308     &     240.53713989  &     36.72050858    &     1682-53173-453      &      4.816  &   0.237  &   0.094    &     Ia  &   2001 bp   \\
   164     &      170925.078     &       221250.45  &     257.35443115  &     22.21404839    &     1688-53462-009      &      2.413  &   0.614  &   0.048    &     Ia  &   2010 ed   \\
   165     &      170007         &       230756     &     255.02867126  &     23.13151169    &     1688-53462-310      &      7.485  &   0.220  &   0.056    &     Ia  &   2008 eq   \\
   166     &      133238         &       114833     &     203.16078186  &     11.80923462    &     1700-53502-359      &      4.738  &   0.265  &   0.150    &     Ia  &   2005 ca   \\
   167     &      145643         &       091942     &     224.18005371  &      9.32679081    &     1715-54212-190      &      7.902  &   0.207  &   0.079    &     Ia  &   2005 ag   \\
   168     &      151836         &       095117     &     229.65385437  &      9.85440731    &     1720-53854-255      &      5.534  &   0.228  &   0.032    &     II  &   2007 dp   \\
   169     &      161412         &       060904     &     243.54829407  &      6.15103197    &     1731-53884-181      &      8.926  &   0.190  &   0.039    &     Ia  &   2000 df   \\
   170     &      082933         &       085205     &     127.38945770  &      8.86817551    &     1758-53084-523      &      4.566  &   0.329  &   0.112    &     Ia  &   2004 cl   \\
   171     &      120940         &       161212     &     182.41508484  &     16.20340538    &     1765-53466-368      &      4.339  &   0.296  &   0.076    &     Ia  &   2013 Y    \\
   172     &      090456.859     &       595558.69  &     136.23130798  &     59.93263245    &     1785-54439-632      &      2.454  &   0.746  &   0.005    &     Ic  &   1995 F    \\
   173     &      152038         &       073932     &     230.15852356  &      7.65969324    &     1818-54539-508      &      4.746  &   0.336  &   0.045    &     Ia  &   2009 eh   \\
   174     &      153452         &       070053     &     233.71890259  &      7.01332378    &     1820-54208-481      &      6.466  &   0.273  &   0.074    &     II  &   2007 ed   \\
   175     &      155113.266     &       254207.41  &     237.80531311  &     25.70191193    &     1850-53786-555      &     12.155  &   0.150  &   0.021    &     Ia  &   2009 dc   \\
   176     &      162034         &       211208     &     245.14274597  &     21.20253754    &     1853-53566-076      &      7.988  &   0.219  &   0.032    &     Ia  &   2011 bk   \\
   177     &      074837         &       521322     &     117.15312958  &     52.22221375    &     1869-53327-355      &      9.383  &   0.197  &   0.064    &     Ia  &   1997 ea   \\
   178     &      092822         &       272640     &     142.09411621  &     27.44466400    &     1940-53383-478      &      3.800  &   0.372  &   0.032    &     Ia  &   2003 ae   \\
   179     &      095847.523     &       344709.59  &     149.69892883  &     34.78640747    &     1948-53388-558      &      2.532  &   0.807  &   0.017    &     Ia  &   2011 hd   \\
   180     &      103610         &       343233     &     159.04226685  &     34.54280472    &     1982-53436-445      &      8.805  &   0.176  &   0.051    &     Ia  &   2007 do   \\
   181     &      124538.609     &       350501.78  &     191.41104126  &     35.08375549    &     1987-53765-329      &     14.577  &   0.173  &   0.032    &     Ia  &   2006 S    \\
   182     &      130502         &       284424     &     196.26057434  &     28.73899651    &     2011-53499-020      &      7.033  &   0.238  &   0.027    &     Ia  &   2006 cg   \\
   183     &      111229         &       312305     &     168.12567139  &     31.38496208    &     2092-53460-516      &      6.309  &   0.176  &   0.027    &     Ib  &   2011 bp   \\
   184     &      120323         &       351933     &     180.84963989  &     35.32583237    &     2103-53467-081      &      4.156  &   0.321  &   0.028    &     II  &   2005 bn   \\
   185     &      131651.141     &       313452.62  &     199.21447754  &     31.57986069    &     2104-53852-427      &     13.090  &   0.191  &   0.029    &     Ia  &   2004 E    \\
   186     &      135806.094     &       282522.12  &     209.52339172  &     28.42267036    &     2118-53820-535      &      8.346  &   0.454  &   0.026    &     II  &   2000 ck   \\
   187     &      144823         &       214751     &     222.09696960  &     21.79764938    &     2144-53770-215      &      5.419  &   0.277  &   0.155    &     Ia  &   2006 ae   \\
   188     &      150030.203     &       235545.91  &     225.12585449  &     23.92932510    &     2152-53874-280      &     10.522  &   0.181  &   0.047    &     Ia  &   2012 af   \\
   189     &      160527.281     &       174951.84  &     241.36349487  &     17.83099174    &     2199-53556-118      &      9.340  &   0.180  &   0.034    &     Ia  &   2000 cp   \\
   190     &      111821         &       281243     &     169.59053040  &     28.20896530    &     2215-53793-088      &      9.862  &   0.173  &   0.068    &     Ia  &   2006 bm   \\
   191     &      111158         &       294205     &     167.99351501  &     29.69836235    &     2215-53793-350      &      3.690  &   0.376  &   0.055    &     Ia  &   2012 ax   \\
   192     &      124937         &       281946     &     192.40368652  &     28.32912064    &     2238-54205-614      &      4.022  &   0.365  &   0.055    &     Ia  &   2008 ad   \\
   193     &      125541.328     &       271502.59  &     193.92207336  &     27.25076485    &     2240-53823-200      &      8.830  &   0.199  &   0.023    &     Ia  &   2001 cg   \\
   194     &      125924         &       282051     &     194.85052490  &     28.34720230    &     2240-53823-566      &      5.998  &   0.247  &   0.067    &     Ia  &   2006 cj   \\
   195     &      125925         &       275948     &     194.85437012  &     27.99674225    &     2241-54169-481      &      7.133  &   0.207  &   0.018    &     Ia  &   2010 ai   \\
   196     &      093606         &       242413     &     144.02465820  &     24.40494728    &     2294-53733-006      &      7.692  &   0.169  &   0.070    &     Ia  &   2005 kw   \\
   197     &      100006         &       281652     &     150.02825928  &     28.28054428    &     2345-53757-472      &     18.021  &   0.157  &   0.089    &     Ia  &   2003 bh   \\
   198     &      103357         &       202025     &     158.48828125  &     20.34046555    &     2376-53770-183      &      4.666  &   0.336  &   0.087    &     Ia  &   2006 af   \\
   199     &      115004         &       211647     &     177.51690674  &     21.27984238    &     2511-53882-115      &      6.521  &   0.274  &   0.025    &     Ia  &   2013 N    \\
   200     &      114328         &       214030     &     175.86759949  &     21.67408752    &     2511-53882-247      &      8.222  &   0.198  &   0.068    &     Ia  &   2004 Y    \\
   201     &      115614         &       252111     &     179.06001282  &     25.35412598    &     2514-53882-549      &      8.475  &   0.176  &   0.032    &     Ia  &   2011 bg   \\
   202     &      162345.031     &       094716.92  &     245.93763733  &      9.78810501    &     2532-54589-360      &      8.316  &   0.216  &   0.034    &     Ia  &   2012 ds   \\
   203     &      100633.289     &       142601.20  &     151.63864136  &     14.43352890    &     2586-54169-158      &      9.686  &   0.180  &   0.029    &     II  &   2013 W    \\
   204     &      100919         &       145932     &     152.33206177  &     14.99094009    &     2586-54169-567      &      9.483  &   0.154  &   0.030    &     Ia  &   2007 ux   \\
   205     &      102311         &       175906     &     155.79826355  &     17.98383331    &     2589-54174-541      &      8.568  &   0.211  &   0.027    &     Ia  &   1999 at   \\
   206     &      123936         &       163516     &     189.89941406  &     16.58782768    &     2599-54234-006      &      3.961  &   0.353  &   0.025    &     Ia  &   2012 G    \\
   207     &      130950         &       204442     &     197.46063232  &     20.74431992    &     2617-54502-531      &      5.778  &   0.210  &   0.095    &     Ia  &   2007 eg   \\
   208     &      145055         &       171312     &     222.73104858  &     17.21818733    &     2777-54554-132      &      7.295  &   0.159  &   0.040    &     II  &   2004 V    \\
   209     &      144747         &       164953     &     221.94726562  &     16.83006668    &     2777-54554-289      &      9.755  &   0.166  &   0.045    &     Ia  &   2007 ee   \\
   210     &      144755.609     &       190326.09  &     221.98185730  &     19.05737495    &     2777-54554-331      &      6.431  &   0.252  &   0.042    &     Ia  &   2010 cs   \\
   211     &      145415         &       185752     &     223.56640625  &     18.96429634    &     2778-54539-336      &      5.738  &   0.316  &   0.058    &     Ia  &   2009 eg   \\
   212     &      140051         &       225728     &     210.21182251  &     22.95813370    &     2784-54529-473      &      7.191  &   0.165  &   0.083    &     Ia  &   2007 dv   \\
   213     &      122248         &       053624     &     185.69837952  &      5.60674953    &     2880-54509-413      &      7.857  &   0.221  &   0.017    &     II  &   2012 ab   \\  \hline
\end{tabular}
\end{center}
\end{table*}

\end{document}